\documentclass[11pt]{article}
\usepackage{amsfonts,amsmath,amssymb}
\usepackage{enumerate}
\usepackage{hyperref}
\usepackage{bbm}
\usepackage{nicefrac}
\usepackage[all]{xy}
\usepackage{graphicx}
\usepackage{bm}

\usepackage{upgreek}

\usepackage{booktabs}

\newcommand{\be}{\begin{equation}}
\newcommand{\ee}{\end{equation}}

\newcommand{\bea}{\begin{eqnarray}}
\newcommand{\eea}{\end{eqnarray}}

\newcommand{\bes}{\begin{subequations}}
\newcommand{\ees}{\end{subequations}}

\newcommand{\cN}{{\cal N}}
\newcommand{\cA}{{\cal A}}

\def\ft#1#2{{\textstyle{{\scriptstyle #1}\over {\scriptstyle #2}}}}
\def\fft#1#2{{#1 \over #2}}

\def\sst#1{{\scriptscriptstyle #1}}
\def\oneone{\rlap 1\mkern4mu{\rm l}}

\def\0{{\sst{(0)}}}
\def\1{{\sst{(1)}}}
\def\2{{\sst{(2)}}}
\def\3{{\sst{(3)}}}
\def\4{{\sst{(4)}}}
\def\5{{\sst{(5)}}}
\def\6{{\sst{(6)}}}
\def\7{{\sst{(7)}}}
\def\8{{\sst{(8)}}}

\def\cA{{{\cal A}}}
\def\cB{{{\cal B}}}
\def\cC{{{\cal C}}}
\def\cH{{{\cal H}}}
\def\cV{{{\cal V}}}
\def\cM{{{\cal M}}}

\usepackage{multirow}
\usepackage{rotating}

\allowdisplaybreaks

\begin{document}

\makeatletter
\renewcommand{\theequation}{\thesection.\arabic{equation}}
\@addtoreset{equation}{section}
\makeatother

\begin{titlepage}

\begin{flushright}
Nikhef-2015-032 \\
CPHT-RR026.0815 
\end{flushright}

\vspace{25pt}

   \begin{center}
   \baselineskip=16pt
   \begin{Large}\textbf{
Consistent $\cN=8$ truncation of massive IIA on $S^6$}
   \end{Large}

\vspace{25pt}
		
{Adolfo Guarino$^{\flat}$ \, and \, Oscar Varela$^{\sharp}$}
		
\vspace{25pt}

	\begin{small}

	{\it ${}^\flat$ Nikhef Theory Group, Science Park 105, 1098 XG Amsterdam, The Netherlands } \\
	aguarino@nikhef.nl

	\vspace{15pt}
          
   {\it ${}^\sharp$	Center for the Fundamental Laws of Nature,\\
	Harvard University, Cambridge, MA 02138, USA } \\
	ovarela@physics.harvard.edu
	
	\vspace{15pt}
	
  {\it ${}^\sharp$	Centre de Physique Th\'eorique, Ecole Polytechnique, CNRS UMR 7644,\\
	91128 Palaiseau Cedex, France }

	\end{small}

\vskip 50pt

\end{center}

\begin{center}
\textbf{Abstract}
\end{center}

\begin{quote}

Massive type IIA supergravity is shown to admit a consistent truncation on the six-sphere to maximal supergravity in four dimensions with a dyonic ISO(7) gauging. We obtain the complete, non-linear embedding of all the $D=4$ fields into the IIA metric and form potentials, and show its consistency. We first rewrite the IIA theory in an $\textrm{SO}(1,3) \times \textrm{SL}(7)$--covariant way. Then, we employ an $\cN=8$ SL(7)--covariant restriction of the  $D=4$ tensor hierarchy in order to find the full embedding. The redundant $D=4$ degrees of freedom introduced by the tensor hierarchy  can be eliminated by writing the embedding in terms of the field strengths and exploiting the restricted duality hierarchy. In particular, closed expressions for the Freund-Rubin term are found using this technique which reveal a pattern valid for other truncations. Finally, we show that the present $\cN=8$ truncation of massive IIA on $S^6$ and the ${\cN=2}$ truncation obtained when $S^6$ is equipped with its nearly-K\"ahler structure, overlap in the $\cN=1$, G$_2$--invariant sector of the former.

\end{quote}

\vfill

\end{titlepage}

\tableofcontents


\section{Introduction}

Gauged supergravities with anti-de-Sitter (AdS) vacua provide a valuable tool to study holographically the strongly coupled behaviour of superconformal field theories at large $N$. In this regard, the SO(8) \cite{deWit:1982ig}, SO(5) \cite{Pernici:1984xx} and SO(6) \cite{Gunaydin:1984qu} gaugings of maximal supergravity in $D=4$, $D=7$ and $D=5$ dimensions occupy a prominent place in that they are holographically related to the M2, M5 and D3 brane worldvolume superconformal field theories. Crucially, these gaugings respectively arise as consistent truncations of $D=11$ supergravity on $S^7$ \cite{deWit:1986iy,deWit:2013ija} and $S^4$ \cite{Nastase:1999cb,Nastase:1999kf}, and type IIB on $S^5$, see  \cite{Lee:2014mla,Ciceri:2014wya,Baguet:2015sma}. By virtue of the consistency of these truncations, these maximal gauged supergravities describe the complete, nonlinear interactions of a fininite number of modes of M-theory or type IIB on the Freund-Rubin backgrounds \cite{Freund:1980xh} $\textrm{AdS}_4 \times S^7$, $\textrm{AdS}_7 \times S^4$ and $\textrm{AdS}_5 \times S^5$, respectively, in a consistently decoupled way from other Kaluza-Klein (KK) modes. The modes that are kept in these maximally supersymmetric truncations lie at the bottom of the KK towers of $D=11$ or type IIB supergravity on these Freund-Rubin AdS vacua or, in fact, any other AdS vacuum, as {\it e.g.}~the spectrum \cite{Klebanov:2008vq} of M-theory on the warped $\textrm{AdS}_4 \times S^7$ $\cN=2$  background of \cite{Corrado:2001nv} suggests. To signify this, these are usually referred to as `massless mode' truncations, although the retained modes do not typically have zero physical mass. 

Various analyses from different perspectives single out, under given assumptions, these $(D,n) = (11,7), (11,4), (\textrm{IIB},5)$ `massless mode' truncations amongst all possible compactifications of $D=10$ or $D=11$ supergravity on $S^n$ \cite{Nastase:2000tu,Cvetic:2000dm,Lee:2014mla,Hohm:2014qga}. These results resonate with well-known facts, including that only the worldvolumes of the M2, M5 and D3 branes can support a maximally supersymmetric conformal field theory, or that $\textrm{AdS}_4 \times S^7$, $\textrm{AdS}_7 \times S^4$ and $\textrm{AdS}_5 \times S^5$ is the complete list of maximally supersymmetric Freund-Rubin solutions in $D=11$ and $D=10$ \cite{FigueroaO'Farrill:2002ft}. These three `massless mode' truncations thus fulfill the expectation that every supersymmetric string/M-theory background containing an $\cN$-supersymmetric AdS$_D$ factor should have an associated consistent truncation to a $D$-dimensional, $\cN$-extended gauged supergravity with that AdS$_D$ as one of its vacua \cite{Gauntlett:2007ma}.

For these reasons, it may come as a surprise that, as we already announced in recent work \cite{Guarino:2015jca}, massive type IIA supergravity \cite{Romans:1985tz} also admits a consistent truncation on the six-sphere $S^6$ to maximal, $\cN=8$, supergravity in four dimensions. The relevant $D=4$ gauge group is the non-semisimple group $\textrm{ISO}(7) = \textrm{SO}(7) \ltimes \mathbb{R}^7$, and the gauging is of the dyonic type recently discussed in \cite{Dall'Agata:2012bb,Dall'Agata:2014ita}. See \cite{Guarino:2015qaa} for the explicit construction of the dyonic ISO(7)-gauged supergravity. Similarly to the $S^7$ truncation of $D=11$ supergravity to the SO(8) gauging, SO(7) gauges electrically the isometries of the internal round $S^6$. In addition, the $\mathbb{R}^7$ translations now gauge dyonically shift symmetries of the IIA forms. Unlike the $\cN=8$ $D=4$ SO(8) \cite{deWit:1982ig} and $D=5$ SO(6) \cite{Gunaydin:1984qu} gaugings, the dyonic ISO(7) gauging does not admit an $\cN=8$ AdS vacuum that can possibly uplift to a maximally supersymmetric Freund-Rubin background $\textrm{AdS}_4 \times S^6$ of massive type IIA. Yet, a  consistent truncation at the level of the supergravities does exist which is, moreover, maximally supersymmetric. Of course, this does not contradict the statement of \cite{Gauntlett:2007ma}.

The intuition that biunivocally associates maximally supersymmetric AdS vacua of string or M-theory to maximally supersymmetric consistent truncations is therefore not always correct. In fact, some maximally supersymmetric truncations with no maximally supersymmetric vacuum are either known or have been conjectured. An example closely related to ours is the conjectured truncation \cite{Hull:1988jw} of massless type IIA on $S^6$ to the $D=4$ purely electric ISO(7) gauging \cite{Hull:1984yy}. The latter theory does not have any vacuum that can possibly uplift on $S^6$ to any AdS$_4$ solution of massless type IIA. Our truncation is valid for all values, finite or zero, of the Romans mass $ \hat F_\0 \equiv m$. The reason for this is that we establish consistency at the level of the supersymmetry variations of the IIA bosons, and these are independent of $m$. Thus, we also give a proof of the conjecture of \cite{Hull:1988jw}. As announced in \cite{Guarino:2015jca}, the $D=10$ Romans mass descends to the $D=4$ magnetic gauge coupling, and this only enters the bosonic field content through the field strengths and covariant derivatives.

In \cite{Guarino:2015jca}, the embedding of $D=4$ $\cN=8$ ISO(7) supergravity into the type IIA metric and all form potentials except the Ramond-Ramond three-form $\hat A_\3$ was given. In this paper, we give the complete embedding, including the embedding into $\hat A_\3$, and offer a detailed derivation. We should stress that we give the full non-linear dependence of the IIA metric, dilaton, and potentials on the entire set of $D=4$ ISO(7) supergravity fields, not only on the $D=4$ scalars. That is, we also obtain the full non-linear dependence of the $D=10$ fields on the $D=4$ vectors and the two-forms required by the magnetic gauging.  In comparison, for the $D=11$ embedding of the purely electric SO(8) gauging \cite{deWit:1982ig}, only the full non-linear dependence of the $D=11$ metric \cite{deWit:1986iy,deWit:1984nz} and three-form \cite{deWit:2013ija} on the $\textrm{E}_{7(7)}/\textrm{SU}(8)$ scalars has been obtained. The $D=11$ embedding of the SO(8) vectors has only been determined at the linear level \cite{deWit:1986iy} (see also \cite{Godazgar:2013pfa}). The linear dependence of the $D=10$ or $D=11$ metric on the $D=4$ (KK) vectors is actually exact, but the latter also enter the $D=10$ or $D=11$ form fields in a non-linear way prescribed by the non-Abelian character of the $D=4$ gauge group, ISO(7) or SO(8). For type IIB on $S^5$, only recently it has been obtained the full non-linear dependence of the IIB fields on the $D=5$ scalars \cite{Lee:2014mla,Ciceri:2014wya,Baguet:2015sma} (confirming previous ansatze \cite{Khavaev:1998fb,Pilch:2000ue}) and vectors and two-forms \cite{Ciceri:2014wya,Baguet:2015sma} of the SO(6) gauging \cite{Gunaydin:1984qu}.

Two steps are crucial, in the approach we follow, in order to determine the full non-linear embedding of the entire ISO(7) supergravity in type IIA. The first is to perform a rewrite of the IIA theory with only $D=4$ local Lorentz symmetry manifest, see section \ref{app:DerivationUplift}. This step is a IIA adaptation of the time-honoured de Wit-Nicolai approach to ${D=11}$ \cite{deWit:1986mz,deWit:2013ija}, and involves exclusively type IIA manipulations independent, in particular, of any $D=4$ gauging. The relevant rewrite involves $\textrm{SO}(1,3) \times \textrm{SL}(7)$--covariant ten-dimensional bosons and $\textrm{SO}(1,3) \times \textrm{SU}(8)$--covariant ten-dimensional fermions, exactly as in the $D=11$ case \cite{deWit:1986mz,deWit:2013ija} and similarly to the type IIB case \cite{Ciceri:2014wya}. As first discussed in a type IIB context \cite{Ciceri:2014wya}, further non-linear field redefinitions are necessary so that the $\textrm{SO}(1,3) \times \textrm{SL}(7)$--covariant bosons acquire supersymmetry transformations compatible with those dictated by (a relevant SL(7)-branching of) the $D=4$ tensor hierarchy \cite{deWit:2007mt,deWit:2008ta,deWit:2008gc}. Accordingly, we work out the hierarchy-compatible supersymmetry variations of the further redefined $\textrm{SO}(1,3) \times \textrm{SL}(7)$--covariant vector, two-form and three-form potentials of type IIA. In this way, we obtain a reformulation of type IIA that is adapted to accommodate the full, non-linear embedding of all fields in any gauging of $D=4$ $\cN=8$ supergravity that can possibly arise by consistent truncation. See  (\ref{eq:SL7fieldcontent}) for the type IIA $\textrm{SO}(1,3) \times \textrm{SL}(7)$--covariant bosonic field content. Of course, determining all possible such truncations and establishing their consistency is an altogether different matter. Here we will focus on the truncation to the ISO(7) gauging, including the non-linear dependence on all $D=4$ fields. 

The second, now gauging-dependent, step is new. The full IIA embedding of the ISO(7) gauging is naturally expressed in terms of a consistent subsector, still $\cN=8$ but only SL(7)--covariant, of the $D=4$ tensor \cite{deWit:2008ta,deWit:2008gc} and duality \cite{Bergshoeff:2009ph} hierarchies, with closed field equations and supersymmetry transformations \cite{Guarino:2015qaa}. See section \ref{sec:S6trunc}. This subsector retains all vectors, only certain two-form and three-form potentials in SL(7) representations and no four-form potentials: see equation (\ref{eq:SL7fieldcontent4D}). This SL(7)--covariant field content is different, however, to that arising in the SL(7)--covariant IIA reformulation of step one above. The relation between both field contents is given by the `KK ansatz', whereby the SL(7)--covariant IIA fields are linearly expressed in terms of the $D=4$ fields in the SL(7) tensor hierarchy and tensors on $S^6$ in suitable intertwining representations of SL(7). Inserting this KK ansatz into the SL(7)--covariant IIA supersymmetry variations, all $S^6$ dependence drops out and the supersymmetry variations of the $D=4$ SL(7) fields arise. This shows the consistency of the truncation. Although the KK ansatz relates linearly the SL(7) fields in $D=10$ and $D=4$, when the non-linear redefinitions that led to the SL(7)--covariant reformulation of type IIA are undone, the original, locally $D=10$-Lorentz-covariant IIA fields develop a full non-linear dependence on the $D=4$ ones: see equation (\ref{KKEmbedding}). 

This construction naturally yields the embedding of the $D=4$ theory at the level, on the one hand, of the $\cN=8$ SL(7) consistent subsector \cite{Guarino:2015qaa}  of the tensor hierarchy  \cite{deWit:2008ta,deWit:2008gc}  for the ISO(7) gauging and, on the other hand, at the level of the IIA metric, dilaton and form potentials. Such embedding is thus expressed in terms of redundant $D=4$ degrees of freedom contained in the tensor hierarchy and is not gauge-independent in $D=10$. It is therefore interesting to also consider the embedding of the $D=4$ theory at the level of the IIA field strengths. This automatically gives gauge-independent expressions in $D=10$. It also allows one to express the embedding in terms of independent $D=4$ degrees of freedom by exploiting the SL(7)--restricted  \cite{Guarino:2015qaa} duality hierarchy \cite{Bergshoeff:2009ph}. For example, the four-form field strengths of the three-form potentials in the $D=4$ hierarchy become dualised into functions on the $\cN=8$ scalar manifold. This dualisation allows us, in particular, to obtain an elegant formula for the Freund-Rubin term, $\hat F_\4 = U \, \textrm{vol}_4 + \ldots$, where
\begin{equation}
\begin{array}{ccl}
\label{U_generalRewriteIntro}
U & = & - \dfrac{g}{84} \,  {X^\prime_{\mathbb{MP}}}^{\mathbb{R}}  {X_{\mathbb{NQ}}}^{\mathbb{S}}  \mathcal{M}^{\mathbb{MN}}   \Big(  \mathcal{M}^{\mathbb{PQ}}  \mathcal{M}_{\mathbb{RS}}  +   7 \,   \delta^\mathbb{P}_\mathbb{S} \,  \delta^\mathbb{Q}_\mathbb{R}  \Big) \ 
\end{array}
\end{equation}
is a function on E$_{7(7)}/$SU(8) (through the scalar matrix $\mathcal{M}_{\mathbb{MN}}$ and its inverse), of the ISO(7) embedding tensor (through the conventional symbols ${X_{\mathbb{NQ}}}^{\mathbb{S}}  $) and on $S^6$ (through the symbols $ {X^\prime_{\mathbb{MP}}}^{\mathbb{R}}$ defined in section \ref{sec:S6truncConsistencyFS}). This expression closely parallels the scalar potential of $D=4$ $\cN=8$ gauged supergravity \cite{deWit:2007mt}
\begin{equation}
\begin{array}{ccl}
\label{V_generalRewriteIntro}
V & = & \dfrac{g^{2}}{168} \,  {X_{\mathbb{MP}}}^{\mathbb{R}}  {X_{\mathbb{NQ}}}^{\mathbb{S}}  \mathcal{M}^{\mathbb{MN}}   \Big(  \mathcal{M}^{\mathbb{PQ}}  \mathcal{M}_{\mathbb{RS}}  +   7 \,   \delta^\mathbb{P}_\mathbb{S} \,  \delta^\mathbb{Q}_\mathbb{R}  \Big) \ ,
\end{array}
\end{equation}
particularised to the ISO(7) gauging. In section \ref{sec:S6truncConsistencyFS} we will consider these dualisations in full generality. In section \ref{app:G2UpliftSubsec} we will work out the explicit embedding and dualisations of the G$_2$--invariant sector of the $\cN=8$ ISO(7) theory, for which an explicit parameterisation of the scalar manifold becomes available \cite{Guarino:2015qaa}.

The $\cN=8$ truncation of massive IIA on $S^6$ can be also regarded as `massless mode', in the sense discussed above. Besides these `massless mode' reductions, other truncations of $D=11$ supergravity on $S^7$, type IIB on $S^5$ and massive IIA on $S^6$ are also known to gauged supergravities with less than maximal supersymmetry. These retain all possible fields along the forms that define the natural constant torsion $G$-structures on the corresponding sphere: weak G$_2$, Sasaki-Einstein \cite{Gauntlett:2009zw} and tri-Sasaki \cite{Cassani:2011fu} in the $S^7$ case, Sasaki-Einstein \cite{Cassani:2010uw,Gauntlett:2010vu,Liu:2010sa} in the $S^5$ case and nearly-K\"ahler \cite{KashaniPoor:2007tr} in the $S^6$ case. See also \cite{Bah:2010yt,Bah:2010cu,Liu:2010pq} for the extension of some of these truncations to the fermion sector. Equivalently, these truncations can be seen to retain all the invariant fields under the homogeneous action of a transitively acting group, see \cite{Cassani:2009ck,Cassani:2011fu,Cassani:2012pj}: for example, G$_2$ in the nearly-K\"ahler truncation on $S^6 = \textrm{G}_2/\textrm{SU}(3)$ \cite{Cassani:2009ck}. Since they only depend  on generic features of the relevant $G$-structure, these truncations are in fact universally valid for all spaces equipped with the same structure, and not only for spheres. In contrast to the $\cN=8$ truncations above, these constant torsion $G$-structure truncations tend to retain some massive modes up the KK towers above their AdS vacua, along with some, but not all, modes at the bottom of the towers. For this reason, these are sometimes referred to as `massive mode' truncations. 

Both types of truncations on $S^7$ and $S^5$, `massless' and `massive', have an overlapping sector: the sector of the $\cN=8$ `massless' theory that is invariant under the same group as the $G$-structure. This is also the sector of the $\cN=8$ theory whose uplift inflicts only homogeneous deformations on $S^7$ and $S^5$. The mass spectra above the AdS vacua in this overlapping sector contain the common modes at the bottom of the KK towers. For example, $D=4$ $\cN=8$ electrically gauged SO(8) supergravity \cite{deWit:1982ig} and the universal $\cN=2$ (skew-whiffed) Sasaki-Einstein truncation of $D=11$ supergravity on $S^7$ \cite{Gauntlett:2009zw} overlap \cite{Bobev:2010ib} in the SU(4)$_-$--invariant sector of the $\cN=8$ theory. Likewise, $D=5$ SO(6)-gauged supergravity \cite{Gunaydin:1984qu} and the universal $\cN=4$ Sasaki-Einstein truncation of type IIB on $S^5$ \cite{Cassani:2010uw,Gauntlett:2010vu} overlap \cite{Suh:2011xc} in the SU(3)-invariant sector of the $\cN=8$ theory. In section \ref{app:mIIAon NK6} we will show that, similarly, $D=4$ $\cN=8$ dyonically gauged ISO(7) supergravity and the universal $\cN=2$ nearly-K\"ahler truncation \cite{KashaniPoor:2007tr} of massive type IIA on $S^6$ overlap in the G$_2$--invariant sector \cite{Guarino:2015qaa} of the $\cN=8$ theory.


\section{\mbox{Type IIA with only $\textrm{SO}(1,3)$ symmetry manifest}}
\label{app:DerivationUplift}

In this section, we rewrite the type IIA bosonic and fermionic field content with manifest $\textrm{SO}(1,3) \times \textrm{SL}(7)$ and $\textrm{SO}(1,3) \times \textrm{SU}(8)$ covariance, respectively. We also perform the non-linear field redefinitions necessary to render the SL(7)--covariant supersymmetry transformations of the bosonic fields compatible with the $D=4$ tensor hierarchy.

\vspace{6pt} 


\subsection{\mbox{Redefinitions that leave $D=4$ spacetime symmetry manifest}} \label{4DLorentzManifest}

We start by splitting the ten-dimensional local Lorentz symmetry as\footnote{The global SO$(1,1)$ of type IIA does not play a significant role in the following. For this reason, we do not keep track of charges under this SO$(1,1)$.}
\begin{eqnarray} \label{LorentzSplit}
\textrm{SO}(1,9) \rightarrow \textrm{SO}(1,3) \times \textrm{SO}(6) \; , 
\end{eqnarray}
so that the ten-dimensional coordinates split as $(x^\mu, y^m)$, $\mu = 0 , 1, 2, 3$, $m = 1, \ldots , 6$. For the bosons, we straightforwardly promote SO(6) to SL(6). Then, the bosonic fields $ d\hat{s}_{10}^2 , \hat A_\3  , \hat B_\2 , \hat A_\1 , \hat \phi $  of the type IIA theory (see appendix \ref{app:mIIAConventions}) give rise, via the standard decompositions
\begin{eqnarray} \label{KKFields}
d\hat{s}_{10}^2 &=&  \Delta^{-1} \, ds_4^2  \, + g_{mn} \big( dy^m + B^m \big)  \big( dy^n + B^n \big) \; , \nonumber \\[8pt]
\hat A_\3 &=&  \tfrac16 A_{\mu\nu\rho} \, dx^{\mu} \wedge dx^{\nu} \wedge dx^{\rho}  + \tfrac12  A_{\mu \nu m } \, dx^\mu \wedge dx^\nu \wedge \big( dy^m + B^m \big) \nonumber \\
&&  + \,  \tfrac12 A_{\mu mn}  \, dx^{\mu}   \wedge \big( dy^m + B^m \big) \wedge \big( dy^n + B^n \big) \nonumber \\
&& + \,  \tfrac16 A_{mnp} \big( dy^m + B^m \big) \wedge \big( dy^n + B^n \big) \wedge \big( dy^p + B^p \big) \; , \nonumber  \\[8pt]
\hat B_\2 &=& \tfrac12 B_{\mu\nu} \, dx^{\mu} \wedge dx^{\nu}  + B_{\mu m }  \, dx^\mu  \wedge \big( dy^m + B^m \big) + \tfrac12 B_{mn} \big( dy^m + B^m \big) \wedge \big( dy^n + B^n \big)  \; , \nonumber  \\[8pt]
\hat A_\1 &=& A_\mu  \, dx^{\mu} + A_{m }  \big( dy^m + B^m \big) \; , 
\end{eqnarray}
to the following local SO$(1,3)$ fields in either singlet or non-trivial representations of SL(6): 
\begin{eqnarray} \label{eq:SL6fieldcontent}
\bm{1} &  \textrm{metric}:  &  \; ds_4^2 \; , \nonumber \\
\bm{21} + \bm{6} + \bm{1}   + \bm{20} + \bm{15} &  \textrm{scalars}:  &   \; g_{mn}  \;, \; A_m \;, \; \hat \phi  \;, \;  A_{mnp}  \;, \;  B_{mn}   \; ,  \nonumber \\
\bm{6}^\prime + \bm{1} + \bm{15} + \bm{6} &  \textrm{vectors}:  &   \; B_\mu{}^m \;, \; A_\mu  \;, \;  A_{\mu mn}  \;, \; B_{\mu m}  \; ,  \nonumber \\
\bm{6} + \bm{1} &  \textrm{two-forms}:  &   \; A_{\mu \nu m} \;, \; B_{\mu \nu} \; , \nonumber \\
\bm{1} &  \textrm{three-form}:  &   \; A_{\mu \nu \rho} \; .
\end{eqnarray}
Following convention, in (\ref{KKFields}) we have introduced
\begin{eqnarray}
\Delta^2 \equiv \frac{ \det \, g_{mn} } {  \det \, \mathring g_{mn}  } \; ,
\end{eqnarray}
for some background internal metric $ \mathring g_{mn}(y)$. 

The field content of $D=4$ maximal supergravity is thus formally recovered after dualising the two-forms into scalars and dropping the three-form on the grounds that it should not carry independent degrees of freedom. Note, however, that all fields in (\ref{eq:SL6fieldcontent}) depend on all ten-dimensional coordinates $(x^\mu , y^m)$, and that both the two-forms and the three-form will play a role in this paper. The field content (\ref{eq:SL6fieldcontent}) can be allocated into SL(7) representations as well, reflecting a rewrite of $D=11$ supergravity with only SO$(1,3)$ symmetry manifest. In any case, neither the SL(6)--covariant fields (\ref{eq:SL6fieldcontent}) nor the SL(7)--covariant fields they can be grouped into, are generally compatible with the $D=4$ tensor hierarchy. We leave this discussion for section \ref{subsec:TensorHierarchy}.

In the fermionic sector, we promote the local SO(6) in the splitting (\ref{LorentzSplit}) (or rather, its cover Spin$(6)$) to SU(8), the R-symmetry group of $D=4$ $\cN=8$ supergravity. In order to do this,  we first introduce the following primed fermions from the IIA fermions $\hat \psi_M $, $\hat \lambda$, $\hat \epsilon$ (see appendix \ref{app:mIIAConventions}):
\begin{eqnarray} \label{primedSpinors}
 \hat{\psi}^\prime_\mu =\Delta^{-\frac14} e_\mu{}^\alpha \big( \hat{\psi}_\alpha +\tfrac12 \hat \Gamma_\alpha \hat \Gamma^a \hat \psi_a \big) \; , \quad 
\hat{\psi}^\prime_a =\Delta^{-\frac14} \hat{\psi}_a \; , \quad 
\hat{\lambda}^\prime =\Delta^{-\frac14} \hat{\lambda} \; , \quad 
\hat{\epsilon}^\prime =\Delta^{\frac14} \hat{\epsilon} \; . \quad 
\end{eqnarray}
Here, $\alpha = 0 , 1,2,3$ and $a = 1 , \ldots , 6$ are tangent space indices in the vector representation of SO$(1,3)$ and SO(6), respectively;  $e_\mu{}^\alpha$ (and  $e_m{}^a$) are vielbeine for the external (and internal) components of the ten-dimensional metric in (\ref{KKFields}); and $\hat \psi_\alpha$, $\hat \psi_a$ are the components of the IIA gravitino, pulled back into tangent space with the ten-dimensional triangular vielbein that can be read off from the metric in (\ref{KKFields}). 

The primed fermions (\ref{primedSpinors}) are still ten-dimensional Majorana, 32-component spinors. In order to incorporate the splitting (\ref{LorentzSplit}), we work without loss of generality in the Cliff$(1,9)$ basis introduced in appendix \ref{ap:Fermions} and write the ten-dimensional gamma matrices $\hat \Gamma_A$, $A=0,1, \ldots, 9$, in terms of four and six-dimensional gamma matrices $\gamma_\alpha$, $\Gamma_a$. The fermions accordingly split into products of four-dimensional, anticommuting, and six-dimensional, commuting, spinors. For the primed supersymmetry parameter we thus write
\begin{eqnarray} \label{split}
\hat{\epsilon}^\prime = e^{i \frac{\pi}{4} \gamma_5} \left( \epsilon_+ \otimes \eta  +\epsilon_- \otimes \eta^{\textrm{c}} \right) = e^{i \frac{\pi}{4} } \, \epsilon_+ \otimes \eta  + e^{-i \frac{\pi}{4}} \, \epsilon_- \otimes \eta^{\textrm{c}} \; , 
\end{eqnarray}
and similarly for the other primed fermions in (\ref{primedSpinors}). The factor of $e^{i \frac{\pi}{4} \gamma_5}$ has been included here by convention, so that the four-dimensional fermions end up with parity-invariant mass terms and Pauli couplings should one be interested in reducing the fermionic Lagrangian. Also, $\epsilon_+$ and $\epsilon_-$ are four-dimensional spinors of positive and negative chirality, $\eta$ is a six-dimensional Dirac spinor and $\eta^\textrm{c}  = C^{-1} \bar{\eta}^{\textrm{T}}$, with $\bar \eta \equiv \eta^\dagger$, denotes charge conjugation in six dimensions. The decomposition (\ref{split}) manifestly preserves the Majorana character of $\hat{\epsilon}^\prime$. Finally, following \cite{deWit:1986mz}, we introduce a convenient notation: while we omit the SO$(1,3)$ spinor indices on $\epsilon_+$ and $\epsilon_-$, we exhibit the SO(6) spinor indices $A=1, \ldots , 8$ on $\eta^A$. The advantage of doing this is that we can omit the symbol for $\eta$ altogether and write the products  $\epsilon_+ \otimes \eta$ and $\epsilon_- \otimes \eta^{\textrm{c}}$ simply as $\epsilon^A$ and $\epsilon_A$, respectively, with the position of the index indicating four-dimensional chirality. Likewise, we will denote $\bar \epsilon_+ \otimes \eta^\dagger$ and $\bar \epsilon_- \otimes \eta^\textrm{T}$ by $\bar \epsilon_A$ and $\bar \epsilon^A$. We thus obtain formally four-dimensional anticommuting supersymmetry parameters $\epsilon^A$, etc., which can be regarded as eight-component, Dirac spinors of the internal Spin(6). 

Following the same steps, we also obtain formally four-dimensional gravitini $\psi_\mu^A$ and fermions $\psi_a^A$, $\lambda^A$, of spin 1/2 with respect to the external SO$(1,3)$, that are Dirac spinors of Spin(6). The fermions $\psi_\mu^A$ and $\epsilon^A$ can thus be straightforwardly declared to lie on the $\bar{\bm{8}}$ of SU(8). The $6 \times 8 + 8$ spin  1/2  fermions $\psi_a^A$, $\lambda^A$ can be packed in the $\overline{\bm{56}}$, totally antisymmetric, representation of SU(8) via
\begin{eqnarray} \label{trispinor1}
\chi^{ABC} = \tfrac{3i}{\sqrt{2}} \,  (\Gamma^a C^{-1})^{[AB} \psi_a^{C]} 
+ i \,  (\Gamma_7 C^{-1})^{[AB} \lambda^{C]} 
 -  \tfrac{i }{4} \,  (\Gamma^a \Gamma_7 C^{-1})^{[AB} (\Gamma_a \lambda)^{C]} \; , 
\end{eqnarray}
or Fierz rearrangements thereof\footnote{ For example, an equivalent expression is
\begin{eqnarray} \label{trispinor2}
\chi^{ABC} = \tfrac{3i}{\sqrt{2}} \,  (\Gamma^a C^{-1})^{[AB}  \Big( \psi_a^{C]} 
-  \tfrac{1 }{6\sqrt{2}} (\Gamma_a \Gamma_7 \lambda)^{C]} \Big)
 + 2i  \,  (\Gamma_7 C^{-1})^{[AB} \lambda^{C]} \; .
\end{eqnarray}
While (\ref{trispinor1}) is more convenient for calculations in our present context, (\ref{trispinor2}) is adapted to the $D=11$ origin of these fermions.}. See appendix \ref{ap:susyvectors} for a derivation of this expression.

\subsection{\mbox{Redefinitions that make the $D=4$ tensor hierarchy manifest}}\label{subsec:TensorHierarchy}

We now rewrite the supersymmetry transformations of the IIA bosonic fields in terms of the 
$ \mathrm{SO}(1,3) \times  \mathrm{SL}(6) $--covariant bosonic fields (\ref{eq:SL6fieldcontent}) and $ \mathrm{SO}(1,3) \times  \mathrm{SU}(8) $--covariant fermions that have introduced in the previous section. The compatibility of the resulting supersymmetry transformations with the $D=4$ tensor hierarchy will require further, non-linear, redefinitions of the vectors and tensors. 

Plugging (\ref{KKFields}) into the type IIA supersymmetry variations (\ref{IIAsusyVars}), and expressing the result in terms of the fermions $\psi_\mu^A$, $\psi_a^A$, $\lambda^A$, $\epsilon^A$ previously introduced, a long calculation allows us to obtain the following supersymmetry variations for the formally $D=4$ vielbein,
\begin{equation} \label{susyvielbein}
\delta e_\mu{}^\alpha = \tfrac{1}{4} \, \bar \epsilon_A \, \gamma^\alpha \,  \psi_\mu^A  + \tfrac{1}{4} \, \bar \epsilon^A \, \gamma^\alpha \,  \psi_{\mu A} \; , 
\end{equation}
\\[-30pt]
vectors,
{\setlength\arraycolsep{2pt}
\begin{eqnarray} \label{susyVectors}
\delta B_\mu{}^m &=& \tfrac{  i }{4}   \Delta^{-\frac12} e_a{}^m (C\Gamma^a)_{AB} \, \bar\epsilon^A  \psi_\mu^B   \, +  \tfrac{1}{4} \, \Delta^{-\frac12} e_a{}^m  \,\bar{\epsilon}_A  \, \gamma_\mu \, (\delta^{ab} +\tfrac12  \Gamma^a \Gamma^b )^A{}_B \,  \psi_b^B   +\textrm{h.c.} \; , \nonumber \\[10pt]
\delta A_\mu &=& \Big[ \tfrac{i}{4} \, e^{-\frac34 \hat \phi} \, \Delta^{-\frac12} (C\Gamma_7)_{AB} \, \bar\epsilon^A  \psi_\mu^B - \tfrac{1}{8}\, e^{-\frac34 \hat \phi} \, \Delta^{-\frac12} \,\bar{\epsilon}_A  \, \gamma_\mu \, ( \Gamma^a \Gamma_7 )^A{}_B \,  \psi_b^B \nonumber \\
&&  \,\,\,\, + \, \tfrac{3}{8\sqrt{2}} \, e^{-\frac34 \hat \phi} \, \Delta^{-\frac12} \,\bar{\epsilon}_A \gamma_\mu \lambda^A +\textrm{h.c.} \Big] - \delta B_\mu{}^m A_m   \; , \nonumber \\[10pt]
\delta A_{\mu mn}  &=&  \Big[-\tfrac{i}{4} \, e^{-\frac14 \hat \phi} \, \Delta^{-\frac12} e_m{}^a e_n{}^b (C\Gamma_{ab})_{AB} \, \bar\epsilon^A  \psi_\mu^B \nonumber \\
&&   \,\,\,\, - \, \tfrac12 \, e^{-\frac14 \hat \phi} \, \Delta^{-\frac12} \, e_m{}^a e_n{}^b \,\bar{\epsilon}_A  \, \gamma_\mu \, \big( \Gamma_{[a} ( \delta_{b]}^c - \tfrac14 \Gamma_{b]} \Gamma^c ) \big)^A{}_B \,  \psi_c^B \nonumber \\
&&  \,\,\,\, + \,  \tfrac{1}{8\sqrt{2}} \, e^{-\frac14 \hat \phi} \, \Delta^{-\frac12} \, e_m{}^a e_n{}^b \, \bar{\epsilon}_A \gamma_\mu (\Gamma_{ab} \Gamma_7 )^A{}_B \lambda^B   +\textrm{h.c.}  \Big]  \nonumber \\
&&  \,\,\,\, - \, \delta B_\mu{}^p A_{mnp} + 2 \, \delta B_\mu{}^p \, A_{[m} \,  B_{n]p} + A_\mu \, \delta B_{mn} - 2 A_{[m|} \, \delta B_{\mu |n]} \; ,  \nonumber \\[10pt]
\delta B_{\mu m}  &=&  \Big[-\tfrac{i}{4} \, e^{\frac12 \hat \phi} \, \Delta^{-\frac12} e_m{}^a (C\Gamma_a\Gamma_7)_{AB} \, \bar\epsilon^A  \psi_\mu^B - \tfrac{1}{4}\, e^{\frac12 \hat \phi} \, \Delta^{-\frac12} \, e_m{}^a \,\bar{\epsilon}_A  \, \gamma_\mu \, ( \delta_a^b - \tfrac12 \Gamma_a \Gamma^b )^A{}_B \,  \psi_b^B \nonumber \\
&&  \,\,\,\, - \,  \tfrac{1}{4\sqrt{2}} \, e^{\frac12 \hat \phi} \, \Delta^{-\frac12} \, e_m{}^a \, \bar{\epsilon}_A \gamma_\mu (\Gamma_a)^A{}_B \lambda^B   +\textrm{h.c.}  \Big] + \delta B_\mu{}^n B_{mn}  \; , 
\end{eqnarray}
}two-forms,
{\setlength\arraycolsep{2pt}
\begin{eqnarray} \label{susyTwoForms}
\delta A_{\mu \nu m}  &=&  \Big[ \tfrac12 \, e^{-\frac14 \hat \phi} \, \Delta^{-1} e_m{}^a \, \bar\epsilon_A \, \gamma_{[\mu} \, (\Gamma_{a})^A{}_B \,  \psi_{\nu]}^B 
  -\tfrac{i}{4}  \, e^{-\frac14 \hat \phi} \, \Delta^{-1} \, e_m{}^a \,\bar{\epsilon}^A  \, \gamma_{\mu \nu} \, \big( C( \delta_a^b - \Gamma_a \Gamma^b) \big)_{AB} \,  \psi_b^B \nonumber \\
&& \,\,  + \tfrac{i}{8\sqrt{2}} \, e^{-\frac14 \hat \phi} \, \Delta^{-1} \, e_m{}^a \, \bar{\epsilon}^A \gamma_{\mu \nu} \, (C \Gamma_a \Gamma_7 )_{AB} \, \lambda^B   +\textrm{h.c.}  \Big]  \nonumber \\
&&   \,\, - \, 2 \,  \delta B_{[\mu}{}^n A_{\nu] mn}   + 2 \,  \delta B_{[\mu}{}^n A_{\nu]} B_{mn}   -2 \,  \delta B_{[\mu}{}^n B_{\nu]n} A_{m}  +A_m \delta B_{\mu \nu} + 2 A_{[\mu} \delta B_{\nu] m}   \; ,  \nonumber \\[10pt]
\delta B_{\mu \nu}  &=&  \Big[ \tfrac12  \, e^{\frac12 \hat \phi} \, \Delta^{-1}  \, \bar\epsilon_A \, \gamma_{[\mu} \,  (\Gamma_7)^A{}_B \, \psi_{\nu]}^B - \tfrac{i}{4}\, e^{\frac12 \hat \phi} \, \Delta^{-1} \,\bar{\epsilon}^A  \, \gamma_{\mu \nu} \, ( C \Gamma^a \Gamma_7 )_{AB} \,  \psi_a^B \nonumber \\
&& \,\,  - \, \tfrac{i}{4\sqrt{2}} \, e^{\frac12 \hat \phi} \, \Delta^{-1} \, \bar{\epsilon}^A \gamma_{\mu \nu} \, C_{AB} \, \lambda^B   +\textrm{h.c.}  \Big] + 2 \, \delta B_{[\mu}{}^m B_{\nu] m }  \; , 
\end{eqnarray}
}and three-form,
{\setlength\arraycolsep{0pt}
\begin{eqnarray} \label{susyThreeForms}
&& \delta A_{\mu \nu \rho}  =  \Big[ -\tfrac{3 i}{4}  \, e^{-\frac14 \hat \phi} \, \Delta^{-\frac32 } \, \bar\epsilon^A \, \gamma_{[\mu \nu} \, C_{AB} \,  \psi_{\rho]}^B 
  +\tfrac{3 }{8}  \, e^{-\frac14 \hat \phi} \, \Delta^{-\frac32} \,\bar{\epsilon}_A  \, \gamma_{\mu \nu \rho} \, (\Gamma^a )^A{}_B \,  \psi_a^B \nonumber \\
&& \qquad \qquad  \,\,\, + \,  \tfrac{1}{8\sqrt{2}} \, e^{-\frac14 \hat \phi} \, \Delta^{-\frac32} \, \bar{\epsilon}_A \gamma_{\mu \nu \rho} \, (\Gamma_7 )^A{}_B \, \lambda^B   +\textrm{h.c.}  \Big]  \nonumber \\
&& \qquad \qquad \,\,\, - \, 3 \,  \delta B_{[\mu}{}^m A_{\nu \rho] m}   + 6 \,  \delta B_{[\mu}{}^n A_{\nu} B_{\rho] m }   +3 A_{[\mu} \delta B_{\nu \rho]}   \; .
\end{eqnarray}
}In these expressions, we have defined $\gamma_{\mu_1 \ldots \mu_p} \equiv e_{\mu_1}{}^{\alpha_1} \ldots e_{\mu_p}{}^{\alpha_p} \,\gamma_{\alpha_1 \ldots \alpha_p} $, $p=1,2,3$, and have omitted the direct product symbol between SO$(1,3)$ and SO$(6)$ gamma matrices, in agreement with the spinor notation $\epsilon^A$, etc., introduced in the previous section. The variation of $B_\mu{}^m$ has been brought to the form given in (\ref{susyVectors}) after performing a local SO$(1,9)$ Lorentz transformations, similarly to \cite{deWit:1986mz}. Since they will not be needed, we omit the supersymmetry variations of the scalars and fermions.

Some of these supersymmetry variations are already compatible with those of $D=4$ supergravity, as dictated by the tensor hierarchy, but others are not. For example, the term $A_\mu \, \delta B_{mn}$ in the expression for $\delta A_{\mu mn}$ in equation (\ref{susyVectors}) is incompatible with the canonical expression for the supersymmetry variations of the $D=4$ vectors: the latter do not contain bare vectors times the variation of scalars. These and similar considerations lead us to perform the following further redefinitions of the vectors,
\begin{eqnarray} \label{redefvectors}
C_\mu{}^{m8} \equiv  B_\mu{}^m  \,\,\,\,\, ,\,\,\,\,\, 
C_\mu{}^{78} \equiv  A_\mu \,\,\,\,\, ,\,\,\,\,\,
\tilde C_{\mu \, m n} \equiv  A_{\mu mn} -A_\mu B_{mn} \,\,\,\,\, , \,\,\,\,\,
\tilde C_{\mu \, m7} \equiv B_{\mu m}   \ ,
\end{eqnarray}
two-forms,
\begin{eqnarray} \label{redefTwoForms}
C_{\mu\nu \, m}{}^{8} \equiv -A_{\mu\nu m} +  C_{[\mu}{}^{n8} \,  \tilde C_{\nu]nm} +  C_{[\mu}{}^{78} \, \tilde C_{\nu]m 7} \,\,\,\, , \,\,\,\, 
C_{\mu \nu \, 7}{}^{8} \equiv -B_{\mu\nu} + C_{[\mu}{}^{m 8} \,  \tilde C_{\nu]m 7} \; , 
\end{eqnarray}
and three-form
\begin{eqnarray} \label{redefThreeForm}
C_{\mu\nu \rho}{}^{88} \equiv  A_{\mu\nu \rho} - C_{[\mu}{}^{m8} \, C_\nu{}^{n8} \,  \tilde{C}_{\rho]mn}+ C_{[\mu}{}^{m8} \, C_\nu{}^{78} \,  \tilde{C}_{\rho]m 7}  + 3 \, C_{[\mu}{}^{78} C_{ \nu \rho] 7 }{}^{8}   \; . 
\end{eqnarray}
Similar redefinitions were first considered in a  type IIB context in \cite{Ciceri:2014wya}; see also \cite{Baguet:2015sma}. See appendix \ref{app:GroupThRedef} for the group theory behind these redefinitions.

The SL(6)--covariant fields (\ref{redefvectors})--(\ref{redefThreeForm}) can be grouped into the SL(7)--covariant combinations
\begin{eqnarray} \label{TensorsSL7}
C_{\mu}{}^{I8} = (C_\mu{}^{m8} \, , \, C_\mu{}^{78}) \, , \,  
\tilde{C}_{\mu \, IJ} = (\tilde C_{\mu \, m n}  \, , \, \tilde C_{\mu \, m7} ) \, , \, 
C_{\mu \nu}{}_I{}^{8} = (C_{\mu \nu \, m}{}^{8} \, , \, C_{\mu \nu \, 7}{}^{8}) \, , \,
C_{\mu \nu \rho}{}^{88}  , 
\end{eqnarray}
with $\tilde C_{\mu \, IJ} \equiv \tilde C_{\mu \, [IJ]}$. We thus obtain (electric) vectors in the $\bm{7^\prime}$ and (magnetic) vectors in the $\bm{21}$ of SL(7), two-forms in the $\bm{7}$ and a singlet three-form. Although these SL(7) representations follow straightforwardly from (\ref{eq:SL6fieldcontent}), the actual dependence (\ref{redefvectors})--(\ref{redefThreeForm})  of the SL(7)--covariant vectors and tensors (\ref{TensorsSL7}) on the fields (\ref{eq:SL6fieldcontent}) involves a non-trivial analysis of the supersymmetry variations. The redefined fields  (\ref{redefvectors})--(\ref{redefThreeForm}) now satisy SL(6)--covariant supersymmetry transformations that are compatible with the $D=4$ tensor hierarchy. Moreover, these SL(6)--covariant supersymmetry variations can be grouped up into SL(7)--covariant transformations, of course also compatible with the $D=4$ tensor hierarchy, in a way consistent with the allocations (\ref{TensorsSL7}). This match provides a consistency check on our calculations. Here we only give the final result --further details can be found in appendix  \ref{sec:SL7covariant}. The SL(7)--covariant supersymmetry transformations of the fields (\ref{TensorsSL7}) are
{\setlength\arraycolsep{0pt}
\begin{eqnarray} 
\label{eq:susySL7Vec1}
&& \delta C_\mu{}^{I8} \, = i\, V^{I8}{}_{AB} \left(  \,  \bar{\epsilon}^A  \psi_\mu{}^B + \tfrac{1}{2\sqrt{2}} \,  \bar{\epsilon}_C  \gamma_\mu \chi^{ABC}\right) + \textrm{h.c.} \; ,  \\[8pt]
\label{eq:susySL7Vec2}
&& \delta \tilde{C}_{\mu \, IJ}  = - i\, V_{IJ \, AB} \left(  \, \bar{\epsilon}^A  \psi_\mu{}^B +  \tfrac{1}{2\sqrt{2}}  \,  \bar{\epsilon}_C  \gamma_\mu \chi^{ABC}\right) + \textrm{h.c.} \; ,  \\[8pt]
\label{eq:susySL72form}
&& \delta C_{\mu \nu \, I }{}^{8} =  \Big[ \tfrac23 \big( V^{J8}{}_{BC} \,  \tilde{V}_{IJ}{}^{AC} +  \tilde{V}_{IJ \, BC}{} \,  V^{J8}{}^{AC} \big) \,  \bar{\epsilon}_{A} \gamma_{[\mu} \psi_{\nu]}^B    \\
&&  \qquad\qquad\quad + \, \tfrac{\sqrt{2}}{3} \,  V^{J8}{}_{AB} \,  \tilde{V}_{IJ \, CD}  \,  \bar{\epsilon}^{[A} \gamma_{\mu \nu} \chi^{BCD]} +\textrm{h.c.} \Big] - C_{[\mu}{}^{J8} \, \delta \tilde{C}_{\nu] IJ} -  \tilde{C}_{[\mu| \, IJ} \,  \delta  C_{|\nu]}{}^{J8} 
 \; , \nonumber  \\[8pt]
\label{eq:susySL73form}
&& \delta C_{\mu \nu \rho}{}^{88}\,  =   \Big[ \tfrac{4i}{7}  \, V^{I8}{}_{BD} \, \big( V^{J8}{}^{DC} \,  \tilde{V}_{IJ}{}_{AC} +  \tilde{V}_{IJ}{}^{DC}{} \,  V^{J8}{}_{AC} \big) \,  \bar{\epsilon}^{A} \gamma_{[\mu \nu} \psi_{\rho]}^B \nonumber \\
&& \qquad\qquad\quad - \, i \tfrac{\sqrt{2}}{3} \, V^{I8 \, AE} \,   V^{J8}{}_{[EB|} \,  \tilde{V}_{IJ \, |CD]}  \,  \bar{\epsilon}_A \gamma_{\mu \nu \rho} \chi^{BCD} +\textrm{h.c.} \Big]    \\
&&  \qquad\qquad\quad + \, 3\, C_{[\mu\nu| \, I}{}^{8} \,  \delta C_{|\rho]}{}^{I8} 
- C_{[\mu}{}^{I8} \,   \big( C_{\nu}{}^{J8} \, \delta \tilde{C}_{\rho] IJ} +  \tilde{C}_{\nu| \, IJ} \,  \delta  C_{|\rho]}{}^{J8} \big) \; . \nonumber 
\end{eqnarray}
}For the vectors and two-forms, these variations coincide with those dictated by the $D=4$ embedding tensor formalism \cite{deWit:2007mt} upon selecting the relevant representations in the branching of their E$_{7(7)}$--covariant formulae under SL(7). The variation of the three-form agrees with a singlet extracted from the E$_{7(7)}$--covariant expression proposed in \cite{Guarino:2015qaa}. All these variations come out naturally written in the SL(8) basis, see {e.g.} (C.3) of \cite{Guarino:2015qaa}. 

We have written the variations (\ref{eq:susySL7Vec1})--(\ref{eq:susySL73form}) in terms of the spinor $\chi^{ABC}$ defined in (\ref{trispinor1}), and have introduced the `generalised vielbeine'
\begin{eqnarray} \label{GenVielbeinSL7}
V^{I8}{}_{AB} = (V^{m8}{}_{AB} \, , \, V^{78}{}_{AB}) \qquad , \qquad 
\tilde{V}_{IJ \, AB} = (\tilde V_{m n \, AB}  \, , \, \tilde V_{m7 \, AB} ) \; , 
\end{eqnarray}
and similarly for their conjugates with upper SU(8) indices $AB$. These can respectively be read off already from the vector variations (\ref{susyVectors}) as the coefficient of the $\bar \epsilon^A \,  \psi_\mu^B$ terms:
\begin{eqnarray} \label{GenVielbein}
V^{m8}{}_{AB} &=&  \tfrac{1}{4} \,  \Delta^{-\frac12} \, e_a{}^m (C\Gamma^a)_{AB} \; ,  \nonumber \\[5pt]
V^{78}{}_{AB} &=&   \tfrac{1}{4} \,  e^{-\frac34 \hat \phi} \, \Delta^{-\frac12} \,  (C\Gamma_7)_{AB} - V^{m8}{}_{AB} \, A_m \; ,  \nonumber \\[5pt]
\tilde{V}_{m7 \, AB} &=&  \tfrac{1}{4} \,  e^{\frac12 \hat \phi} \, \Delta^{-\frac12} \, e_m{}^a (C\Gamma_a\Gamma_7)_{AB} + V^{n8}{}_{AB} \, B_{nm} \; ,  \nonumber \\[5pt]
\tilde{V}_{mn \, AB} &=&   \tfrac{1}{4} \,  e^{-\frac14 \hat \phi} \, \Delta^{-\frac12} \, e_m{}^a e_n{}^b (C\Gamma_{ab})_{AB} + V^{p8}{}_{AB} ( A_{pmn} - 2B_{p[m} A_{n]} ) \nonumber \\
&& + \, V^{78}{}_{AB} \,  B_{mn} + 2\, \tilde{V}_{ [m|7  \, AB} A_{|n]} \ , 
\end{eqnarray}
and the $\bar \epsilon_A \,  \psi_{\mu \, B}$ terms inside the h.c.~contributions:
\begin{eqnarray}  \label{GenVielbeinUpperAB}
V^{m8 \, AB} &=&  - \tfrac{1}{4} \, \Delta^{-\frac12} \, e_a{}^m (\Gamma^a C^{-1})^{AB} \; ,  \nonumber \\[5pt]
V^{78 \, AB} &=&  - \tfrac{1}{4} \,  e^{-\frac34 \hat \phi} \, \Delta^{-\frac12} \,  (\Gamma_7 C^{-1})^{AB} - V^{m8 \, AB} A_m \; ,  \nonumber \\[5pt]
\tilde{V}_{m7}{}^{AB} &=&   \tfrac{1}{4} \, e^{\frac12 \hat \phi} \, \Delta^{-\frac12} \, e_m{}^a (\Gamma_a\Gamma_7 C^{-1} )^{AB} + V^{n8 \, AB} B_{nm} \; ,   \nonumber \\[5pt]
\tilde{V}_{mn}{}^{AB} &=&   \tfrac{1}{4} \,  e^{-\frac14 \hat \phi} \, \Delta^{-\frac12} \, e_m{}^a e_n{}^b (\Gamma_{ab} C^{-1})^{AB} + V^{p8 \, AB} ( A_{pmn} - 2B_{p[m} A_{n]} ) \nonumber \\
&& + \, V^{78 \, AB} B_{mn} + 2\, \tilde{V}_{[m|7}{}^{AB} \,  A_{|n]} \; . 
\end{eqnarray}

To summarise, the $\mathrm{SL}(7) $--covariant bosonic field content that arises when type IIA supergravity is rewritten with only $D=4$ local Lorentz symmetry manifest includes
\begin{eqnarray} \label{eq:SL7fieldcontent}
\bm{1} &  \textrm{metric}:  &  \; ds_4^2 \, (x,y)  \; ,\nonumber \\
\bm{7}^\prime  + \bm{21} &  \textrm{generalised vielbeine}:  &   \; V^{I8}{}_{AB} (x,y)   \;, \;  \tilde{V}_{IJ \, AB} (x,y)   \; , \nonumber \\
\bm{7}^\prime  + \bm{21} &  \textrm{vectors}:  &   \;  C_\mu{}^{I8} (x,y)  \;, \;  \tilde{C}_{\mu \, IJ} (x,y)    \; ,  \nonumber \\
\bm{7} &  \textrm{two-forms}:  &   \;  C_{\mu \nu \,  I}{}^{8} (x,y)  \; , \nonumber \\
\bm{1} &  \textrm{three-form}:  &   \; C_{\mu \nu \rho}{}^{88} (x,y)   \; .
\end{eqnarray}
These depend on the SL(6)--covariant fields (\ref{eq:SL6fieldcontent}) that enter the IIA fields (\ref{KKFields}) through (\ref{GenVielbein}) and (\ref{redefvectors})--(\ref{redefThreeForm}). The representations shown for the generalised vielbeine correspond to their SL(7) indices. Their (antisymmetric) indices $AB$ label the $\bm{28}$ of SU(8). This is in agreement with the fact that the generalised vielbeine only take values along the antisymmetric combinations (\ref{AntiSymGammaMat6D}) of six-dimensional gamma matrices. Finally, the SU(8)--covariant IIA fermionic field content includes
\begin{eqnarray} \label{eq:SU8fieldcontent}
\overline{\bm{8}} &  \textrm{gravitini}:  &  \; \psi_\mu^A (x,y) \; , \nonumber \\
\overline{\bm{56}} &  \textrm{spin $1/2$}:  &   \; \chi^{ABC} (x,y)   \; .
\end{eqnarray}
In (\ref{eq:SL7fieldcontent}), (\ref{eq:SU8fieldcontent}), we have explicitly restored the $(x,y)$ dependence of these fields in order to emphasise their ten-dimensional character.


\section{Truncation on $S^6$ to $D=4$ $\cN=8$  ISO(7) supergravity} \label{sec:S6trunc}


Until now, all the quantities that we have introduced depend on all ten-dimensional coordinates $(x^\mu , y^m)$. We have merely rewritten the field content and supersymmetry variations of type IIA supergravity in terms of the ten-dimensional $ \mathrm{SO}(1,3) \times  \mathrm{SL}(7) $--covariant bosons (\ref{eq:SL7fieldcontent}) and $ \mathrm{SO}(1,3) \times  \mathrm{SU}(8) $--covariant fermions (\ref{eq:SU8fieldcontent}). Now, this reformulation is especially suitable to determine truncations of type IIA supergravity down to $D=4$ $\cN=8$ gauged supergravity. In this paper, we will focus on the $S^6$ truncation down to the ISO(7) gauging. Sections \ref{sec:KKansatze} and \ref{sec:FullNonLinEmb} derive the KK embedding formulae at the level of the gauge potentials. Consistency is then shown at the level of the supersymmetry transformations in section \ref{sec:S6truncConsistencySusy} and at the level of the field strengths in section \ref{sec:S6truncConsistencyFS}.


\subsection{Kaluza-Klein ansatze} \label{sec:KKansatze}


As we will now show, the full embedding of $\cN=8$ ISO$(7)$ supergravity into type IIA becomes naturally determined in terms of the restricted tensor hierarchy introduced in \cite{Guarino:2015qaa}. The latter contains, besides the metric and scalars, all 56 electric and magnetic vectors, the two-forms associated to the generators of $\textrm{SL}(7) \ltimes \mathbb{R}^7 \subset \textrm{E}_{7(7)}$ and the three-forms related to the electric components of the ISO(7) embedding tensor. All of these fields are allocated in SL(7) representations, namely,
\begin{eqnarray} \label{eq:SL7fieldcontent4D}
\bm{1} &  \textrm{metric}:  &  \; ds_4^2 \, (x) \; ,\nonumber \\
\bm{21}^\prime  + \bm{7}^\prime  + \bm{21}+ \bm{7} &  \textrm{coset representatives}:  &   \; {\cal V}^{IJ \, ij} (x) \;, \; {\cal V}^{I8 \, ij}   (x)  \;, \;  \tilde{{\cal V}}_{IJ}{}^{ij}  (x) \;, \; \tilde{{\cal V}}_{I8}{}^{ij}  (x) \; ,   \nonumber \\
\bm{21}^\prime  + \bm{7}^\prime  + \bm{21}+ \bm{7} &  \textrm{vectors}:  &   \; \cA_\mu{}^{IJ} (x) \;, \; \cA_\mu{}^{I}  (x)  \;, \;  \tilde{\cA}_{\mu \, IJ}  (x) \;, \; \tilde{\cA}_{\mu \, I}  (x)  \; ,  \nonumber \\
\bm{48} + \bm{7}^\prime &  \textrm{two-forms}:  &   \; \cB_{\mu \nu \, I}{}^J  (x) \;, \; \cB_{\mu \nu}{}^I  (x) \; , \nonumber \\
\bm{28}^\prime  &  \textrm{three-forms}:  &   \; \cC_{\mu \nu \rho}{}^{IJ}  (x) \; ,
\end{eqnarray}
along with the SU(8)--covariant fermions
\begin{eqnarray} \label{eq:SU8fieldcontent4D}
\overline{\bm{8}} &  \textrm{gravitini}:  &  \; \psi_\mu^i (x) \; , \nonumber \\
\overline{\bm{56}} &  \textrm{spin $1/2$}:  &   \; \chi^{ijk} (x)   \; .
\end{eqnarray}
We have explicitly written the $x$ dependence in order to emphasise that these are four-dimensional fields. 
The supersymmetry transformations and field equations of these fields close among themselves for the ISO(7) gauging \cite{Guarino:2015qaa}. Thus, for this gauging, these fields define a consistent subsector, still $\cN=8$ but only SL(7)--covariant, of the full $\cN=8$ E$_{7(7)}$--covariant tensor \cite{deWit:2008ta,deWit:2008gc} and duality \cite{Bergshoeff:2009ph} hierarchies of four-dimensional maximal supergravity.

The SL(7) representations that appear in the $D=4$ field content (\ref{eq:SL7fieldcontent4D}) differ from those that arise in the rewrite (\ref{eq:SL7fieldcontent}) of type IIA with only $D=4$ local Lorentz symmetry manifest. For example, while (\ref{eq:SL7fieldcontent4D}) contains all 56 (electric and magnetic) vectors of maximal supergravity, the IIA rewrite (\ref{eq:SL7fieldcontent}) only contains $\bm{7}^\prime$ (electric) and $\bm{21}$ (magnetic) vectors. More subtle is the fact that the IIA rewrite (\ref{eq:SL7fieldcontent}) contains two-forms in the $\bm{7}$ of SL(7), while the $D=4$ field content (\ref{eq:SL7fieldcontent4D}) includes two-forms in the $\bm{7}^\prime$, along with the $\bm{48}$. The $\bm{7}$ and $\bm{7}^\prime$ conjugate representations of SL(7) are inequivalent. Accordingly, the respective two-forms have different supersymmetry transformations: compare (\ref{eq:susySL72form}) above to the second equation in (2.38) of \cite{Guarino:2015qaa}. 

Both sets of fields are related by a KK ansatz, namely, a linear relation between the IIA fields (\ref{eq:SL7fieldcontent}), the $D=4$ fields (\ref{eq:SL7fieldcontent4D}) and tensors on $S^6$ in $\mathrm{SL} (7) \times \mathrm{SL} (7)$ representations, with the representations of the left and right SL(7) factors respectively matching those of the IIA and $D=4$ fields. For the type IIA case at hand, the KK ansatz can equivalently, but more naturally, be given for the SL(6)--covariant IIA fields (\ref{redefvectors})--(\ref{redefThreeForm}) in terms of the $D=4$ SL(7) fields (\ref{eq:SL7fieldcontent4D}) and tensors  on $S^6$ in $\mathrm{SL} (6) \times \mathrm{SL} (7)$ representations. We will proceed this way. Besides trivial constants, these tensors turn out to be exclusively given by combinations of the coordinates $\mu^I(y)$ (in the $(\bm{1}, \bm{7}^\prime)$) that embed $S^6$ into $\mathbb{R}^7$ and their derivatives $\partial_m \mu^I$ (in the  $(\bm{6}, \bm{7}^\prime)$) with respect to the $S^6$ angles $y^m$. Further combinations of those include {\it e.g.}~the Killing vectors $K^m_{IJ}$ (in the $(\bm{6}^\prime, \bm{21})$) of the round metric $\mathring{g}_{mn}(y)$ on $S^6$; and their derivatives $K_{mn}^{IJ}$  (in the $(\bm{15}, \bm{21}^\prime)$). See appendix \ref{subset:GeneralSn} for our conventions for these quantities. Of course, these tensors on $S^6$ can also be regarded as  $\textrm{SO} (6) \times \textrm{SO}(7)$ tensors, for which indices are raised and lowered with the $S^6$ and $\mathbb{R}^7$ metrics $\mathring{g}_{mn}(y)$ and $\delta_{IJ}$ so that conjugate representations become equivalent (and possibly reducible). For this reason, we will write interchangeably $\mu^I$ and $\mu_I$, etc. In contrast, the indices $m,n, \ldots$ in the IIA fields  (\ref{redefvectors})--(\ref{redefThreeForm}) and $I, J, \ldots$ in the $D=4$ fields (\ref{eq:SL7fieldcontent4D}) are exclusively SL(6) and SL(7) indices. Thus, they cannot be raised or lowered (conjugate representations are not equivalent, see the comment above about the two-forms). Finally, a (singular) five-form potential for the $S^6$ volume form would appear in the KK ansatze for the dual fields \cite{Godazgar:2013pfa, Baguet:2015sma} of the democratic formulation \cite{Bergshoeff:2001pv} of type IIA. Since we do not use such formulation, that form does not play a role in this paper.

Following this discussion, for the embedding of the $D=4$ metric  (\ref{eq:SL7fieldcontent4D}) into the metric in (\ref{eq:SL7fieldcontent}) we simply declare
\begin{equation} \label{eq:KKmetric}
ds_4^2 (x,y) = ds_4^2 (x) \; ,
\end{equation}
and similarly for the vielbein $e_\mu{}^\alpha$. For the $\bm{6}^\prime + \bm{1} + \bm{15} + \bm{6}$ SL(6) vectors (\ref{redefvectors}), we write the following ansatz, 
\begin{eqnarray} \label{KKvectors}
C_\mu{}^{m8} (x,y)  = \tfrac12 \, g \, K_{IJ}^m (y) \, \cA_\mu{}^{IJ} (x) 
& \,\,\, , \,\,\, &
C_\mu{}^{78}  (x,y)  = -\mu_I (y)  \, \cA_\mu{}^{I} (x) \ , \nonumber \\[5pt]
\tilde C_{\mu \, mn} (x,y)  = \tfrac14 \, K_{mn}^{IJ} (y) \, \tilde \cA_{\mu \, IJ} (x)
& \,\,\, , \,\,\, &
\tilde C_{\mu \, m7} (x,y)  = -g^{-1}   \, (\partial_m \mu^I) (y)  \, \tilde \cA_{\mu \, I} (x) \ ,
 \end{eqnarray}
which preserves the electric and magnetic character on both sides of the equations. The ansatz for $C_\mu{}^{m8} (x,y) $ is the well-known KK expression that relates, in the case at hand, the electric SO(7) gauge fields $\cA_\mu{}^{IJ}$ to the isometries of the compatifying $S^6$ generated by the Killing vectors $K_{IJ}^m (y)$. The ansatz for the magnetic $\tilde C_{\mu \, mn}$  has recently appeared, in a $D=11$ on $S^7$ context, in \cite{deWit:2013ija}. 

Moving to the $\bm{6} + \bm{1}$ two-forms (\ref{redefTwoForms}), we write
 \begin{eqnarray} \label{KKTwoForms}
C_{\mu \nu \, m}{}^{8}  (x,y) = -g^{-1}  \, ( \mu_I \partial_m \mu^J) (y) \, \cB_{\mu \nu \, J}{}^I (x)  
\,\,\,\,\,\,  , \,\,\,\,\,\, 
C_{\mu \nu \, 7}{}^{8}  (x,y) =  \mu_I (y) \, \cB_{\mu \nu}{}^I (x) \; .
 \end{eqnarray}
Note that $ \mu_I \partial_m \mu^J$ can be assigned to the $(\bm{6}, \bm{48})$ of $\textrm{SL} (6) \times \textrm{SL}(7)$ since it is traceless in $IJ$. To conclude with the ansatze for the purely bosonic fields, we write
 \begin{eqnarray} \label{KKThreeForms}
  C_{\mu \nu \rho}{}^{88}  (x,y) = ( \mu_I \mu_J) (y) \, \cC_{\mu \nu \rho}{}^{IJ} (x)  \ ,
  \end{eqnarray}
for the three-form (\ref{redefThreeForm}). 

The generalised vielbeine  (\ref{GenVielbeinUpperAB}) have both SL(6) indices $m,n = 1, \ldots, 6$ and SU(8) indices $A, B = 1, \ldots , 8$. The former are rotated into the SL(7) indices of the $D=4$ E$_{7(7)}/$SU(8) coset representatives (\ref{eq:SL7fieldcontent4D}) with the same $\textrm{SL} (6) \times \textrm{SL}(7)$ tensors on $S^6$ that appear in the vector ansatze (\ref{KKvectors}). The SU(8) indices are rotated into the $D=4$ SU(8) indices $i,j = 1, \ldots , 8$ in (\ref{eq:SU8fieldcontent4D}) with the Killing spinors $ \eta_i^A (y)$ on $S^6$:
 \begin{eqnarray} \label{KKGenVielbein}
V^{m8 \, AB} (x,y) &=& \tfrac12 \, g \, K_{IJ}^m (y) \, \eta_i^A (y) \,  \eta_j^B (y) \, {\cal V}^{IJ \, ij} (x)  \; , \nonumber \\[7pt]
V^{78 \,  AB} (x,y)  &=&  - \mu_I (y)  \, \eta_i^A (y) \,  \eta_j^B (y) \, {\cal V}^{I8 \, ij}  (x)  \; ,
 \nonumber \\[7pt]
\tilde V_{mn}{}^{AB} (x,y)  &=&  \tfrac14 \, K_{mn}^{IJ} (y)  \, \eta_i^A (y) \,  \eta_j^B (y) \, \tilde {\cal V}_{IJ}{}^{ij} (x)  \; , 
 \nonumber \\[7pt]
\tilde V_{m7}{}^{AB} (x,y)   &=&   - g^{-1}  \, (\partial_m \mu^I) (y) \, \eta_i^A (y) \,  \eta_j^B (y) \,  \tilde {\cal V}_{I8}{}^{ij} (x)   \; ,
 \end{eqnarray}
and similarly for the conjugates  (\ref{GenVielbein}) with lower SU(8) indices. We have omitted on all four right-hand-sides in (\ref{KKGenVielbein}) additional scalar-dependent SU(8) rotations. These have been discussed at length in \cite{deWit:1986iy,Nicolai:2011cy}. Finally, the KK ansatz for the fermions is 
\begin{equation} \label{eq:KKfermions}
\psi_\mu^A (x,y) = \eta^A_i (y) \,   \psi_\mu^i (x) \qquad , \qquad
\chi^{ABC} (x,y) = \eta^A_i (y) \, \eta^B_j (y) \,  \eta^C_k (y)  \, \chi^{ijk} (x) \; ,
\end{equation}
and similarly for the supersymmetry parameter $\epsilon^A (x,y)$.

While the electric coupling constant $g$ of dyonic ISO(7) supergravity appears in the KK ansatze (\ref{eq:KKmetric})--(\ref{eq:KKfermions}), the magnetic coupling $m$ does not. In particular, the generalised vielbein $V^{m8} (x,y)$ in (\ref{KKGenVielbein}) is independent of $m$. The `Clifford property' issue that prevented \cite{deWit:2013ija} the embedding of the dyonic SO(8) gauging \cite{Dall'Agata:2012bb} (at least within the SL(8) frame\footnote{See \cite{Lee:2015xga} for a more general no-go result.}) in $D=11$ is thus circumvented by our construction. Also, one might have naively expected the magnetic vector $ \tilde \cA_{\mu \, I}$ to descend from $\tilde C_{\mu \, m7} $ with strength $m$, rather than $g^{-1}$. Incidentally, the non-analytic dependece  of this and other of the above expressions on $g$ restricts the validity of these KK ansatze to $g \neq 0$. This is related to the fact that $g^{-1}$ is related to the radius of the compactifying $S^6$ and therefore needs to be non-vanishing. See section \ref{sec:Discussion} for further comments on the $g=0$ case. Finally, we have fixed the coefficients in (\ref{eq:KKmetric})--(\ref{KKGenVielbein}) by solving the field equations in various invariant sectors, including the G$_2$ (see section \ref{app:G2UpliftSubsec}) and SU(3) \cite{Varela:2015uca} sectors, and then imposing the IIA equations of motion. In particular, the $\textrm{SU}(3) \times \textrm{U}(1)$--invariant $\cN=2$  AdS$_4$ solution of \cite{Guarino:2015jca} fixes some of these coefficients.


\subsection{The full non-linear embedding} \label{sec:FullNonLinEmb}


In order to find the full non-linear embedding of ISO(7) supergravity into type IIA, we need to bring the KK ansatze that we have just proposed into the ten-dimensional bosons (\ref{KKFields}) and fermions (\ref{primedSpinors}). We will focus on the bosonic fields. Equations (\ref{redefvectors})--(\ref{redefThreeForm}) can be easily inverted to solve for the fields that enter (\ref{KKFields}) in terms of the tensor-hierarchy-compatible fields. Expressing the latter through the KK ansatze  (\ref{KKvectors})--(\ref{KKThreeForms}), we obtain the following expressions for the vectors,
{\setlength\arraycolsep{4pt}
\begin{eqnarray} \label{eq:KKintermediatestepVec}
B_\mu{}^m &=& \tfrac12 \, g \, K_{IJ}^m \, \cA_\mu{}^{IJ} \; , \nonumber \\[5pt]
A_\mu &=& -\mu_I \, \cA_\mu{}^{I} \; , \nonumber \\[5pt] 
A_{\mu mn} &=&  \tfrac14  \, K_{mn}^{IJ} \, \tilde{\cA}_{\mu \, IJ} -\mu_I \, B_{mn} \, \cA_\mu{}^I \; , \nonumber \\[5pt] 
B_{\mu m} &=&  -g^{-1} \, ( \partial_m \mu^I)  \, \tilde \cA_{\mu \, I} \; ,
\end{eqnarray}
}two-forms,
{\setlength\arraycolsep{4pt}
\begin{eqnarray} \label{eq:KKintermediatestep2forms}
A_{\mu \nu m} &=&  g^{-1} \, (\mu_I \partial_m \mu^J)  \, \Big(  \cB_{\mu\nu \, J}{}^I +\cA_{[\mu}{}^{IK} \tilde{\cA}_{\nu]  KJ} + \cA_{[\mu}{}^I \tilde{\cA}_{\nu]  J}  \Big) \; ,  \nonumber  \\[5pt]
B_{\mu \nu} &=&  - \mu_I  \, \big( \cB_{\mu\nu}{}^I + \cA_{[\mu}{}^{IJ} \tilde{\cA}_{\nu]  J}  \big) \; , 
\end{eqnarray}
}and three-form,
{\setlength\arraycolsep{4pt}
\begin{eqnarray} \label{eq:KKintermediatestep3form}
A_{\mu \nu \rho} \,=\,  \mu_I \mu_J  \, \Big( \cC_{\mu\nu\rho}{}^{IJ} + 3 \,  \cA_{[\mu}{}^{I} \cB_{\nu\rho]}{}^J  + \cA_{[\mu}{}^{IK} \cA_\nu{}^{JL} \tilde{\cA}_{\rho] KL}  +\cA_{[\mu}{}^{I} \cA_\nu{}^{JK} \tilde{\cA}_{\rho] K} \Big)  \ .
\end{eqnarray}
}In these expressions we have again dropped the labels $(x,y)$ on the left-hand-sides and $(x)$ and $(y)$ on the right-hand-sides. In order to simplify them, we have used some tensorial identities on $S^6$, including (\ref{eq:relationsSn}), (\ref{eq:propsS6}). Now, bringing (\ref{eq:KKmetric}) and (\ref{eq:KKintermediatestepVec})--(\ref{eq:KKintermediatestep3form}) to (\ref{KKFields}) and performing some further simplifications of the same type, we finally obtain the full non-linear embedding of ISO(7) supergravity into type IIA:
{\setlength\arraycolsep{0pt}
\begin{eqnarray} \label{KKEmbedding}
&& d\hat{s}_{10}^2 =  \Delta^{-1} \, ds_4^2  \, + g_{mn}  \, Dy^m \, Dy^n  \; , \nonumber \\[9pt]
&& \hat A_\3 \,= \mu_I \mu_J  \, \big( \cC^{IJ}  + \cA^I \wedge \cB^J +\tfrac16 \cA^{IK} \wedge \cA^{JL} \wedge \tilde{\cA}_{KL}  +\tfrac16 \cA^{I} \wedge \cA^{JK} \wedge \tilde{\cA}_{K}   \big)   \nonumber \\[3pt]
&& \qquad \quad + \, g^{-1}  \, \big( \cB_J{}^I  +\tfrac12 \cA^{IK} \wedge \tilde{\cA}_{KJ}  + \tfrac12 \cA^I \wedge \tilde{\cA}_J \big)  \wedge \mu_I D\mu^J +\tfrac12 g^{-2}\, \tilde \cA_{IJ} \wedge D\mu^I \wedge D \mu^J  \nonumber \\[3pt]
&& \qquad \quad  - \, \tfrac12 \, \mu_I  \,  B_{mn} \, \cA{}^{I}  \wedge Dy^m \wedge Dy^n 
+ \tfrac16 A_{mnp} \, Dy^m \wedge Dy^n \wedge Dy^p \; , \nonumber  \\[9pt]
&& \hat B_\2 = -\mu_I \, \big( \cB^I  +\tfrac12 \cA^{IJ} \wedge \tilde{\cA}_J \big)   - g^{-1}  \, \tilde \cA_{I}   \wedge D \mu^I  + \tfrac12  B_{mn} \,  Dy^m \wedge Dy^n   \; , \nonumber  \\[9pt]
&& \hat A_\1 = - \mu_I  \, \cA^{I}  + A_{m}  \, Dy^m\; .
\end{eqnarray}
}Here, we have defined the covariant derivatives
\begin{eqnarray} \label{DyDmu}
Dy^m \equiv  dy^m + \tfrac12 \, g \, K_{IJ}^m \, \cA^{IJ}  \qquad , \qquad 
D\mu^I  \equiv d \mu^I -g \, \cA^{IJ} \mu_J \; ,
\end{eqnarray}
which feature only the vectors $ \cA^{IJ}$ that gauge SO(7) electrically. The expressions for $d\hat{s}_{10}^2$, $\hat B_\2$ and  $\hat A_1$ have already appeared in \cite{Guarino:2015jca}. The expression for the Ramond-Ramond three-form $\hat A_\3$ appears here for the first time. Here we have also provided a detailed derivation of these formulae, and will show their consistency in sections \ref{sec:S6truncConsistencySusy} and \ref{sec:S6truncConsistencyFS}. Although the KK ansatze (\ref{KKvectors})--(\ref{KKThreeForms}) relate linearly the tensor-hierarchy-compatible IIA fields (\ref{eq:SL7fieldcontent}) to the $D=4$ fields (\ref{eq:SL7fieldcontent4D}), the dependence (\ref{KKEmbedding}) of the usual, locally ten-dimensional-Lorentz-covariant IIA fields becomes non-linear in the $D=4$ fields.

We still need to specify the embedding of the $D=4$ scalars into the ten-dimensional fields. In order to do this, we follow \cite{deWit:1984nz,deWit:2013ija,Ciceri:2014wya}. We first substitute the ansatz (\ref{KKGenVielbein}) for the generalised vielbeine and the analogue for its conjugates into (\ref{GenVielbeinUpperAB}), (\ref{GenVielbein}). Then, we take all possible products between generalised vielbeine and their conjugates, and trace over SU(8) indices. The last step cancels the dependence on the $S^6$ Killing spinors by virtue of their orthogonality, as well as the dependence on the SU(8) rotations which we omitted. These products become naturally written in terms of SL(7)--covariant blocks of the matrix ${\cal M}_{\mathbb{MN}}$ (see {\it e.g.}~(2.15) of \cite{Guarino:2015qaa}) quadratic in the SL(7)-branched-out E$_{7(7)}/$SU(8) coset representative ${\cal V}_\mathbb{M}{}^{ij}$ in (\ref{eq:SL7fieldcontent4D}). The minimal set of these relations that fully specifies the complete, non-linear embedding of the $D=4$ scalars into the type IIA fields is 
{\setlength\arraycolsep{4pt}
\begin{eqnarray} 
{\cal M}^{IJ \, KL} \,  K^m_{IJ} \, K^n_{KL} &=&  4 g^{-2}   \, \Delta^{-1} \,  g^{mn}  \; , \label{eq:internalmetric} \\[7pt]
{\cal M}^{IJ \, K8} \, K^m_{IJ} \,  \mu_K  &=& 2 g^{-1}  \, \Delta^{-1} g^{mn} A_n   \; , \label{eq:internalAm} \\[7pt]
{\cal M}^{IJ}{}_{K8} \, K^m_{IJ} \, \partial_{n} \mu^K  &=&   -2 \,  \Delta^{-1} \,  g^{mp}  B_{pn}  \; , \label{eq:internalAmn}   \\[7pt]
{\cal M}^{IJ}{}_{KL}  \, K^m_{IJ}  \, K_{np}^{KL}  &=&  8 g^{-1}  \,  \Delta^{-1} \,   g^{mq} \, \left( A_{qnp}  - A_q B_{np}  \right)  \; , \label{eq:internalAmnp} \\[7pt]
{\cal M}^{I8 \, J8} \, \mu_I \, \mu_J  &=&  \Delta^{-1}  \left( e^{-\frac32 \hat  \phi} +g^{mn} A_m A_n \right)    \; .   \label{eq:dilaton}  
\end{eqnarray}
}Equations (\ref{eq:internalmetric})--(\ref{eq:dilaton}) allow one to solve sequentially for the IIA embedding of the $D=4$ scalars: (\ref{eq:internalmetric}) determines the embedding of the scalars into the inverse internal metric $g^{mn}$; inserting this result into (\ref{eq:internalAm}), the embedding of the scalars into the internal components $A_m$ of the Ramond-Ramond one-form can be worked out; etc. A straightforward rearrangement of (\ref{eq:internalmetric})--(\ref{eq:dilaton}) has already appeared in \cite{Guarino:2015jca}. Together with these equations, the remaining relations that can be obtained from products of generalised vielbeine are
{\setlength\arraycolsep{0pt}
\begin{eqnarray} 
&& {\cal M}^{I8}{}_{JK} \, \mu_I \, K_{mn}^{JK} =  -4 \, \Delta^{-1}  \left( B_{mn} \, e^{-\frac32 \hat  \phi} -g^{pq} A_p ( A_{qmn} - A_q B_{mn} ) \right)    \;  , \label{eq:other1st} \\[7pt]
&& {\cal M}^{I8}{}_{J8} \, \mu_I \, \partial_m \mu^J  =   g \, \Delta^{-1}  g^{np} \, B_{mp} \, A_n  \; ,  \\[7pt]
&& {\cal M}_{IJ \, KL} \, K_{mn}^{IJ} \, K_{pq}^{KL} =   16 \, \Delta^{-1}  \Big( 2 e^{-\frac12 \hat  \phi} g_{m[p} g_{q]n }   +g^{rs} ( A_{rmn} - 2B_{r[m} A_{n]} )  ( A_{spq} - A_s B_{pq} ) \nonumber  \\
&& \qquad \qquad \qquad \qquad   \qquad   \qquad  \qquad  
 + e^{-\frac32 \hat  \phi} B_{mn} B_{pq}   - g^{rs} B_{mn}  A_r ( A_{spq} - A_s B_{pq} ) \
   \;  \nonumber   \\
&& \qquad \qquad \qquad \qquad   \qquad   \qquad  \qquad  
 -4 e^{\hat \phi} A_{[m} g_{n][p} A_{q]}   +2 g^{rs} A_{[m} A_{n]s}   ( A_{rpq} - A_r B_{pq} ) \Big)  ,   \\[7pt]
&& {\cal M}_{I8 \, JK} \, \partial_m \mu^I \, K_{np}^{JK} =  -4g  \, \Delta^{-1}  \left( 2 e^{\hat  \phi} g_{m[n} A_{p]}  +g^{qr} B_{rm} ( A_{qnp} - A_q B_{np} ) \right)    \; ,   \\[7pt]
&& {\cal M}_{I8 \, J8} \, \partial_m \mu^I \, \partial_n \mu^J =  g^2 \, \Delta^{-1}  \left( e^{\hat  \phi} g_{mn} + g^{pq} B_{pm} B_{qn} \right)    \; . \label{eq:otherlast}
\end{eqnarray}
}Only the first block of equations, (\ref{eq:internalmetric})--(\ref{eq:dilaton}), should be regarded as independent. Equations in the second block, (\ref{eq:other1st})--(\ref{eq:otherlast}), should be redundant identities obtainable from the first block. We have indeed rederived some of the equations in the second block from equations in the first for the G$_2$--invariant sector.  It would be interesting to study the relation between the two blocks more generally.


\subsection{Consistency: supersymmetry transformations} \label{sec:S6truncConsistencySusy}


We will now establish the consistency of the truncation formulae (\ref{KKEmbedding}) at the level of the supersymmetry transformations of the bosons. The goal is to show that, when the supersymmetry transformations (\ref{susyvielbein}), (\ref{eq:susySL7Vec1})--(\ref{eq:susySL73form}) of the $\textrm{SO}(1,3) \times \textrm{SL}(7)$--covariant type IIA bosons  are evaluated on the truncation ansatze (\ref{eq:KKmetric})--(\ref{eq:KKfermions}), all $S^6$ dependence factorises and the supersymmetry transformations of the bosons of $D=4$ ISO(7) supergravity arise. In the bosonic supersymmetry transformations, all fermionic indices $A, B, \ldots$ and  $i,j, \ldots$ are contracted. Thus, the dependence on the $S^6$ Killing spinors $\eta^A_i$ and scalar-dependent matrices \cite{deWit:1986iy,Nicolai:2011cy} that the KK ansatze for the vielbeine (\ref{KKGenVielbein}) and fermions (\ref{eq:KKfermions}) introduce, automatically cancels due to the orthogonality properties of the former and the SU(8) character of the latter. Consistency thus boils down to showing that the re-allocation of SL(7) representations, from (\ref{eq:SL7fieldcontent}) to (\ref{eq:SL7fieldcontent4D}), that the KK ansatze (\ref{eq:KKmetric})--(\ref{eq:KKfermions}) induces, is compatible with the supersymmetry transformations. 

The consistency of the supersymmetry transformation of the vielbein, an SL(7)--singlet, is thus straightforward: (\ref{susyvielbein}) trivially reduces on (\ref{eq:KKmetric}), (\ref{eq:KKfermions}) to the supersymmetry variation for the $D=4$ vielbein, see (2.36) of \cite{Guarino:2015qaa} for our conventions. Consistency is also automatic for the transformations of the vectors, given that their dependence (\ref{KKvectors}) on $S^6$ quantities matches that of the generalised vielbeine, (\ref{KKGenVielbein}). Bringing (\ref{KKvectors}), (\ref{KKGenVielbein}), (\ref{eq:KKfermions}) to the SL(6)--covariant transformations (\ref{VectorsBranchingSL6}) or, equivalently, to the SL(7)--covariant  transformations (\ref{eq:susySL7Vec1}), (\ref{eq:susySL7Vec2}), all $S^6$ dependence trivially drops out and we are left with the supersymmetry transformations of the $D=4$ vectors,
\begin{equation}
\label{VectorsBranchingSL7}
\begin{array}{llrl}
 \delta \cA_\mu{}^{IJ} &=&  i\,  \cV^{IJ}{}_{ij} \left(  \, \bar{\epsilon}^i  \psi_\mu{}^j + \tfrac{1}{2\sqrt{2}} \,  \bar{\epsilon}_k  \gamma_\mu \chi^{ijk}\right) + \textrm{h.c.} & , \\[3mm]
\delta \cA_\mu{}^{I} &=&  i\,  \cV^{I8}{}_{ij} \left(  \, \bar{\epsilon}^i  \psi_\mu{}^j + \tfrac{1}{2\sqrt{2}}  \,  \bar{\epsilon}_k  \gamma_\mu \chi^{ijk}\right) + \textrm{h.c.} & , \\[3mm]
\delta \tilde{\cA}_{\mu \, IJ} &=&  -i\,  \tilde{\cV}_{IJ \, ij} \left(  \, \bar{\epsilon}^i  \psi_\mu{}^j + \tfrac{1}{2\sqrt{2}} \,  \bar{\epsilon}_k  \gamma_\mu \chi^{ijk}\right) + \textrm{h.c.} & ,  \\[3mm]
\delta \tilde{\cA}_{\mu \, I} &=&  -i\,  \tilde{\cV}_{I8 \, ij} \left(  \, \bar{\epsilon}^i  \psi_\mu{}^j + \tfrac{1}{2\sqrt{2}}  \,  \bar{\epsilon}_k  \gamma_\mu \chi^{ijk}\right) + \textrm{h.c.} & ,
\end{array}
\end{equation}
see (2.37) of \cite{Guarino:2015qaa}. 

Some work is needed to show consistency for the supersymmetry transformations of the two- and three-forms. Also in these cases, the $S^6$ dependences on the left and right-hand-sides of the transformations eventually match and thus drop out. For example, evaluating the l.h.s.~of (\ref{susytensors4dSL6m}) on the first expression in (\ref{KKTwoForms}) straightforwardly gives $\delta C_{\mu \nu \, m}{}^{8}  = -g^{-1}   \mu_I \partial_m \mu^J \, \delta \cB_{\mu \nu \, J}{}^I $. On the other hand, the r.h.s~of (\ref{susytensors4dSL6m}) evaluated on the vielbeine (\ref{KKGenVielbein}), vector (\ref{KKvectors}) and fermion ansatze (\ref{eq:KKfermions}) also yields a factorised $g^{-1}   \mu_I \partial_m \mu^J$ dependence after using the sphere identities (\ref{eq:propsS6}). Repeating a similar exercise for $\delta C_{\mu \nu \, 7}{}^{8} $ in (\ref{KKTwoForms}) and consistently dropping the $S^6$ dependence, we arrive at
{\setlength\arraycolsep{1pt}
\begin{eqnarray} \label{susytensors4dSL7bis}
\delta \cB_{\mu \nu \, J}{}^I & = &   \Big[ - \tfrac{2}{3} \big( {\cal V}^{IK}{}_{jk} \,  \tilde{{\cal V}}_{JK}{}^{ik} + {\cal V}^{I8}{}_{jk} \,  \tilde{{\cal V}}_{J8}{}^{ik}  +  \tilde{{\cal V}}_{JK \, jk}{} \,  {\cal V}^{IK}{}^{ik}   +  \tilde{{\cal V}}_{J8 \, jk}{} \,  {\cal V}^{I8}{}^{ik} \big)  \bar{\epsilon}_{i} \gamma_{[\mu} \psi_{\nu]}^j \nonumber \\
&& \quad -  \tfrac{\sqrt{2}}{3} \, \big(  {\cal V}^{IK}{}_{ij} \,  \tilde{{\cal V}}_{JK \, kl}  +  {\cal V}^{I8}{}_{ij} \,  \tilde{{\cal V}}_{J8 \, kl}   \big) \,  \bar{\epsilon}^{[i} \gamma_{\mu \nu} \chi^{jkl]} +\textrm{h.c.} \Big]  \nonumber \\
&&  \quad  +  \big( \cA_{[\mu}^{IK} \, \delta \tilde{\cA}_{\nu] JK} + \cA_{[\mu}^{I} \, \delta \tilde{\cA}_{\nu] J} +  \tilde{\cA}_{[\mu| \, JK} \,  \delta  \cA_{|\nu]}{}^{IK}+  \tilde{\cA}_{[\mu| \, J} \,  \delta  \cA_{|\nu]}{}^{I} \big) 
-\tfrac17 \,  \delta_J^I \,  (\textrm{trace}) \; , \nonumber  \\[12pt]
\delta \cB_{\mu \nu}{}^I & = & \Big[  \tfrac{2}{3}  \big( {\cal V}^{IJ}{}_{jk} \,  \tilde{{\cal V}}_{J8}{}^{ik} +  \tilde{{\cal V}}_{J8 \, jk}{} \,  {\cal V}^{IJ}{}^{ik} \big)  \, \bar{\epsilon}_{i}  \gamma_{[\mu} \psi_{\nu]}^j + \tfrac{\sqrt{2}}{3} \,  {\cal V}^{IJ}{}_{ij} \,  \tilde{{\cal V}}_{J8 \, kl}  \,  \bar{\epsilon}^{[i} \gamma_{\mu \nu} \chi^{jkl]} +\textrm{h.c.} \Big]   \nonumber \\
&&  \quad -  \big( \cA_{[\mu}^{IJ} \, \delta \tilde{\cA}_{\nu] J} +  \tilde{\cA}_{[\mu| \, J} \,  \delta  \cA_{|\nu]}{}^{IJ} \big) \; . 
\end{eqnarray}
}These expressions coincide with the supersymmetry transformations for the two-forms that were computed in (2.38) of \cite{Guarino:2015qaa} from the $D=4$ embedding tensor formalism. In particular, the subtraction of the trace in the first equation of (\ref{susytensors4dSL7bis}) is enforced in the present approach upon dropping the $S^6$ dependence --as we have just discussed, this equation arises contracted with the traceless tensor $\mu_I \, \partial_m \mu^J$. 

The consistency of the three-form transformations is shown similarly. Inserting the KK ansatz (\ref{KKThreeForms}) into the l.h.s.~of the three-form variation (\ref{eq:susySL73form}), and the vielbeine (\ref{KKGenVielbein}), vector (\ref{KKvectors}), two-form (\ref{KKTwoForms}) and fermion ansatze (\ref{eq:KKfermions}) into the r.h.s., a factored-out  dependence on $\mu_I \mu_J$ is found on both sides of the equation after using, on the r.h.s., the $S^6$ relations (\ref{eq:propsS6}). Consistently dropping $\mu_I \mu_J$ and symmetrising accordingly, we are left with
{\setlength\arraycolsep{1pt}
\begin{eqnarray} \label{susy3forms4dSL7bis}
\delta \cC_{\mu \nu \rho}{}^{IJ} & = &  \Big[ -  \tfrac{4i}{7} \,\Big( 
 \cV^{K(I}{}_{jl} \big(  \cV^{J)L \, lk} \,  \tilde \cV_{KL \, ik}  + \tilde \cV_{ KL}{}^{ lk} \,  \cV^{J)L}{}_{ ik}   \big) \nonumber \\
&& \qquad  \quad  + \, \cV^{K(I}{}_{jl} \big(  \cV^{J)8 \, lk} \,  \tilde \cV_{K8 \, ik}  + \tilde \cV_{K8}{}^{ lk} \,  \cV^{J)8}{}_{ ik}   \big) \nonumber \\
&& \qquad  \quad + \, \cV^{(I|8}{}_{jl} \big(  \cV^{|J)K \, lk} \,  \tilde \cV_{K8 \, ik}  + \tilde \cV_{K8}{}^{ lk} \, \cV^{|J)K}{}_{ik}  \big) \Big)  \, \bar{\epsilon}^{i}  \gamma_{[\mu \nu} \psi_{\rho]}^j  \nonumber \\
&& + \, i \tfrac{\sqrt{2}}{3}  \,\Big( 
 \cV^{K(I|}{}^{hi}  \, \cV^{|J)L}{}_{[ij|} \,  \tilde \cV_{KL}{}_{|kl]}   + \,  \cV^{K(I}{}^{hi} \,  \cV^{J)8}{}_{[ij|} \,  \tilde \cV_{K8}{}_{|kl]}  \\ 
&& \qquad  \quad  + \, \cV^{(I|8}{}^{hi}  \, \cV^{|J)K}{}_{[ij|} \,  \tilde \cV_{K8}{}_{|kl]}   \Big)  \, \bar{\epsilon}_{h}  \gamma_{\mu \nu \rho } \chi^{jkl} \; + \textrm{h.c.} \Big] \nonumber \\
&& - \, 3 \,  \Big(  \cB_{[\mu \nu| K}{}^{(I} \, \delta \cA^{J)K}_{|\rho]} +   \cB_{[\mu \nu}{}^{(I} \, \delta \cA^{J)}_{\rho]}   \Big) \nonumber \\
&& + \,  \cA^{K(I}_{[\mu} \big(  \cA^{J)L}_\nu \,  \delta \tilde \cA_{\rho] KL}  + \tilde \cA_{\nu KL} \, \delta \cA^{J)L}_{\rho]}   \big)  + \cA^{K(I}_{[\mu} \big(  \cA^{J)}_\nu \,  \delta \tilde \cA_{\rho] K}  + \tilde \cA_{\nu K} \, \delta \cA^{J)}_{\rho]}   \big) \nonumber \\
&& \qquad  \quad + \,   \cA^{(I}_{[\mu} \big(  \cA^{J)K}_\nu \,  \delta \tilde \cA_{\rho] K}  + \tilde \cA_{\nu K} \, \delta \cA^{J)K}_{\rho]}   \big)  \; . \nonumber
\end{eqnarray}
}These are the supersymmetry variations of the $D=4$ three-forms given in (2.39) of \cite{Guarino:2015qaa}.

As we have already noted, the consistent embedding (\ref{KKEmbedding}) is independent both of the magnetic gauge coupling $m$ and the Romans mass, which we denote in this paper with the same symbol, $m$. These formulae are thus insensitive to whether $m =0$ or $m \neq 0$. The consistency proof at the level of the supersymmetry variations of the bosons that we have given in this subsection is independent of $m$ as well. This is ultimately related to the fact that the variations (\ref{IIAsusyVars}) of the IIA bosons are independent of the Romans mass, and the variations of the $D=4$ bosons are independent of the embedding tensor. In conclusion, (\ref{KKEmbedding}) simultaneously gives the IIA embedding (at the level of metric and form potentials) of the purely electric, $m =0$  \cite{Hull:1984yy}, or dyonic, $m \neq 0$, ISO(7) gauging into massless or massive type IIA. Likewise, the consistency proof in this subsection pertains simulatenously to both truncations. The consistent truncation of massless IIA to the purely electric gauging \cite{Hull:1984yy} had been conjectured in \cite{Hull:1988jw}.


\subsection{Consistency: field strengths and duality hierarchy} \label{sec:S6truncConsistencyFS}


The different embeddings, in either massless or massive type IIA, of the electric or dyonic ISO(7) gauging are distinguished by gauging-dependent quantities, for example the field strengths. This was already discussed in \cite{Guarino:2015jca} so we will be brief. Our main, two-fold purpose here to compute the field strengths is actually different. On the one hand, we will show the consistency of the truncation at the level of the Bianchi identities. On the other hand, we will use the restricted duality hierarchy discussed in \cite{Guarino:2015qaa} to express the embedding in terms of independent $D=4$ degrees of freedom. This will allow us, in particular, to find  closed expressions for the Freund-Rubin term.

The field strengths corresponding to the IIA form potentials (\ref{KKEmbedding}) can be computed using the definitions (\ref{romfields}). A lengthy calculation shows that these are given by 
\begin{eqnarray} \label{eq:10DFS}
&& \hat F_\4 = \mu_I \mu_J  \, {\cal H}_\4^{IJ} + g^{-1} \,   {\cal H}_{\3 \, J}{}^I \wedge \mu_I D\mu^J + \tfrac12 \, g^{-2}  \, \tilde{{\cal H}}_{\2 IJ} \wedge D\mu^I \wedge D\mu^J+ \ldots  \; ,  \nonumber \\[6pt]
&& \hat H_\3 =  - \mu_I \,  {\cal H}_\3^I -g^{-1} \, \tilde{{\cal H}}_{\2 I} \wedge D\mu^I + \ldots  \; ,   \nonumber \\[6pt]
&& \hat F_\2 = - \mu_I {\cal H}^I_\2 + g^{-1} \, \big(\, g \, \delta_{IJ} \, \cA^J - m \, \tilde{\cA}_I  \, \big) \wedge D \mu^I +  \ldots \; ,
\end{eqnarray}
The dots stand for terms containing $D=4$ scalars: either scalars multiplying four-dimensional vector field strengths or covariant derivatives of scalars. In particular, the terms in $\hat F_\2$ that depend on bare vectors $\cA^J$, $\tilde \cA_I$ combine into covariant derivatives of scalars with other terms not shown. Here, $ {\cal H}_\4^{IJ}$, $ {\cal H}_{\3  J}{}^I $, etc., turn out to coincide with the field strengths of the restricted $D=4$ $\cN=8$ tensor hierarchy (\ref{eq:SL7fieldcontent4D}),  given in (2.7)--(2.9) of \cite{Guarino:2015qaa}, of either electric (if $m = 0$) or dyonic (if $m \neq 0$) ISO(7) supergravity, provided the $D=10$ Romans mass and the $D=4$ magnetic coupling are indeed identified \cite{Guarino:2015jca} as in  (\ref{RomansMass}). These field strengths (except $ {\cal H}_\4^{IJ}$) were computed in \cite{Guarino:2015qaa} using the $D=4$ embedding tensor formalism \cite{deWit:2007mt,Bergshoeff:2009ph}. The fact that we can now reproduce them from a different, $D=10$ route provides a self-consistency check.

The Bianchi identities can now be seen to hold consistently: when the field strengths (\ref{eq:10DFS}) are inserted into the type IIA Bianchi identities (\ref{d10Bianchis}), the terms contained in the ellipses cancel, all $S^6$ dependence drops out, and one is left with
\begin{equation}
\label{eq:TruncBianchis}
\begin{array}{l}
D {\cal H}_{\2}^{IJ} = 0 \hspace{2mm} , \hspace{2mm}
D {\cal H}_\2^{I} = m \, \cH^{I}_{\3} \hspace{2mm} , \hspace{2mm}
D {\tilde{\cal H}}_{\2 IJ} = - 2 \, g \, {\cH_{\3 [I}}^{K}\,\delta_{J]K} \hspace{2mm} , \hspace{2mm}
D {\tilde{\cal H}}_{\2 I} = g \, \delta_{IJ} \,  \cH^{J}_{\3} \ , \\[3mm]
D {\cal H}_{\3 I}{}^J = \cH_\2^{JK} \wedge \tilde{\cH}_{\2 IK}  + \cH_\2^{J} \wedge \tilde{\cH}_{\2 I}   -2 g\, \delta_{IK} \, \cH_\4^{JK} - \tfrac17 \, \delta_I^J \,  (\textrm{trace})   \ , \\[3mm]
D {\cal H}_\3^{I} = -\cH_\2^{IJ} \wedge \tilde{\cH}_{\2 J}  \hspace{3mm} , \hspace{5mm}
D {\cal H}_{\4}^{IJ} \equiv 0 \ .
\end{array}
\end{equation}
These equations coincide with the Bianchi identities of the restricted, SL(7)--covariant $D=4$ tensor hierarchy (\ref{eq:SL7fieldcontent4D}), see equation (2.13) of \cite{Guarino:2015qaa}. The covariant derivatives were defined in (2.11), (2.12) therein. The results of this and the previous subsection establish the consistency of the truncation of (massive) type IIA supergravity on $S^6$ to (dyonic) ISO(7) supergravity, at the level of the supersymmetry transformations of the bosons and the Bianchi identities of the restricted tensor hierarchy (\ref{eq:SL7fieldcontent4D}). In $\cN=8$ supergravity, the former should imply the equations of motion. 

In any case, the Bianchi identities (\ref{eq:TruncBianchis}) do already give rise explicitly to some of the four-dimensional equations of motion upon using the $D=4$ duality hierarchy, as discussed in section 2.4 of \cite{Guarino:2015qaa} following \cite{Bergshoeff:2009ph}. For example, the magnetic field strengths $\tilde{\cH}_{\2 IJ}$, $\tilde{\cH}_{\2 I}$ become equivalent, upon dualisation, to scalar-dependent combinations of electric field strengths and their Hodge duals, see below. Thus, this approach establishes partially the consistency of the $\cN=8$ truncation of IIA on $S^6$ at the level of the equations of motion as well. In section \ref{sec:G2uplift}, we will explicitly verify consistency at the level of the equations of motion for the G$_2$-invariant sector.

More generally, the power of the duality hierarchy resides in that it allows one to transfer the independent degrees of freedom among the fields in the $D=4$ hierarchy. In our present context, it allows one to give the full embedding of $D=4$ ISO(7) supergravity into type IIA in terms of independent degrees of freedom. Namely, the embeddings (\ref{KKEmbedding}) and (\ref{eq:10DFS}) are written, at face value, in terms of the redundant $D=4$ degrees of freedom contained in the hierarchy (\ref{eq:SL7fieldcontent4D}). When the duality relations discussed in \cite{Guarino:2015qaa} are used, the full non-linear embedding (at the level of the field strengths) (\ref{eq:10DFS}) becomes expressed in terms of independent $D=4$ degrees of freedom: electric field strengths and their Hodge duals, scalars, their ISO(7)--covariant derivatives and their Hodge duals. Thus, the non-linear embedding of (dyonic) ISO(7) supergravity into massive type IIA is given by (\ref{eq:10DFS}) with $\tilde{\cH}_{\2 IJ}$, $\tilde{\cH}_{\2 I}$, $\cH_{\3 I}{}^J$, $\cH_\3^I$, ${\cal H}_\4^{IJ}$, regarded as short-hand notations for the corresponding dual expressions in terms of independent $D=4$ degrees of freedom \cite{Guarino:2015qaa}: 
{\setlength\arraycolsep{2pt}
\begin{eqnarray}
\label{H2IJDuality}
\tilde{\cH}_{\2 IJ} &\equiv& \tfrac{1}{2} {\cal I}_{[IJ][KL]} \, *{\cal H}_\2^{KL} + {\cal I}_{[IJ][K8]} \, *{\cal H}_\2^{K}  + \tfrac{1}{2} {\cal R}_{[IJ][KL]} \, {\cal H}_\2^{KL} +  {\cal R}_{[IJ][K8]} \, {\cal H}_\2^{K}  \ , \\[7pt]
\label{H2IDuality}
\tilde{\cH}_{\2 I} &\equiv& \tfrac{1}{2} {\cal I}_{[I8][KL]} \, *{\cal H}_\2^{KL} + \, {\cal I}_{[I8][K8]} \, *{\cal H}_\2^{K}  + \tfrac{1}{2} {\cal R}_{[I8][KL]} \, {\cal H}_\2^{KL} + {\cal R}_{[I8][K8]} \, {\cal H}_\2^{K} \  , \\[7pt]
\label{H3IJDuality}
\cH_{\3 I}{}^J &\equiv& \tfrac{1}{12}  (t_I{}^J)_{\mathbb{M}}{}^\mathbb{P}  \, \cM_{\mathbb{N}\mathbb{P}} *D \cM^{\mathbb{M} \mathbb{N}} - \tfrac17 \, \delta_I^J \, (\textrm{trace}) \ , \\[7pt]
\label{H3IDuality}
\cH_{\3 }{}^I &\equiv& \tfrac{1}{12}  (t_8{}^I)_{\mathbb{M}}{}^\mathbb{P} \, \cM_{\mathbb{N}\mathbb{P}} *D \cM^{\mathbb{M} \mathbb{N}} \  , \\[7pt]
\label{H4Duality}
\cH_{\4}^{IJ} &\equiv &  \tfrac{1}{84}  {X_{\mathbb{NQ}}}^{\mathbb{S}} \big( (t_K{}^{(I|})_{\mathbb{P}}{}^{\mathbb{R}}    \mathcal{M}^{|J)K \, \mathbb{N}} +  (t_8{}^{(I|})_{\mathbb{P}}{}^{\mathbb{R}}    \mathcal{M}^{|J)8 \, \mathbb{N}} \big)  \big(  \mathcal{M}^{\mathbb{PQ}}  \mathcal{M}_{\mathbb{RS}}  +   7 \,   \delta^\mathbb{P}_\mathbb{S} \,  \delta^\mathbb{Q}_\mathbb{R}  \big)  \textrm{vol}_4  \ . \,\, \qquad 
\end{eqnarray}
}Here, ${\cal I}_{[IJ][KL]}$, ${\cal R}_{[IJ][KL]}$ , etc. are the SL(7)--covariant components of the scalar dependent $D=4$ gauge kinetic matrices (see {\it e.g.} (2.15) of \cite{Guarino:2015qaa}) and ${X_{\mathbb{NQ}}}^{\mathbb{S}}$ is the usual \cite{deWit:2007mt} constant tensor constructed by contraction  ${X_{\mathbb{N}\mathbb{Q}}}^{\mathbb{S}}={\Theta_{\mathbb{N}}}^{\alpha}\,{(t_{\alpha})_{\mathbb{Q}}}^{\mathbb{S}}$ of the embedding tensor ${\Theta_{\mathbb{N}}}^{\alpha}$ of dyonic ISO(7) supergravity and the generators ${(t_{\alpha})_{\mathbb{Q}}}^{\mathbb{S}}$ of E$_{7(7)}$ (in {\it e.g.} the SL(8) basis, see (C.3), (C.4) of \cite{Guarino:2015qaa}). The terms shown in  (\ref{eq:10DFS}) contain, upon dualisation (\ref{H2IJDuality})--(\ref{H4Duality}), all terms in scalars with no derivatives (that is, the complete Freund-Rubin term, $\mu_I \mu_J  \, {\cal H}_\4^{IJ} \,$, in $\hat F_\4$), all terms containing Hodge duals of electric vector field strengths (coming from (\ref{H2IJDuality}), (\ref{H2IDuality})), and all terms containing Hodge duals of covariant derivatives of scalars (coming from (\ref{H3IJDuality}), (\ref{H3IDuality})). The ellipses in (\ref{eq:10DFS}) contain further terms in the electric vector fields strengths and in covariant derivatives of scalars, but not in their Hodge duals.

The $D=4$ Bianchi identities (\ref{eq:TruncBianchis}) can be used to simplify the embedding (\ref{eq:10DFS}), in particular the Freund-Rubin term. Also useful for this purpose are the identities \cite{Guarino:2015qaa}
\begin{equation}
\label{DH_singlets}
\begin{array}{lll}
D {\cal H}_{\3} &=& \cH_\2^{IJ} \wedge \tilde{\cH}_{\2 IJ}  + \cH_\2^{I} \wedge \tilde{\cH}_{\2 I}   -2 g\, \delta_{IJ} \, \cH_\4^{IJ}  - 14 \, m \, \tilde{\cH}_{\4} \ , \\[2mm]
D {\tilde{\cal H}}_{\4} & \equiv & 0 \ ,
\end{array}
\end{equation}
corresponding to the singlet two-form potential $\cB$ (with three-form field strength ${\cal H}_{\3}$) that would render $\cB_I{}^J$ in (\ref{eq:SL7fieldcontent4D}) traceful, and to the singlet three-form potential $\tilde{\cC}$ (with four-form field strength $\tilde{\cH}_{\4}$) related to the magnetic component of the $D=4$ ISO(7) embedding tensor. The dualisation conditions for ${\cal H}_{\3}$ and $\tilde{\cH}_{\4}$ were given in (2.25) of \cite{Guarino:2015qaa}. The four-form field strengths $\cH_{\4}^{IJ}$,  $\tilde{\cH}_{\4}$ and the scalar potential $V$ of ISO(7) supergravity are related through \cite{Guarino:2015qaa}
\begin{eqnarray}
\label{H4/Potential2}
 g \, \delta_{IJ} \, \cH_{\4}^{IJ} + m \, \tilde{\cH}_{\4} = - 2 \, V \, \textrm{vol}_4 \; . \; 
\end{eqnarray}
Now, some manipulation of the Bianchi identities (\ref{eq:TruncBianchis}), (\ref{DH_singlets}) and the relation (\ref{H4/Potential2}) allows us to derive the following expression for the Freund-Rubin term $\cH_{\4}^{IJ}  \, \mu_{I}  \mu_{J}$,
{\setlength\arraycolsep{1pt}
\begin{equation} \label{FRForms}
\begin{array}{lll}
\cH_{\4}^{IJ}  \, \mu_{I} \, \mu_{J}  &=& -\frac{1}{3} \, g^{-1} \, V \,  \textrm{vol}_{4} + \frac{1}{84} \, g^{-1} \, \big(D\cH_{\3} - 7 \, \cH_\2^{IJ} \wedge \tilde{\cH}_{\2 IJ}  - 7 \, \cH_\2^{I} \wedge \tilde{\cH}_{\2 I} \big)  \\[2mm]
& & - \frac{1}{2} \, g^{-1} \, \big(D\cH_{\3 I}{}^{J} -\cH_\2^{JK} \wedge \tilde{\cH}_{\2 IK}  - \cH_\2^{J} \wedge \tilde{\cH}_{\2 I}\big) \, \mu^{I}  \mu_{J}  \ ,
\end{array}
\end{equation}
}in terms of the scalar potential, the vector field strengths and the covariant derivatives of the three-form field strengths. Upon dualisation with (\ref{H2IJDuality})--(\ref{H4Duality}) above and  (2.25) of \cite{Guarino:2015qaa}, the latter become equivalent to (projections of) the scalar equations of motion \cite{Bergshoeff:2009ph} and, thus, to derivatives of the scalar potential, as in \cite{Godazgar:2015qia}. The expression (\ref{FRForms}) is valid, upon dualisation, for arbitrary values of the $D=4$ metric, electric vector field strengths and scalars. It is of course an exact expression --it does not rely on any approximation whatsoever.       

Some terms in this expression depend on the coordinates of $S^6$ through the combination $\mu_I \mu_J$. At a critical point of the scalar potential, $\cH_{\3 I}{}^{J} = \cH_{\3} = \cH_\2^{IJ} =  \cH_\2^{I} = \tilde{\cH}_{\2 IJ} =  \tilde{\cH}_{\2 I} =0$ and (\ref{FRForms}) reduces to 
\begin{equation} \label{FRFormsCritPoint}
\cH_{\4}^{IJ} |_0  \, \mu_{I} \, \mu_{J}  = - \tfrac{1}{3} \, g^{-1} \, V_0 \,  \textrm{vol}_{4}  \; , 
\end{equation}
where $|_0$ and $V_0$ denote evaluation at a critical point. The r.h.s.~of (\ref{FRFormsCritPoint}) becomes a constant, independent of the $S^6$ coordinates, and so must be the l.h.s. This is indeed the case: at a critical point, one has
\begin{eqnarray}
\label{Extreme27}
\cH_\4^{IJ} |_0 =  \tfrac17 \, \delta^{IJ} \,  \delta_{KL} \, \cH_\4^{KL}|_0  \;  , 
\end{eqnarray}
(see (2.31) of \cite{Guarino:2015qaa}) and the contraction with $\mu_I \mu_J$ becomes independent of the sphere coordinates upon using the defining relation (\ref{eqSn}). Thus, at a critical point of the $D=4$ scalar potential, the Freund-Rubin term becomes constant (as required by the IIA Bianchi identities) and in fact proportional to the cosmological constant at that critical point: from (\ref{FRFormsCritPoint}), (\ref{Extreme27}),
\begin{eqnarray}
U_0 \, \textrm{vol}_4 \equiv  \tfrac17 \, \delta_{IJ} \,  \cH_\4^{IJ}|_0  = - \tfrac{1}{3} \, g^{-1} \, V_0 \, \textrm{vol}_4 \; .
\end{eqnarray}
This relation can be seen to hold identically using the Bianchi identities  (\ref{eq:TruncBianchis}), (\ref{DH_singlets}) and the relation (\ref{H4/Potential2}), evaluated at a critical point.

The alternative, though equivalent, rewrite of the Freund-Rubin term that we brought to the introduction, (\ref{U_generalRewriteIntro}), utilises explicitly the dualisation (\ref{H4Duality}) of the four-form $\cH_\4^{IJ}$ into scalars. For this rewrite, it is appealing to introduce some notation as follows. We find it useful to introduce a(n $S^6$-dependent) `primed embedding tensor' ${\Theta^\prime_{\mathbb{M}}}^{\alpha}$ with, for the case at hand, only active components in the $\bm{36} + \bm{36}^\prime$ of SL(8) (in fact, only the  $\bm{28}$ of SL(7) active, see (\ref{theta_Prime})),
\begin{equation}
\label{Theta_Def_primed}
\Theta_{[AB]\,\,\,\,D}^{\prime \phantom{[AB]}C}= 2 \, \delta_{[A}^{C} \, \theta^\prime_{B]D} 
\hspace{10mm} , \hspace{10mm} 
 {\Theta^{\prime [AB]C}}_{D}  = 2 \, \delta^{[A}_{D} \,  \xi^{\prime B]C} \ ,
\end{equation}
with $A, B, \ldots = 1 , \ldots , 8$. We also introduce the corresponding `primed $X$-tensor', ${X^\prime_{\mathbb{M}\mathbb{N}}}^{\mathbb{P}}={\Theta^\prime_{\mathbb{M}}}^{\alpha}\,{(t_{\alpha})_{\mathbb{N}}}^{\mathbb{P}}$, with ${(t_{\alpha})_{\mathbb{N}}}^{\mathbb{P}}$ the generators of E$_{7(7)}$ in the SL(8) basis. In the present case, these correspond to the generators of the SL(8) subgroup only, and thus 
\begin{eqnarray}
\label{X_primed}
&& {X^\prime_{[AB] [CD]}}^{[EF]} = - X^\prime_{[AB]}{}^{[EF]}{}_{[CD]} = -8 \, \delta_{[A}^{[E} \theta^\prime _{B][C} \delta_{D]}^{F]}  \ , \nonumber \\[7pt]
&& X^{\prime[AB]}{}_{[CD]}{}^{[EF]} = - {X^{\prime [AB] [EF]}}_{[CD]} = -8 \, \delta_{[C}^{[A} \xi^{\prime \,  B][E} \delta_{D]}^{F]} \ .
\end{eqnarray}
The $\bm{36}^\prime$ magnetic components $\xi^{\prime \, AB}$ of the primed embedding tensor are zero (we introduced them merely to give a formal symmetric appearance to the formulae). Only the electric components $\theta^\prime_{AB}$, in the $\bm{28}$ of SL(7), are non-vanishing. These depend on the coordinates of $S^6$ through the tensor  $\mu_I \mu_J$, which also sits in the $\bm{28}$ of SL(7):
\begin{equation} \label{theta_Prime}
\theta^\prime_{IJ} = \mu_I \mu_J \; , \quad \theta^\prime_{I8} = 0 \; , \quad \ \theta^\prime_{88} = 0 \; ; \qquad \quad \xi^{\prime \, AB} = 0 \; .
\end{equation}
Now, contracting the duality relation (\ref{H4Duality}) with $\mu_I \mu_J$ and employing the definitions (\ref{Theta_Def_primed})--(\ref{theta_Prime}), the Freund-Rubin term acquires the expression brought to the introduction,
\begin{equation}
\label{U_generalRewrite}
U \, \textrm{vol}_4 \,  \equiv \,  \cH_{\4}^{IJ}  \, \mu_{I} \, \mu_{J}   \, =  - \dfrac{g}{84} \,  {X^\prime_{\mathbb{MP}}}^{\mathbb{R}}  {X_{\mathbb{NQ}}}^{\mathbb{S}}  \mathcal{M}^{\mathbb{MN}}   \Big(  \mathcal{M}^{\mathbb{PQ}}  \mathcal{M}_{\mathbb{RS}}  +   7 \,   \delta^\mathbb{P}_\mathbb{S} \,  \delta^\mathbb{Q}_\mathbb{R}  \Big) \, \textrm{vol}_4 \ .
\end{equation}
This expression depends on the $S^6$ coordinates through ${X^\prime_{\mathbb{MP}}}^{\mathbb{R}}$; on the electric (if $m=0$) or dyonic (if $m \neq 0$) ISO(7) embedding tensor through the constants ${X_{\mathbb{NQ}}}^{\mathbb{S}}$;  and on the $D=4$ scalars through the square $\mathcal{M}_{\mathbb{MN}}$ of the E$_{7(7)}/$SU(8) coset representative and its inverse $\mathcal{M}^{\mathbb{MN}}$. Note that ${X^\prime_{\mathbb{MP}}}^{\mathbb{R}}$ is always purely electric, regardless of whether ${X_{\mathbb{NQ}}}^{\mathbb{S}}$ is purely electric or dyonic.

The preceeding discussion makes apparent the asymmetric role of the $D=4$ three-form potential $\,\tilde{\cC}\,$ with four-form field strength $\tilde{\cH}_\4$ that was mentioned in \cite{Guarino:2015qaa}. Recall that these fields are related to the SL(7)--singlet magnetic component of the dyonic ISO(7) embedding tensor. In \cite{Guarino:2015qaa}, they were excluded from the restricted $D=4$ tensor hierarchy (\ref{eq:SL7fieldcontent4D}), even though $\tilde{\cH}_\4$ contributes to the $D=4$ scalar potential if $m \neq 0$, see (\ref{H4/Potential2}). The reason for their exclusion is that these fields do not enter the consistent embedding of ISO(7) supergravity into type IIA, at least directly: $\tilde{\cC}$ is absent from (\ref{KKEmbedding}) and $\tilde{\cH}_\4$ from (\ref{eq:10DFS}). For the purely electric gauging, $\tilde{\cC}$,   $\tilde{\cH}_\4$ do not play any role whatsoever: neither they appear in the consistent embedding formulae (into massless IIA) nor does $\tilde{\cH}_\4$ contribute to the $D=4$ scalar potential. For the dyonic, $m \neq 0$, ISO(7) gauging, $\tilde{\cH}_\4$ does not contribute either to the Freund-Rubin term (\ref{U_generalRewrite}). This is related to the fact that the magnetic components $\xi^{\prime \, AB}$ of the primed embedding tensor (\ref{Theta_Def_primed}) vanish, see (\ref{theta_Prime}). Only when the Bianchi identities are used to rewrite the Freund-Rubin factor in terms of the $D=4$ potential, as in (\ref{FRForms}), does $\tilde{\cH}_\4$ appear when $m \neq 0$. This (derived) role of  $\tilde{\cH}_\4$ in the embedding of dyonic ISO(7) supergravity into massive type IIA should be put down to the presence of a scalar potential (the last term of (\ref{10DLagrangian})) already in ten dimensions.

\section{\mbox{Truncation to the G$_2$ sector}} \label{sec:G2uplift}

We can now use the consistent truncation formulae (\ref{KKEmbedding})--(\ref{eq:dilaton}) to work out the explicit embedding of specific sectors of $\cN=8$ ISO(7)--gauged supergravity into massive type IIA. We do this in section \ref{app:G2UpliftSubsec} for the G$_2$--invariant  sector of the ISO(8) theory. In section \ref{app:mIIAon NK6} we show the overlap of this sector with the universal nearly  K\"ahler truncation of \cite{KashaniPoor:2007tr}.

\subsection{Explicit embedding formulae} 
\label{app:G2UpliftSubsec}


The bosonic field content of the G$_2$--invariant sector of ISO(7) supergravity includes, besides the metric, two real scalars $\varphi$, $\chi$ with dynamics governed by the Lagrangian (4.4), (4.5) of \cite{Guarino:2015qaa}. In addition, the restricted tensor hierarchy (\ref{eq:SL7fieldcontent4D}) gives rise in this sector to a singlet three-form $C$ with four-form field strength $H_\4 = d C$, so that
\begin{eqnarray} \label{ThreeFormPotG2}
\cC^{IJ} \equiv C \, \delta^{IJ}  \qquad , \qquad \cH_\4^{IJ} \equiv H_\4 \, \delta^{IJ} \; . 
\end{eqnarray}
The duality relation satisfied by $H_\4 $ reads \cite{Guarino:2015qaa}
\begin{equation} \label{H4DualityG2}
H_{\4} = \big[ g \, e^{\varphi } \, \big(1+ e^{2 \varphi }\, \chi ^2 \big)^2 \,  \big(5 - 7 \, e^{2 \varphi }\, \chi ^2 \big)  + m \, e^{7 \varphi } \, \chi ^3   \big] \, \textrm{vol}_4 \ .
\end{equation}
The $\cN=8$ scalar matrix ${\cal M}_{\mathbb{M}\mathbb{N} }$ evaluated in this sector has only 
$(\varphi, \chi)$-dependent components along the G$_2$--invariant tensors in $\mathbb{R}^7$: the delta symbol $\delta_{IJ}$, the associative three-form $\psi_{IJK}$ and the co-associative four-form $\tilde \psi_{IJKL}$. See appendix D.2 of  \cite{Guarino:2015qaa} for the explicit expressions.

Since the G$_2$--invariant sector does not contain vectors or two-forms, the embedding formulae (\ref{KKEmbedding}) simplify accordingly: only the three-form (\ref{ThreeFormPotG2}) and components of the IIA forms along $S^6$ are activated. Inserting the G$_2$--invariant scalar matrix (D.16) of \cite{Guarino:2015qaa} into (\ref{eq:internalmetric}), we find that the dependence of the internal metric in the $D=4$ scalars $\varphi$, $\chi$ factorises into an $S^6$--independent warp factor in front of the homogeneous metric $ds^2 (S^6)$ on $S^6 = \textrm{G}_2/\textrm{SU}(3)$ (see appendix \ref{subset:NKS6}). Moving to the expressions (\ref{eq:internalAm})--(\ref{eq:internalAmnp}) for the internal components of the IIA potentials, $A_m$, $B_{mn}$, $A_{mnp}$, we find that $\mu^I$  appear always contracted with the G$_2$--invariant forms $\psi_{IJK}$ and  $\tilde \psi_{IJKL}$ through the combinations (\ref{JOmegaintermsofmu}) that define the homogeneous nearly-K\"ahler forms ${\cal J}$, $\Upomega$ on $S^6 = \textrm{G}_2/\textrm{SU}(3)$. Finally, from (\ref{eq:dilaton}) we learn that  the IIA dilaton becomes a function of $\varphi$, $\chi$ only, and exhibits no $S^6$ dependence. The embedding formulae (\ref{KKEmbedding}) therefore reduce to
{\setlength\arraycolsep{0pt}
\begin{eqnarray} \label{G2embeddingGeom}
&& d \hat{s}_{10}^2 = e^{\frac{3}{4} \varphi} \big( 1+e^{2 \varphi} \chi^2 \big)^{\frac{3}{4}}  ds^2_4 + g^{-2} e^{-\frac{1}{4} \varphi} \big( 1+e^{2 \varphi} \chi^2 \big)^{-\frac{1}{4}}   ds^2(S^6) \; , \nonumber \\[5pt]
&& e^{\hat \phi} = e^{\frac{5}{2} \varphi} \big( 1+e^{2 \varphi} \chi^2 \big)^{-\frac32}  \; , \nonumber \\[5pt]
&&  \hat A_\3 = C + g^{-3} \chi \, \textrm{Im} \, \Upomega  \; , \qquad
 \hat B_\2 = g^{-2} \, e^{2 \varphi} \chi  \big( 1+e^{2 \varphi} \chi^2 \big)^{-1}   \,  {\cal J}  \; , \qquad   
 \hat A_\1 = 0 \; .
\end{eqnarray}
}The $S^6$ dependence in the term $\mu_I \mu_J  \, \cC^{IJ}$ of $ \hat A_\3$ also cancels upon substitution of the first relation in (\ref{ThreeFormPotG2}) and use of (\ref{eqSn}). Thus, the truncation of type IIA down to the G$_2$--invariant sector of ISO(7) supergravity retains only  deformations on $S^6 = \textrm{G}_2/\textrm{SU}(3)$ that respect the homogeneous G$_2$--invariant nearly-K\"ahler structure.

The IIA field strengths corresponding to the form potentials in (\ref{G2embeddingGeom}) can be computed from  (\ref{romfields}) using the nearly-K\"ahler differential relations (\ref{SU3strDif}). We obtain
\begin{eqnarray} \label{G2embeddingGeomFieldStr}
\hat F_{(4)} &=&   \left[ g \, e^\varphi \big( 1+ e^{2\varphi} \chi^2 \big)^2 \big( 5 - 7 e^{2\varphi} \chi^2  \big) 
+m \,e^{7\varphi} \chi^3 \right] \textrm{vol}_4  
+ g^{-3}  d \chi \wedge \textrm{Im} \,  \Upomega   \nonumber \\
&& \qquad \quad \;\; +   \left[ \tfrac{1}{2} m g^{-4} \,  e^{4 \varphi}  \, \chi^2  \, \big( 1+e^{2 \varphi} \chi^2 \big)^{-2}  -2 g^{-3} \,  \chi  \right] {\cal J} \wedge {\cal J}  \; ,\nonumber \\[5pt]
\hat H_{(3)} &=& g^{-2}  d \Big(  e^{2 \varphi}  \, \chi  \, \big( 1+e^{2 \varphi} \chi^2 \big)^{-1} \Big) \wedge {\cal J} +3 g^{-2} \,   e^{2 \varphi}  \, \chi  \, \big( 1+e^{2 \varphi} \chi^2 \big)^{-1}  \, \textrm{Re} \, \Upomega \; , \nonumber \\[5pt]
\hat F_{(2)} &=& m g^{-2} \,   e^{2 \varphi}  \, \chi  \, \big( 1+e^{2 \varphi} \chi^2 \big)^{-1}  \, {\cal J} \; . 
\end{eqnarray}
Of the terms explicitly shown in the $\cN=8$ expressions (\ref{eq:10DFS}), only (the G$_2$--invariant truncation of) the Freund-Rubin term, $ \cH_{\4}^{IJ}  \, \mu_{I} \, \mu_{J}  $, is present here. We have used the second equation in (\ref{ThreeFormPotG2}), equation (\ref{eqSn}) and the dualisation relation (\ref{H4DualityG2}) to bring it to the form presented in (\ref{G2embeddingGeomFieldStr}). Equivalently, this expression for the Freund-Rubin term can be also obtained from the master formula  (\ref{U_generalRewrite}). The remaining terms shown in (\ref{eq:10DFS}) vanish in the G$_2$--invariant sector and thus do not appear in (\ref{G2embeddingGeomFieldStr}). All terms in (\ref{G2embeddingGeomFieldStr}) except Freund-Rubin hide behind the ellipses of (\ref{eq:10DFS}). 

We have explicitly verified the consistency of the truncation in this sector at the level of the equations of motion, including the Einstein equation. When the metric and dilaton in (\ref{G2embeddingGeom}) and the field strengths (\ref{G2embeddingGeomFieldStr}) are inserted into the type IIA equations of motion (\ref{d10eom}), all $S^6$ dependence drops out, and the $D=4$ field equations that derive from the G$_2$--invariant Lagrangian (4.4), (4.5) of \cite{Guarino:2015qaa} arise.

Given that they depend on the natural nearly-K\"ahler structure on $S^6$ only, the truncation formulae (\ref{G2embeddingGeom}), (\ref{G2embeddingGeomFieldStr}) from massive IIA to the G$_2$--invariant sector of dyonic ISO(7) supergravity are still valid if $S^6$ is replaced with an arbitrary nearly-K\"ahler six-manifold\footnote{The theory (4.4), (4.5) of \cite{Guarino:2015qaa} will no longer arise  necessarily as a consistent subsector of a larger $D=4$ supergravity.}. In \cite{KashaniPoor:2007tr}, a consistent truncation of massive IIA also valid for any nearly-K\"ahler six-manifold was constructed. We now turn to show that a further subtruncation of the $D=4$ theory of \cite{KashaniPoor:2007tr} coincides with the G$_2$--invariant sector of dyonic ISO(7) supergravity.

\subsection{Overlap with the universal nearly-K\"ahler truncation} 
\label{app:mIIAon NK6}


Massive type IIA supergravity can be consistently truncated on any nearly K\"ahler six-manifold $M_6$ by expanding the IIA fields along the nearly K\"ahler SU(3)-structure forms on $M_6$ \cite{KashaniPoor:2007tr}. See appendix \ref{subset:NKS6} for a brief description of nearly-K\"ahler geometry. The resulting $D=4$ supergravity is a `massive mode' truncation, in the sense discussed in the introduction. This $D=4$ theory corresponds to $\cN=2$ supergravity, coupled to one vector multiplet and one hypermultiplet, and with an abelian, dyonic SO$(1,1)^2$ gauging in the hyper sector. Here we will only review the ingredients of this theory and its IIA origin necessary to exhibit the overlap with $\cN=8$ ISO(7) dyonic supergravity, in a notation  that is useful for that purpose. Please refer to \cite{KashaniPoor:2007tr} for further details, including the $\cN=2$ special geometry of theory, and to \cite{Cassani:2009ck} for a rederivation of this truncation specific to the homogeneous nearly-K\"ahler structure on $S^6 = \textrm{G}_2/\textrm{SU}(3)$.

The universal nearly-K\"ahler truncation to $D=4$  \cite{KashaniPoor:2007tr} contains, besides the metric $ds_4^2$, the following bosonic fields, all of them real: five  scalars $U$, $\phi$, $b$, $\xi$, $\tilde \xi$, two vectors $A_1$, $B_1$, one two-form $B_2$, and one three-form $C_3$. In addition, the theory is characterised by two coupling constants, $g$ and $m$. The former descends from an overall scale in $M_6$ ({\it i.e.} its `inverse radius') and the latter descends directly from the Romans mass\footnote{ 
The $g \rightarrow 0$ limiting theory can be obtained by first rescaling 
\begin{eqnarray}
&& ds_4^2 = g^{6} ds_4^{\prime \, 2} \; , \quad 
e^U = g \, e^{U^\prime} \; , \quad 
e^\phi =  e^{\phi^\prime} \; , \quad 
b =  g^{2} \, b^\prime \; , \quad 
\tilde \xi =  g^{3} \, \tilde \xi^\prime \; , \nonumber \\
&& B_2 = B_2^\prime \; , \quad 
A_1 = A_1^\prime \; , \quad 
B_1 = g^{2}  B_1^\prime \; , 
\end{eqnarray}
and the flux $e_0 \equiv 5g$ entering the first term in the scalar potential (\ref{PotNK6}) as $e_0 = g^{5} e_0^\prime$, and then letting $g \rightarrow 0$. This limit is smooth at the level of the equations of motion and, if (\ref{eq:NKLagrangian}) is also rescaled as ${\cal L} = g^{6} {\cal L}^\prime$, also at the level of the Lagrangian. In this limit, $M_6$ becomes Calabi-Yau and  the resulting $D=4$ $g \rightarrow 0$, $m \neq 0$ theory describes a consistent truncation of massive type IIA valid for any Calabi-Yau manifold. The $D=4$ $g \neq 0$, $m=0$ theory arises upon truncation of massless IIA on any nearly-K\"ahler manifold.
}. The field strengths are given by 
\begin{eqnarray} \label{eq:NKPotForms}
F_2 = dA_1 + m B_2 \; , \quad 
H_2 = dB_1 +A_1 \wedge db +mb B_2  \; , \quad
H_3 = dB_2 \; , \quad
H_4 = dC_3 \; , 
\end{eqnarray}
and are subject to the Bianchi identities
\begin{eqnarray} \label{eq:NKBianchis}
d F_2 - m H_3 = 0 \; , \qquad 
d H_2 - F_2 \wedge db - mb H_3 =0 \; , \qquad 
dH_3 = 0 \; , \qquad 
dH_4 \equiv  0 \; .
\end{eqnarray}
The equations of motion derive from the Lagrangian
\begin{eqnarray} \label{eq:NKLagrangian}
{\cal L} & =& R \, \textrm{vol}_4 +24 \,  dU \wedge * dU +\tfrac12 \, d\phi \wedge * d\phi +\tfrac32 e^{-\phi-4U}db \wedge * db \nonumber \\
&&   +\tfrac12 e^{\frac12\phi -6U} D\xi \wedge *D\xi +\tfrac12 e^{\frac12 \phi-6U} d\tilde \xi \wedge * d\tilde \xi +\tfrac12 e^{-\phi +12U} H_3 \wedge * H_3 \nonumber \\
&& -\tfrac12 e^{\frac32 \phi +6U} F_2 \wedge *F_2  -\tfrac32 e^{\frac12 \phi +2U} H_2 \wedge *H_2 + {\cal L}_{\textrm{top}} 
 -V \textrm{vol}_4 \; . 
\end{eqnarray}
Here, $D\xi = d \xi -6g B_1+6g bA_1 $, the contribution ${\cal L}_{\textrm{top}}$ is a topological term we do not need to specify, and the scalar potential $V$ is given by 
\begin{eqnarray} \label{PotNK6}
V &=& \tfrac12 e^{-\frac12\phi-18U} \big(5g+m b^3+6 g  b  \tilde\xi \, \big)^2 
+ \tfrac32 e^{\frac12 \phi -14U} \big( mb^2 + 2g \tilde \xi \big)^2 \nonumber \\
&&  -30 g^2  e^{-8U} +18 g^2 \,  b^2 e^{-\phi-12U} +\tfrac32 m^2 b^2 e^{\frac32 \phi -10U}  +\tfrac12 m^2 e^{\frac52 \phi -6U}   \; .
\end{eqnarray}

The embedding of this $D=4$ theory into massive type IIA utilises exclusively the nearly-K\"ahler structure ${\cal J}$, $\Upomega$ on $M_6$ and its associated metric $ds^2(M_6)$. More concretely, using the type IIA conventions of appendix \ref{app:mIIAConventions}, the four-dimensional metric $ds_4^2$ and the scalar $U$ enter the ten-dimensional metric as
\begin{eqnarray} \label{eq:MetricNK}
d\hat{s}_{10}^2 = e^{-6U} ds_4^2 + g^{-2} \,   e^{2U} ds^2(M_6) \; , 
\end{eqnarray}
while the scalars $\phi$, $b$, $\xi$, $\tilde \xi$ and the forms $C_3$, $B_2$, $A_1$, $B_1$ enter the ten-dimensional fields as
\begin{eqnarray} \label{eq:NKForms}
&&  \hat A_\3 = C_3 + g^{-2} B_1 \wedge {\cal J} + \tfrac12 \, g^{-3} \, \xi \,  \textrm{Re} \, \Upomega - \tfrac12 \,g^{-3} \, \tilde \xi \,  \textrm{Im} \, \Upomega \; , \qquad \qquad  \qquad  \nonumber \\[5pt]
&&  \hat B_\2 = B_2 + g^{-2} \, b \,  {\cal J} \; , \qquad  \qquad 
 \hat A_\1 = A_1\; , \qquad \qquad
\hat \phi = \phi \  .\qquad \qquad
\end{eqnarray}
The field strengths can be computed using (\ref{romfields}). With the help of the differential relations (\ref{SU3strDif}), these can be written as
{\setlength\arraycolsep{1pt}
\begin{eqnarray} \label{eq:NKFieldStr} 
 \hat F_\4 & =  & e^{-\frac12 \phi -18U} \big( 5g +mb^3+6g b\tilde\xi \, \big) \, \textrm{vol}_4 + g^{-2} H_2 \wedge {\cal J} +  \tfrac12 \, g^{-3} D \xi \wedge \textrm{Re} \, \Upomega -\tfrac12 \,  g^{-3} d\tilde{\xi} \wedge \textrm{Im} \, \Upomega   \nonumber \\
&& + \tfrac12 g^{-4} \big( mb^2+ 2g  \tilde \xi  \big) \, {\cal J} \wedge {\cal J} \; ,  \nonumber \\[5pt]
\hat H_\3 &=& H_3 +g^{-2} db \wedge {\cal J} +3g^{-2} b \, \textrm{Re} \, \Upomega \; ,  \\[5pt]
\hat F_\2 &=& F_2 + m g^{-2} b \, {\cal J} \; . \nonumber 
\end{eqnarray}
}Here $H_3$, $H_2$ and $F_2$ are the four-dimensional field strengths defined in (\ref{eq:NKPotForms}), $H_4$ in the Freund-Rubin term has been dualised into scalars by using the type IIA field equations, and $D\xi$ has been defined under (\ref{eq:NKLagrangian}). Equations (\ref{eq:MetricNK})--(\ref{eq:NKFieldStr}) define the consistent truncation discussed in \cite{KashaniPoor:2007tr} of massive type IIA to the $D=4$ theory (\ref{eq:NKLagrangian}), (\ref{PotNK6}). Some algebra indeed allows one to verify the consistency of the truncation. When (\ref{eq:MetricNK})--(\ref{eq:NKFieldStr}) are inserted into the Bianchi identities (\ref{d10Bianchis}) and equations of motion (\ref{d10eom}) of massive type IIA, all $M_6$ dependence drops out and the $D=4$  Bianchi identities (\ref{eq:NKBianchis}) and the equations of motion that derive from the Lagrangian (\ref{eq:NKLagrangian}) arise.

Now, we find that the following identifications
{\setlength\arraycolsep{0pt}
\begin{eqnarray} \label{eq:SubtrunctoG2}
&&
e^{\phi} = e^{\frac{5}{2} \varphi } \big( 1+e^{2\varphi} \chi^2 \big)^{-\frac32}  ,\ 
e^{-8U} =  e^{ \varphi } \big( 1+e^{2\varphi} \chi^2 \big)  , \ 
b =e^{ 2\varphi } \chi  \big( 1+e^{2\varphi} \chi^2 \big)^{-1}   , \
\xi = 0  , \
\tilde{\xi} = -2\chi  \nonumber  \\
&& B_2= 0 \; , \qquad A_1 = B_1 = 0   \; , \qquad C_3 = C \; ,   
\end{eqnarray}
}define a further consistent truncation of the $D=4$ theory (\ref{eq:NKLagrangian}) to a theory containing only the metric $ds_4^2$ and two real scalars $\varphi$, $\chi$, along with the three-form potential $C$. We have indeed verified that the equations of motion that derive from the Lagrangian   (\ref{eq:NKLagrangian}) consistently truncate under (\ref{eq:SubtrunctoG2}) to equations of motion for  $ds_4^2$, $\varphi$, $\chi$ that can be integrated into the Lagrangian
\begin{eqnarray} \label{G2sectorGeomFrom NK6}
{\cal L} = R \, \textrm{vol}_4 +\tfrac{7}{2} \, d \varphi \wedge * d \varphi  + \tfrac{7}{2} \,e^{2\varphi} d \chi \wedge * d \chi 
-V \textrm{vol}_4 \ , 
\end{eqnarray}
with scalar potential
\begin{equation} \label{PotG2mgGeomFromNK6}
V = \tfrac72 \, g^2 \, e^\varphi \big( 1+ e^{2\varphi} \chi^2 \big)^2 \big( -5 + 7 e^{2\varphi} \chi^2  \big) 
-7gm \,e^{7\varphi} \chi^3
 +\tfrac12 m^2 e^{7\varphi}  \, .
\end{equation}
This turns out to coincide with the G$_2$--invariant sector of $\cN=8$ dyonically-gauged ISO(7) supergravity: see equations (4.4), (4.5) of \cite{Guarino:2015qaa} with the couplings $g$, $m$ there identified with those here. We have also checked that this truncation works out at the level of the Lagrangian as well: (\ref{eq:NKLagrangian}), (\ref{PotNK6}) evaluated on (\ref{eq:SubtrunctoG2}) reduce to (\ref{G2sectorGeomFrom NK6}), (\ref{PotG2mgGeomFromNK6}). It is also easy to verify that the nearly-K\"ahler truncation formulae (\ref{eq:MetricNK})--(\ref{eq:NKFieldStr}) evaluated on (\ref{eq:SubtrunctoG2}) coincide with those, (\ref{G2embeddingGeom}), (\ref{G2embeddingGeomFieldStr}), corresponding to the truncation of massive IIA down to the G$_2$--invariant sector of dyonic ISO(7) supergravity.

\begin{table}[t!]
\centering
\footnotesize{
\begin{tabular}{ccc}
\hline
	susy  	& bos. sym. & $M^2 L^2$
\\[2pt]
\hline
\\[-10pt]
 $ \cN=1 $ &	G$_2$ &	$ \tfrac{47}{3} \pm \sqrt{\tfrac{53}{3}}  \ , \;    \bm{4 \pm \sqrt{6} }   \ , \;  0 \ , \;  0 $ 
\\[10pt]
 $ \cN=0 $ &	G$_2$   &	$ 20  \ , \;   20  \ , \;  \bm{6}  \ , \;   \bm{6}  \ , \;   0   \ , \;   0 $ 
\\[10pt]
 $ \cN=0 $ &	SO$(7)_+$  &	$ 20  \ , \;   \tfrac{64}{5}  \ , \;  \bm{6}  \ , \;   \mathbf{- \tfrac{6}{5}}   \ , \;   0   \ , \;   0 $ 
\\[5pt]
\hline
\end{tabular}
\\[10pt]	
\caption[4DN=2Sols]{  \small{Summary of critical points of the potential (\ref{PotNK6}) of the nearly-K\"ahler truncation of massive type IIA. For each critical point it is shown the supersymmetry $\cN$, the bosonic symmetry of the massive type IIA uplift and the scalar spectrum within the theory (\ref{eq:NKLagrangian}), (\ref{PotNK6}). Masses in bold overlap with masses in the corresponding critical points of $\cN=8$ dyonic ISO(7) supergravity. The zeroes correspond to Goldstone bosons that give mass to the vectors of the theory (\ref{eq:NKLagrangian}). }}\label{Table:4DN=2Sols} 
}\normalsize
\end{table}

The theory (\ref{eq:NKLagrangian}), (\ref{PotNK6}) contains three AdS vacua \cite{Cassani:2009ck}. All three can be checked to survive the subtruncation (\ref{eq:SubtrunctoG2}); thus, they are also vacua of $\cN=8$ dyonic ISO(7) supergravity. When regarded as vacua of the latter, these correspond to the critical points with G$_2$ residual symmetry and $\cN=1$, $\cN=0$ supersymmetry, and with SO$(7)_+$, $\cN=0$. These are also the symmetries of the corresponding massive IIA uplifts. We have computed the scalar mass spectrum of these critical points within the nearly-K\"ahler truncation (\ref{eq:NKLagrangian}), (\ref{PotNK6}) and have brought the result to table \ref{Table:4DN=2Sols} above. The spectrum of these points within the $\cN=8$ ISO(7) theory can be found in table 1 of \cite{Guarino:2015qaa}. Comparison of both tables makes apparent the overlap between the two theories. Indeed, for the common critical points, all spectra have two scalars (corresponding to $\varphi$, $\chi$) with the same mass in both tables. These appear in bold in table \ref{Table:4DN=2Sols} above.  In the $\cN=8$ table, these correspond to the G$_2$ singlets for the G$_2$ points. For the SO$(7)_+$ point, these scalars correspond to the explicit singlet and the singlet that arises in the 
branching $\mathbf{35} \rightarrow \mathbf{1}  + \mathbf{7}  + \mathbf{27}$ of SO(7) under G$_2$ (the other SO(7) representations, $\bm{27}$ and $\bm{7}$, in the spectrum of this point are irreducible under G$_2$). The masses in table \ref{Table:4DN=2Sols} above which do not overlap with the $\cN=8$ table are larger than those which do overlap. Thus, they correspond to higher KK modes about the corresponding massive IIA solutions. This discussion endorses the interpretation put forward in the introduction of $\cN=8$ dyonic ISO(7) supergravity and the nearly-K\"ahler truncation of  \cite{KashaniPoor:2007tr} as `massless' and `massive' truncations of massive type IIA on $S^6$.

\section{Discussion} \label{sec:Discussion}

We have given a detailed derivation of the consistent truncation of massive type IIA supergravity on the six-sphere down to ISO(7)-dyonically-gauged $D=4$ $\cN=8$ supergravity that was recently announced in \cite{Guarino:2015jca}. In order to do this, we first rewrote the field content and supersymmetry variations of (massive) type IIA supergravity with only manifest $D=4$ local Lorentz covariance. We then employed a suitable restriction \cite{Guarino:2015qaa} of the $D=4$ $\cN=8$ tensor \cite{deWit:2008ta,deWit:2008gc} and duality hierarchies  \cite{Bergshoeff:2009ph} to obtain the complete non-linear embedding of {\it all} $D=4$ fields into type IIA. The bosonic sector of type IIA supergravity develops a manifest SL(7)--covariance when rewritten with only explicit local Lorentz-four-dimensional symmetry. In section \ref{app:DerivationUplift} we provided an analysis of this $\textrm{SO}(1,3) \times \textrm{SL}(7)$ reformulation of type IIA supergravity, including the resulting formally four-dimensional two-form and three-form potentials. 

The results of section \ref{app:DerivationUplift} are adapted to accommodate any possible consistent truncation of either massless or massive type IIA supergravity to $D=4$ $\cN=8$ gauged supergravity. Any possible such consistent embedding is naturally determined in this formalism at the level of the metric and IIA form potentials. These are independent of the Romans mass, $\hat F_\0$. The proof of consistency at the level of the supersymmetry transformations of the bosons is also independent of $\hat F_\0$, since the latter does not appear in their IIA variations: see (\ref{IIAsusyVars}). In the bosonic sector, the Romans mass only appears in the field strengths and the covariant derivatives. Its role as a magnetic coupling upon reduction to four dimensions has long been known \cite{Polchinski:1995sm,Louis:2002ny}. These observations  imply that, whenever a consistent truncation of massless IIA down to a $D=4$ (gauged) supergravity exists, a truncation of massive IIA to some dyonic counterpart of that $D=4$ gauged supergravity is guaranteed to exist. The consistent embedding formulae at the level of metric and IIA potentials and the proof of consistency are the same for $\hat F_\0 =0$  and $\hat F_\0  \neq 0$. Of course, this discussion applies regardless of the compactification manifold and regardless of the number of supercharges preserved by the truncation.

For example, it is well known that the truncation of massless type IIA (or $D=11$ supergravity) on $T^6$ to ungauged $D=4$ $\cN=8$ supergravity is consistent. By the preceeding arguments, the consistency of the truncation of massive IIA on $T^6$ to the $g=0$, $m \neq 0$ purely magnetic, non-semisimple gauging briefly touched upon in section 2.3 of \cite{Guarino:2015qaa} is also consistent. The consistency of these toroidal truncations is actually guaranteed by group theory, but there exist more non-trivial examples which lack a priori such protection. For example, the truncation of massless type IIA on $\mathbb{CP}^3$ down to to $\cN=6$ $\textrm{SU}(4) \times \textrm{U}(1)$--gauged supergravity is consistent (and the embedding of an $\cN=4$ subsector is explicitly known \cite{Cvetic:2000yp}). The above arguments predict a similar consistent truncation of massive type IIA to a dyonic deformation of that  $D=4$ theory (where the U(1) will no longer be necessarily compact). These arguments presumably also apply to reductions to other dimensions, with $\hat F_\0$ still appearing only at the level of the field strengths and turning on magnetic couplings with respect to higher-rank tensors (in even dimensions), or altogether released from that duty (in odd dimensions). For example, maximally supersymmetric consistent truncations of massive type IIA on $S^3$ and $S^4$ should exist associated to the massless IIA truncations of \cite{Cvetic:2000ah}. In particular, this $S^4$ truncation should extend the half-maximal truncation of \cite{Cvetic:1999un} into maximal supersymmetry. An interesting question would be whether the corresponding gauge groups, ISO(4) and ISO(5), remain the same in the deformed theory. 

Our results are based on an SL(7)--covariant rewrite of the field content and supersymmetry transformations of type IIA. It was not our intention to extend this SL(7) to a full E$_{7(7)}$ covariance. Extensions to exceptional symmetries have been recently studied, following the same approach, in $D=11$ \cite{Godazgar:2013dma} and type IIB \cite{Ciceri:2014wya} and, following other approaches, in \cite{Coimbra:2011ky,Coimbra:2012af,Hohm:2013pua,Hohm:2013vpa,Hohm:2013uia}. In sight of \cite{Godazgar:2013dma,Ciceri:2014wya}, the realisation of E$_{7(7)}$ covariance in type IIA would require to start from the democratic formulation \cite{Bergshoeff:2001pv} containing both regular and dual fields, rather than the conventional formulation  \cite{Romans:1985tz} that we have used in this work.

Although extremely interesting in itself, the realisation of exceptional symmetries in the higher dimensional supergravities is not a prerequisite to establish consistent truncations at the full non-linear level in all the lower-dimensional fields, as our results show. At least this is so for lower dimensional theories, like $D=4$, where the dimension of the adjoint of any gauge group must be less that the dimension of the fundamental of the relevant exceptional group. This is indeed the case provided the higher-rank forms are also incorporated into the analysis, as we have done in this paper with the two- and three-form potentials. A crucial step is the study of the non-linear field redefinitions which all forms, including those of higher rank, must undergo in order for their supersymmetry variations to be compatible with the lower-dimensional tensor hierarchy. When this formalism is used to study consistent truncations, the resulting embeddings become naturally written, at the level of the metric and form potentials, in terms of a restricted tensor hierarchy in the lower dimension with closed field equations and supersymmetry transformations. This restricted tensor hierarchy necessarily contains redundant degrees of freedom carried by the forms of higher rank. However, this redundancy can always be removed at the level of the field strengths, by exploting the (restricted) duality hierarchy \cite{Bergshoeff:2009ph}. One can still insist in writing the KK embedding in terms of (redundant) dual fields in the higher dimension (see {\it e.g.}~\cite{Godazgar:2013pfa} for the $D=11$ on $S^7$ case). This approach should be equivalent to using the lower-dimensional restricted tensor and duality hierarchies.

Using the restricted duality hierarchy method we were able to find, in particular, simple formulae for the Freund-Rubin term of the $\cN=8$ truncation of massive IIA on $S^6$. The formula (\ref{U_generalRewriteIntro}) and its formal resemblance to the $D=4$ $\cN=8$ scalar potential (\ref{V_generalRewriteIntro}) (particularised to the ISO(7) gauging) makes the pattern obvious for the Freund-Rubin term corresponding to other consistent truncations, spherical or otherwise. For example, similar steps lead to a Freund-Rubin term for the $S^7$ truncation of $D=11$ supergravity given by  (\ref{U_generalRewriteIntro}) with the primed embedding tensor constructed via (\ref{Theta_Def_primed}), (\ref{X_primed}) with $\theta^\prime_{AB} = \mu_{A} \mu_{B}$ now in the $\bm{36}$ of SL(8), like the embedding tensor for the electric SO(8) gauging, and $\xi^{\prime AB} = 0$ as well. Likewise, the Freund-Rubin term $\hat F_\5 = U \, \textrm{vol}_5 +\ldots$ for the compactification of type IIB on $S^5$ to $D=5$ $\cN=8$ SO(6)-gauged supergravity is given by
\begin{equation}
\begin{array}{ccl}
\label{U_generalIIB}
U & = & - \dfrac{g}{15} \,  {X^\prime_{\mathbb{MP}}}^{\mathbb{R}}  {X_{\mathbb{NQ}}}^{\mathbb{S}}  \mathcal{M}^{\mathbb{MN}}   \Big(  \mathcal{M}^{\mathbb{PQ}}  \mathcal{M}_{\mathbb{RS}}  +   5 \,   \delta^\mathbb{P}_\mathbb{S} \,  \delta^\mathbb{Q}_\mathbb{R}  \Big) \; , 
\end{array}
\end{equation}
paralelling the scalar potential of $D=5$ $\cN=8$ supergravity\footnote{This form of the $D=5$ $\cN=8$ scalar potential, which is a rewrite of that in \cite{deWit:2004nw}, was given in \cite{Baguet:2015sma}.},
\begin{equation}
\begin{array}{ccl}
\label{V_generalIIB}
V & = & \dfrac{g^{2}}{30} \,  {X_{\mathbb{MP}}}^{\mathbb{R}}  {X_{\mathbb{NQ}}}^{\mathbb{S}}  \mathcal{M}^{\mathbb{MN}}   \Big(  \mathcal{M}^{\mathbb{PQ}}  \mathcal{M}_{\mathbb{RS}}  +   5 \,   \delta^\mathbb{P}_\mathbb{S} \,  \delta^\mathbb{Q}_\mathbb{R}  \Big) \ ,
\end{array}
\end{equation}
particularised to the SO(6)-gauging. By a construction analogue to (\ref{Theta_Def_primed})--(\ref{theta_Prime}), ${X^\prime_{\mathbb{MP}}}^{\mathbb{R}}$ in (\ref{U_generalIIB}) depends on the $\mu^I$ of section \ref{subset:GeneralSn} with $n=5$ through the combination $\theta^\prime_{IJ} \equiv \mu_I \mu_J$, which now sits in the $(\bm{21} , \bm{1})$ of $\textrm{SL}(6) \times \textrm{SL}(2)$, like the embedding tensor of SO(6)-gauged supergravity. The rest of the symbols here are the usual ones of $D=5$ $\cN=8$ supergravity \cite{deWit:2004nw}. In particular,  $\mathcal{M}_{\mathbb{MN}}$ here is the square of the E$_{6(6)}/$Sp(8) coset representative. The type IIB Freund-Rubin term (\ref{U_generalIIB}) can be checked to reduce to that of the SL(2)--invariant truncation given in \cite{Cvetic:2000nc}. An analogue construction holds for the remaining well-known spherical truncation, $D=11$ supergravity on $S^4$ \cite{Nastase:1999cb,Nastase:1999kf}, as is perhaps more apparent from the notation used in \cite{Cvetic:2000ah}. The Freund-Rubin terms of other truncations, including the massive IIA truncations mentioned above and {\it e.g.}~the hyperbolic truncations of \cite{Baron:2014bya,Hohm:2014qga} to non-compact gaugings, should all follow similar patterns.

As discussed in the introduction, maximal supergravities in dimensions 4, 7 and 5 with SO(8) \cite{deWit:1982ig}, SO(5) \cite{Pernici:1984xx} and SO(6) \cite{Gunaydin:1984qu} gaugings are very useful to study holographically consistent sectors of the M2, M5 and D3 brane superconformal field theories. Similarly, the dyonic ISO(7) gauging in $D=4$ will be relevant to the study of superconformal phases, with reduced supersymmetry, of the D2-brane field theory. The Romans mass induces a Chern-Simons term on the D2-brane worldvolume, and this triggers flows from $\cN=8$ three-dimensional super-Yang-Mills to the type of superconformal Chern-Simons-matter theories discussed in \cite{Schwarz:2004yj}. A precise holographic match was given in \cite{Guarino:2015jca}. See \cite{Fluder:2015eoa} for further related work. It would be interesting to construct these flows holographically in dyonically gauged ISO(7) supergravity.

We will return to these interesting questions in future research.

\section*{Acknowledgements}

We thank Daniel Jafferis for collaboration in related projects and \mbox{Franz Ciceri,} Bernard de Wit and Gianluca Inverso for discussions. AG is supported in part by the ERC Advanced Grant no. 246974, {\it Supersymmetry: a window to non-perturbative physics}.  OV is supported by the Marie Curie fellowship PIOF-GA-2012-328798, managed from the CPHT of \'Ecole Polytechnique, and partially by the Fundamental Laws Initiative at Harvard.

\appendix

\addtocontents{toc}{\setcounter{tocdepth}{1}}

\section{Massive type IIA supergravity} 
\label{app:mIIAConventions}

We follow the conventions of \cite{Cvetic:1999un} for massive type IIA supergravity \cite{Romans:1985tz}, with the Neveu-Schwarz fields denoted here as $\hat H_3 = d \hat B_\2$.  In these conventions, the Einstein frame Lagrangian for the the bosonic sector reads
%
\begin{eqnarray}
2 \hat{\kappa}^2 {\cal L}_{10} \!\! &=& \!\! \hat R\,  \hat{\textrm{vol}}_{10} +
\ft12 d\hat \phi \wedge {\hat *d\hat \phi}
- \ft12 e^{\fft32\hat\phi}\,  \hat F_\2 \wedge   {\hat *\hat F_\2} + 
\ft12 e^{-\hat \phi}\, \hat H_\3 \wedge {\hat *\hat H_\3} - 
\ft12 e^{\fft12\hat\phi}\, \hat F_\4 \wedge  {\hat *\hat F_\4}  \nonumber \\
&& \!\!
 -\ft12 (d\hat A_\3 )^2  \wedge \hat B_\2 - \ft16 m\, 
 d\hat A_\3 \wedge (\hat B_\2)^3
 -\ft1{40} m^2\, (\hat B_\2)^5 -\ft12 m^2\, e^{\fft52\hat \phi}\, 
  \hat{\textrm{vol}}_{10} \,,\label{10DLagrangian}
\end{eqnarray}
%
where $\hat R$, $\hat{\textrm{vol}}_{10}$ and $\hat *$ are the ten-dimensional Ricci scalar, volume form and Hodge dual, the subscripts indicate the degree of the forms,  $(d\hat A_\3 )^2 \equiv d\hat A_\3  \wedge d\hat A_\3 $, etc., and hats over all quantities are used to emphasise their 10-dimensional character. The ten-dimensional gravitational coupling constant is given in terms of the string length $\ell_s = \sqrt{\alpha^\prime}$ by $2\hat\kappa^2 = (2\pi)^7 \ell_s^8$. The dilaton is denoted by $\hat \phi$ and the Romans mass by 
\begin{equation} \label{RomansMass}
\hat F_\0 \equiv m \; .
\end{equation}

The field strengths $ \hat F_\4$,  $\hat H_\3$, $ \hat F_\2$ are subject to the Bianchi identities
\begin{eqnarray}
d\hat F_\4 - \hat F_\2\wedge \hat H_\3 = 0 \,,\qquad
d\hat H_\3=0\,,\qquad d\hat F_\2 - m\,\hat H_\3 = 0 \,,
\label{d10Bianchis}
\end{eqnarray}
which can be integrated in terms of gauge potentials, $ \hat A_\3$,  $\hat B_\2$ entering the above Lagrangian,  and  $ \hat A_\1$ as
%
\begin{eqnarray}
\hat F_\4 = d\hat A_\3 + \hat A_\1\wedge d\hat B_\2 + \ft12 m\, 
\hat B_\2\wedge \hat B_\2 \, , \quad 
\hat H_\3 =  d\hat B_\2\,, \quad 
\hat F_\2 = d\hat A_\1 + m\, \hat B_\2  .
\label{romfields}
\end{eqnarray}
The equations of motion that derive from the Lagrangian (\ref{10DLagrangian}) read
%
\begin{eqnarray} \label{d10eom}
&&d(e^{\fft12\hat \phi}\, {\hat *\hat F_\4}) + \hat H_\3\wedge \hat
F_\4 =0 \ ,\qquad d(e^{\fft32\hat \phi}\, {\hat * \hat F_\2}) +
e^{\fft12\hat \phi}\, \hat H_\3 \wedge {\hat *\hat F_\4} =0  \,,\nonumber \\[8pt]
&&d(e^{-\hat\phi}\, {\hat * \hat H_\3}) + \ft12 \hat F_\4\wedge \hat
F_\4 + m\, e^{\fft32\hat\phi}\, {\hat * \hat F_\2} +
e^{\fft12\hat \phi}\, \hat F_\2\wedge  {\hat * \hat F_\4}  = 0 \,,   \\[8pt]
&&d{\hat * d\hat \phi} + \ft54m^2\, e^{\fft52\hat\phi}\, 
\hat{\textrm{vol}}_{10}+ \ft34 e^{\fft32\hat \phi}\, F_\2 \wedge
 {\hat *\hat F_\2}  + \ft12 e^{-\hat \phi}\,  \hat H_\3 \wedge {\hat *\hat H_\3}
+ \ft14 e^{\fft12\hat\phi}\, \hat F_\4 \wedge {\hat *\hat F_\4} = 0 \,, \nonumber \\[8pt]
&& \hat R_{MN} = \tfrac12 \partial_M \hat \phi \,   \partial_N \hat \phi + \tfrac{1}{16} m^2 \,e^{\frac52 \hat \phi} \,  \hat g_{MN} 
+\tfrac{1}{12} e^{\frac12 \hat \phi} \big( \hat F_{MPQR} \hat F_N{}^{PQR} - \tfrac{3}{32} \hat g_{MN} \hat F_{PQRS} \hat F^{PQRS} \big) \nonumber \\
&& +\tfrac{1}{4} e^{- \hat \phi} \big( \hat H_{MPQ} \hat H_N{}^{PQ} - \tfrac{1}{12} \hat g_{MN} \hat H_{PQR} \hat H^{PQR} \big)
+\tfrac{1}{2} e^{ \frac32 \hat \phi} \big( \hat F_{MP} \hat F_N{}^{P} - \tfrac{1}{16} \hat g_{MN} \hat F_{PQ} \hat F^{PQ} \big) \; , 
\nonumber
\end{eqnarray}
%
where we have introduced local indices, $M, N, \ldots = 0 , 1, \ldots , 9$, in order to write the Einstein equation. Introducing also tangent space indices  $A, B, \ldots = 0 , 1, \ldots , 9$,  the supersymmetry variations of the bosonic fields to first order in fermions read
\begin{eqnarray} \label{IIAsusyVars}
&& \delta   \hat{e}_M{}^A  = \tfrac{1}{4} \,  \bar{\hat{\epsilon}} \, \hat{\Gamma}^A \,  \hat{\psi}_M \; , \nonumber \\[6pt]
&& \delta   \hat{\phi} = \tfrac{\sqrt{2}}{4}  \,   \bar{\hat{\epsilon}} \, \hat{\Gamma}_{11} \,  \hat{\lambda} \;,  \nonumber \\[6pt]
&& \delta   \hat{A}_M  = \tfrac{1}{4} \, e^{-\frac34 \hat{\phi}} \,  \bar{\hat{\epsilon}} \, \hat{\Gamma}_{11} \,  \hat{\psi}_M +  \tfrac{3}{8\sqrt{2}} \,  e^{-\frac34 \hat{\phi}} \,  \bar{\hat{\epsilon}} \, \hat{\Gamma}_{M} \,  \hat{\lambda} \; ,  \\[6pt]
&& \delta   \hat{B}_{MN} =  - \tfrac{1}{2} \,   e^{\frac12 \hat{\phi}} \,  \bar{\hat{\epsilon}} \, \hat{\Gamma}_{11}  \hat{\Gamma}_{[M} \,  \hat{\psi}_{N]}  -  \tfrac{1}{4 \sqrt{2}} \,   e^{\frac12 \hat{\phi}} \,  \bar{\hat{\epsilon}} \, \hat{\Gamma}_{MN} \,  \hat{\lambda} \; , \nonumber \\[6pt]
&&  \delta   \hat{A}_{MNP} = - \tfrac{3}{4} \,  e^{-\frac14 \hat{\phi}} \,  \bar{\hat{\epsilon}} \, \hat{\Gamma}_{[MN} \,  \hat{\psi}_{P]}  - \tfrac{1}{8\sqrt{2}} \,  e^{-\frac14 \hat{\phi}} \,  \bar{\hat{\epsilon}} \,  \hat{\Gamma}_{11}  \hat{\Gamma}_{MNP} \,  \hat{\lambda} + 3 \, \hat{A}_{[M} \, \delta   \hat{B}_{NP]}  \nonumber  \; .
\end{eqnarray}
The gravitino, dilatino and supersymmetry parameter, $\hat{\psi}_M$, $\hat \lambda$ and $\hat{\epsilon}$, are Majorana. We have introduced the vielbein  $ \hat{e}_M{}^A$, the antisymmetrised products $\hat{\Gamma}_{ABC\ldots}$ of the ten-dim\-en\-sio\-nal gamma matrices (see appendix \ref{ap:Fermions}), have employed the conventional definition $\hat{\Gamma}_M \equiv \hat{e}_M{}^A \hat{\Gamma}_A$, and have denoted by $\hat{\Gamma}_{11}$ the ten-dimensional chirality matrix. It is easy to see that these variations are manifestly real provided the charge conjugation matrix $\hat C$ is chosen such that (\ref{TransposeC10D}) holds. 
Observe that the Romans mass $m$ does not enter (\ref{IIAsusyVars}) --it only appears in the supersymmetry variations of the fermions.

The global SO$(1,1)$ symmetry of type IIA supergravity is preserved by the mass deformation, provided the Romans mass is allowed to rescale under SO$(1,1)$. Indeed, the Lagrangian (\ref{10DLagrangian}), the Bianchi identities (\ref{d10Bianchis}) and the supersymmetry variations  (\ref{IIAsusyVars}) are invariant under the transformation
\begin{equation} \label{SO11inIIA}
\hat \phi^\prime = \hat \phi +k \; , \quad 
\hat A_\1^\prime = e^{-\frac34 k} \hat A_\1 \; , \quad 
\hat B_\2^\prime = e^{\frac12 k} \hat B_\2 \; , \quad 
\hat A_\3^\prime = e^{-\frac14 k} \hat A_\3 \; , \quad 
 m^\prime = e^{-\frac54 k}  m \; , \quad 
\end{equation}
for any real constant $k$, with the metric and fermions inert. Alternatively, this SO$(1,1)$ symmetry can be fixed if used to set the Romans mass to unity. To see this, first note that $m$ can always be taken positive: were it negative, the flip
\begin{eqnarray}
m^\prime = -m \; , \qquad 
\hat A_\1^\prime = - \hat A_\1 \; , \qquad 
\hat A_\3^\prime = - \hat A_\3 \; , \qquad 
\end{eqnarray}
with all other fields unchanged, is a discrete symmetry of the theory that could be used to make $m^\prime >0$. Then, one can set $m^\prime =1$ just by fixing $k = \tfrac45 \log m$ in (\ref{SO11inIIA}). Thus, all non-vanishing values of the Romans mass are classically equivalent. This parallels the fact that all non-vanishing values of the dyonically gauging parameter in $D=4$ ISO(7) supergravity are equivalent \cite{Dall'Agata:2014ita}.
Finally, the field equations are invariant under the following rescaling, similar to that of $D=11$ supergravity:
\begin{eqnarray} \label{RescalinginIIA}
\hat \phi^\prime = \hat \phi +k \; , \; 
\hat A_\1^\prime = e^{-2 k} \hat A_\1 \; , \; 
\hat B_\2^\prime = e^{-2 k} \hat B_\2 \; , \;
\hat A_\3^\prime = e^{-4 k} \hat A_\3 \; , \;
  \hat{e^\prime}_M{}^A = e^{-\frac54 k}    \hat{e}_M{}^A  \; . \;
\end{eqnarray}
This,  unlike the SO$(1,1)$ transformation (\ref{SO11inIIA}), does involve the Einstein frame metric.

\section{Spinor conventions} 
\label{ap:Fermions}

The ten-dimensional gamma matrices $\hat{\Gamma}_A$, $A=0,1, \ldots, 9$, are subject to the $\textrm{Cliff}(1,9)$ anticommutation relations
\begin{eqnarray} \label{Cliff19}
\{ \hat{\Gamma}_A , \hat{\Gamma}_B \} = 2 \, \eta_{AB} \, \oneone_{32} \; . 
\end{eqnarray}
We choose the ten-dimensional charge conjugation matrix $\hat C$ to obey
\begin{eqnarray} \label{TransposeC10D}
\hat \Gamma_A^{\textrm{T}} = - \hat C \Gamma_A \hat C^{-1} \; , \qquad \hat{C}^{\textrm{T}} = -\hat C \; .
\end{eqnarray}

In this paper, we find it convenient to make use of the Clifford algebra isomorphism $\textrm{Cliff}(1,9) = \textrm{Cliff}(1,3) \times \textrm{Cliff}(6)$, adapted to the splitting (\ref{LorentzSplit}), and work without loss of generality in the $\textrm{Cliff}(1,9)$ basis
\begin{eqnarray} \label{Cliff10}
\hat \Gamma_\alpha = \gamma_\alpha \otimes \oneone_8 \; , \qquad 
\hat \Gamma_a = \gamma_5 \otimes \Gamma_a  \; , \qquad 
\hat \Gamma_{11} = \gamma_5 \otimes \Gamma_7 \; , 
\end{eqnarray}
where $\gamma_\alpha$, $\alpha =0,1, \ldots, 3$,  and $\Gamma_a$, $a =1, \ldots, 6$,  respectively are $\textrm{Cliff}(1,3)$ and $\textrm{Cliff}(6)$ gamma matrices,
\begin{eqnarray}
\{ \gamma_\alpha \, , \,  \gamma_\beta \}  = 2 \, \eta_{\alpha \beta} \, \oneone_{4} \; , \qquad 
\{ \Gamma_a \, , \,  \Gamma_b \}  = 2 \, \delta_{ab} \, \oneone_{8} \; ,
\end{eqnarray}
and $\gamma_5$, $\Gamma_7$ and $\hat{\Gamma}_{11}$ are the chirality matrices in four, six and ten dimensions.

We employ charge conjugation matrices $C_4$ and $C$ in four and  six dimensions such that $\hat C = C_4 \otimes C$ and 
\begin{eqnarray} \label{TransposeC4D6D}
\gamma_\alpha^{\textrm{T}} = -C_4 \gamma_\alpha C_4^{-1} \; , \quad C_4^{\textrm{T}} = -C_4 \; , \qquad
\textrm{and} \qquad
\Gamma_a^{\textrm{T}} = -C \Gamma_a C^{-1} \; , \quad C^{\textrm{T}} = C \; . \; 
\end{eqnarray}
With this choice of six-dimensional charge conjugation, the antisymmetrised products of gamma matrices that are symmetric and antisymmetric in spinor indices ($A=1 , \ldots , 8$, omitted in this appendix and not to be confused with global SO$(1,9)$ indices) are
\begin{eqnarray} \label{SymGammaMat6D}
\textrm{symmetric:}  &&  \quad C  \; , \quad C \Gamma_{ab} \Gamma_7  \; , \quad C \Gamma_{abc} \; ,  \\
\label{AntiSymGammaMat6D}
\textrm{antisymmetric:}  &&  \quad C \Gamma_7 \; , \quad  C \Gamma_a \; , \quad C \Gamma_{a} \Gamma_7 \; , \quad C \Gamma_{ab}  \; .
\end{eqnarray}
Thus, the symmetric and antisymmetric six-dimensional gamma matrices respectively realise the branchings
\begin{eqnarray}
  \label{eq:36and28branching}
&&  \mathbf{36}\;\stackrel{\mathrm{SO}(8)}{\longrightarrow} \; \mathbf{1} + \mathbf{35}
  \;\stackrel{\mathrm{SO}(7)}{\longrightarrow} \;
  \mathbf{1} + \mathbf{35}
   \;\stackrel{\mathrm{SO}(6)}{\longrightarrow} \;
   \mathbf{1} + \mathbf{15} + \mathbf{10} + \overline{ \mathbf{10}} \; , \\
&&   \mathbf{28}\;\stackrel{\mathrm{SO}(8)}{\longrightarrow} \; \mathbf{28} \quad \;\;\;
  \;\stackrel{\mathrm{SO}(7)}{\longrightarrow} \; 
  \mathbf{7} + \mathbf{21}
   \;\stackrel{\mathrm{SO}(6)}{\longrightarrow} \;
   \mathbf{1} + \mathbf{6} + \overline{\mathbf{6}} + \mathbf{15} \; , 
\end{eqnarray}
of the $ \mathbf{36}$ and $ \mathbf{28}$ of SU(8) under SO(6) via  $\mathrm{SU}(8) \supset \mathrm{SO}(8) \supset \mathrm{SO}(7) \supset \mathrm{SO}(6)$.

\section{SL(7)--covariant supersymmetry transformations} \label{sec:SL7covariant}

Here we give some details about how the supersymmetry transformations (\ref{susyVectors})--(\ref{susyThreeForms}) give rise to the SL(7)--covariant and tensor-hierarchy-compatible supersymmetry transformations (\ref{eq:susySL7Vec1})--(\ref{eq:susySL73form}) for the redefined fields (\ref{TensorsSL7}). The easiest way to verify this is to work backwards and reproduce the former from the latter. That is how we will proceed.

\subsection{Supersymmetry transformations of the vectors} \label{ap:susyvectors}

Splitting the index $I=(m,7)$, the supersymmetry transformations (\ref{eq:susySL7Vec1}), (\ref{eq:susySL7Vec2}) for the vectors become
\begin{eqnarray} \label{VectorsBranchingSL6}
&& \delta C_\mu{}^{m8} =  i\,  V^{m8}{}_{AB} \left(   \, \bar{\epsilon}^A  \psi_\mu{}^B + \tfrac{1}{2\sqrt{2}} \,  \bar{\epsilon}_C  \gamma_\mu \chi^{ABC}\right) + \textrm{h.c.} \; , \nonumber \\[6pt]
&& \delta C_\mu{}^{78} =  i\,  V^{78}{}_{AB} \left(   \, \bar{\epsilon}^A  \psi_\mu{}^B + \tfrac{1}{2\sqrt{2}} \,  \bar{\epsilon}_C  \gamma_\mu \chi^{ABC}\right) + \textrm{h.c.} \; , \nonumber \\[6pt]
&& \delta \tilde{C}_{\mu \, mn} =  -i\,  \tilde{V}_{mn \, AB} \left(   \, \bar{\epsilon}^A  \psi_\mu{}^B + \tfrac{1}{2\sqrt{2}} \,  \bar{\epsilon}_C  \gamma_\mu \chi^{ABC}\right) + \textrm{h.c.} \; , \nonumber \\[6pt]
&& \delta \tilde{C}_{\mu \, m7} =  -i\,  \tilde{V}_{m7 \, AB} \left(  \, \bar{\epsilon}^A  \psi_\mu{}^B + \tfrac{1}{2\sqrt{2}} \,  \bar{\epsilon}_C  \gamma_\mu \chi^{ABC}\right) + \textrm{h.c.}
\end{eqnarray}
When the generalised vielbeine (\ref{GenVielbein}) are introduced in (\ref{VectorsBranchingSL6}), the $ \bar{\epsilon}^A  \psi_\mu{}^B$ terms in (\ref{susyVectors}) are straightforwardly reproduced. 

Turning to the spin 1/2 contributions, we first propose an ansatz for the trispinor in terms of the fermions introduced in section \ref{4DLorentzManifest},
\begin{equation} \label{trispinor1ansatz}
\chi^{ABC} = (\Gamma^a C^{-1})^{[AB} \Big( y \,  \psi_a^{C]} + z  \,  ( \Gamma_a \Gamma_7 \lambda)^{C]} \Big) 
+ x \,  (\Gamma_7 C^{-1})^{[AB} \lambda^{C]} 
 + t \,   (\Gamma^a \Gamma_7 C^{-1})^{[AB} (\Gamma_a \lambda)^{C]} \; , 
\end{equation}
where $y$, $z$, $x$, $t$ are constants and $\Gamma_a$ are the Cliff$(6)$ gamma matrices (see appendix \ref{ap:Fermions}).  We find this parameterisation of $\chi^{ABC}$ in terms of  $\psi_a^A$, $\lambda^A$ useful, despite being overdetermined: one of the terms can be eliminated by a Fierz rearrangement\footnote{Further redundant terms, $(\Gamma^{ab} C^{-1})^{[AB}  \,  (\Gamma_a \psi_b)^{C]} $ and $(\Gamma^{ab} C^{-1})^{[AB}  \,  (\Gamma_{ab} \lambda)^{C]}$, could be added to $\chi^{ABC}$, given that $C \Gamma_{ab}$ is antisymmetric with our six-dimensional charge conjugation conventions (see (\ref{AntiSymGammaMat6D})). These contributions  would be Fierzable too into terms that already appear in (\ref{trispinor1ansatz}).}. Now, introducing (\ref{GenVielbein}) and (\ref{trispinor1ansatz}) into (\ref{VectorsBranchingSL6}), and using the relations
\begin{eqnarray} 
&& (C \Gamma_a )_{BC} \, \chi^{ABC} = -\tfrac43 \,  y\,  \big( ( \delta_a^b + \tfrac12 \Gamma_a \Gamma^b ) \psi_b \big)^A -\tfrac23  (x+8z+4t) (\Gamma_a \Gamma_7 \lambda)^A \; ,  \nonumber \\[6pt]
&& (C \Gamma_7 )_{BC} \, \chi^{ABC} = \tfrac23 \,  y\,  ( \Gamma^a \Gamma_7 \psi_a)^A -2 (x+2z-2t) \lambda^A \; , \nonumber \\[6pt]
&& (C \Gamma_{ab})_{BC} \, \chi^{ABC} = \tfrac23 \,  y\,  (\Gamma^c \Gamma_{ab} \psi_c \big)^A +\tfrac23  (x+2z-2t) (\Gamma_{ab} \Gamma_7 \lambda)^A \; , \nonumber \\[6pt]
&& (C \Gamma_a \Gamma_7)_{BC} \, \chi^{ABC} = \tfrac43 \,  y\,  \big( ( \delta_a^b - \tfrac12 \Gamma_a \Gamma^b ) \Gamma_7 \psi_b \big)^A -\tfrac23  (x-4z-8t) (\Gamma_a \lambda)^A \; , 
\end{eqnarray}
it can be verified that the spin 1/2 terms in (\ref{susyVectors}) are reproduced provided that $y=\frac{3i}{\sqrt{2}}$ and  the following system of linear equations holds
\begin{eqnarray}
 x+8z+4t = 0 \; , \qquad 
x+2z-2t =  \tfrac{ 3i}{ 2 } \; , \qquad
 x -4z -8t =  3i \; . 
\end{eqnarray}
This system is overdetermined, like (\ref{trispinor1ansatz}), and admits a line of solutions. The solution
\begin{equation}
x= i \; , \qquad 
z =0  \; , \qquad
t = -\tfrac{ i}{ 4} \; 
\end{equation}
leads to the expression for $\chi^{ABC}$ in (\ref{trispinor1}) and the solution
\begin{equation}
x= 2i \; , \qquad 
z = - \tfrac{i}{4} \; , \qquad
t = 0 \;  
\end{equation}
leads to (\ref{trispinor2}). We will use $\chi^{ABC}$ in (\ref{trispinor1}) in the remainder.

\subsection{Supersymmetry transformations of the two-forms} \label{ap:susy2forms}

We will next retrieve the supersymmetry transformations (\ref{susyTwoForms}) for the two-forms from the SL(7)--covariant and tensor-hierarchy-compatible expressions (\ref{eq:susySL72form}). Splitting the SL(7) indices as $I=(m,7)$, the latter gives
{\setlength\arraycolsep{2pt}
\begin{eqnarray} \label{susytensors4dSL6m}
\delta C_{\mu \nu \, m }{}^{8} & = &  \Big[ \tfrac23  \big( V^{n8}{}_{BC} \,  \tilde{V}_{mn}{}^{AC} + V^{78}{}_{BC} \,  \tilde{V}_{m7}{}^{AC} +  \tilde{V}_{mn \, BC}{} \,  V^{n8}{}^{AC}+  \tilde{V}_{m7 \, BC}{} \,  V^{78}{}^{AC} \big) \,  \bar{\epsilon}_{A} \gamma_{[\mu} \psi_{\nu]}^B \nonumber \\[4pt]
&& \quad + \tfrac{\sqrt{2}}{3} \big(  V^{n8}{}_{AB} \,  \tilde{V}_{mn \, CD}  + V^{78}{}_{AB} \,  \tilde{V}_{m7 \, CD} \big)  \,  \bar{\epsilon}^{[A} \gamma_{\mu \nu} \chi^{BCD]}  +\textrm{h.c.} \Big]  \nonumber \\[4pt]
&&  \quad  -  \,  C_{[\mu}^{n8} \, \delta \tilde{C}_{\nu] mn} - C_{[\mu}^{78} \, \delta \tilde{C}_{\nu] m7} -  \tilde{C}_{[\mu| \, mn} \,  \delta  C_{|\nu]}{}^{n8} -  \tilde{C}_{[\mu| \, m7} \,  \delta  C_{|\nu]}{}^{78}  \; ,  \\[12pt]
\label{susytensors4dSL67}
\delta C_{\mu \nu \, 7 }{}^{8} & = & -\Big[ \tfrac23  \big( V^{n8}{}_{BC} \,  \tilde{V}_{n7}{}^{AC} +  \tilde{V}_{n7 \, BC}{} \,  V^{n8}{}^{AC} \big)  \, \bar{\epsilon}_{A} \gamma_{[\mu} \psi_{\nu]}^B  + \tfrac{\sqrt{2}}{3} \,  V^{n8}{}_{AB} \,  \tilde{V}_{n7 \, CD}  \,  \bar{\epsilon}^{[A} \gamma_{\mu \nu} \chi^{BCD]}  \nonumber \\[4pt]
&&  \quad +\textrm{h.c.} \Big]  + C_{[\mu}^{n8} \, \delta \tilde{C}_{\nu] n7} +  \tilde{C}_{[\mu| \, n7} \,  \delta  C_{|\nu]}{}^{n8} \;  .
\end{eqnarray}
}

Let us first focus on the $\bar{\epsilon}_{A} \gamma_{[\mu} \psi_{\nu]}^B$ terms. Taking combinations of the 
generalised vielbeine (\ref{GenVielbein}) and their conjugates (\ref{GenVielbeinUpperAB}), and using six-dimensional gamma matrix algebra we can compute 
\begin{eqnarray}
\label{eq1twoforms}
V^{n8}{}_{BC} \,  \tilde{V}_{mn}{}^{AC} +   \tilde{V}_{mn \, BC}{} \,  V^{n8}{}^{AC}  &=& -\tfrac58 \, e^{-\frac14 \hat \phi} \,  \Delta^{-1} e_m{}^a (\Gamma_a)^A{}_B  -\tfrac58 \, e^{\frac12 \hat \phi} \,  \Delta^{-1} A_m (\Gamma_7)^A{}_B \nonumber \\[4pt]
&&  -\tfrac18 \, \Delta^{-1} \,g^{np}A_n B_{mp} \,  \delta^A_B \; ,  \\[10pt]
\label{eq2twoforms}
V^{78}{}_{BC} \,  \tilde{V}_{m7}{}^{AC} +   \tilde{V}_{m7 \, BC}{} \,  V^{78}{}^{AC}  &=& -\tfrac18  \, e^{-\frac14 \hat \phi} \,  \Delta^{-1} e_m{}^a (\Gamma_a)^A{}_B  -\tfrac18 \, e^{\frac12 \hat \phi} \,  \Delta^{-1} A_m (\Gamma_7)^A{}_B \nonumber \\[4pt]
&&  +\tfrac18 \, \Delta^{-1} \,g^{np}A_n B_{mp} \,  \delta^A_B \; ,   \\[10pt]
\label{eq3twoforms}
 V^{n8}{}_{BC} \,  \tilde{V}_{n7}{}^{AC} +  \tilde{V}_{n7 \, BC}{} \,  V^{n8}{}^{AC}   &=& \tfrac{3}{4} \, e^{\frac12 \hat \phi} \,  \Delta^{-1}  (\Gamma_7)^A{}_B \; .
\end{eqnarray}
Adding (\ref{eq1twoforms}) and (\ref{eq2twoforms}), the $\bar{\epsilon}_{A}\gamma_{[\mu} \psi_{\nu]}^B$ terms of $\delta C_{\mu \nu \, m }{}^{8}$ in (\ref{susytensors4dSL6m}) reproduce those of $\delta A_{\mu\nu m}$ in (\ref{susyTwoForms}). The relation (\ref{eq3twoforms}) similarly ensures that the $\bar{\epsilon}_{A} \gamma_{[\mu} \psi_{\nu]}^B$ terms of $\delta B_{\mu \nu}$ are reproduced from $\delta C_{\mu \nu \, 7 }{}^{8} $ in (\ref{susytensors4dSL67}). 

In order to match the spin 1/2 terms, we first compute, from the generalised vielbeine (\ref{GenVielbein}) and $\chi^{ABC}$ in (\ref{trispinor1}):

\newpage

{\setlength\arraycolsep{0pt}
\begin{eqnarray}
&& 2 \, V^{n8}{}_{AB} \,  \tilde{V}_{mn \, CD}  \,\,  \bar{\epsilon}^{[A}  \chi^{BCD]}  =
\tfrac{5i}{4\sqrt{2}} \, e^{-\frac14 \hat \phi} \, \Delta^{-1} e_m{}^a \, \bar \epsilon^A \big( C( \delta_a^b-\Gamma_a\Gamma^b) \big)_{AB} \psi^B_b \nonumber \\[3pt]
&& \qquad -\tfrac{5i}{16} \, e^{-\frac14 \hat \phi} \, \Delta^{-1} e_m{}^a \, \bar \epsilon^A ( C \Gamma_a\Gamma_7)_{AB} \lambda^B
-\tfrac{i}{4\sqrt{2}}  \, e^{-\frac34 \hat \phi} \, \Delta^{-1} \delta^{ab} e_a{}^n B_{mn}  \, \bar \epsilon^A (C\Gamma_7)_{AB} \psi^B_b \nonumber \\[3pt]
&& \qquad -\tfrac{3i}{16}\, e^{-\frac34 \hat \phi} \, \Delta^{-1} e_a{}^n B_{mn}  \, \bar \epsilon^A (C\Gamma^a)_{AB} \lambda^B
+\tfrac{i}{4}\, e^{\frac12 \hat \phi} \, \Delta^{-1} e_a{}^n A_{[m} e_{n]}{}^b \, \bar \epsilon^A (C\Gamma^a \Gamma_b)_{AB} \lambda^B \nonumber \\[3pt]
&& \qquad -\tfrac{i}{2\sqrt{2}} \, e^{\frac12 \hat \phi} \, \Delta^{-1} e_a{}^n A_{[m} e_{n]}{}^b  \, \bar \epsilon^A \big(C (\Gamma^a \delta_b^c - \Gamma_b \delta^{ac} - \delta^a_b \Gamma^c ) \Gamma_7 \big)_{AB} \psi_c^B \nonumber \\[3pt]
&& \qquad -\tfrac{i}{\sqrt{2}} \, \Delta^{-1} e_a{}^n  e_b{}^p A_{[m} B_{n]p}  \, \bar \epsilon^A \big(C \Gamma^{(a} ( \delta^{b)c} +\tfrac12 \Gamma^{b)} \Gamma^c \big)_{AB} \psi_c^B \; , 
\end{eqnarray}
}and
{\setlength\arraycolsep{0pt}
\begin{eqnarray}
&& 2 \, V^{78}{}_{AB} \,  \tilde{V}_{m7 \, CD}  \,\,  \bar{\epsilon}^{[A}  \chi^{BCD]}  =
\tfrac{i}{4\sqrt{2}} \, e^{-\frac14 \hat \phi} \, \Delta^{-1} e_m{}^a \, \bar \epsilon^A \big( C( \delta_a^b-\Gamma_a\Gamma^b) \big)_{AB} \psi^B_b \nonumber \\[2pt]
&& \qquad -\tfrac{i}{16} \, e^{-\frac14 \hat \phi} \, \Delta^{-1} e_m{}^a \, \bar \epsilon^A ( C \Gamma_a\Gamma_7)_{AB} \lambda^B
+\tfrac{i}{4\sqrt{2}}  \, e^{-\frac34 \hat \phi} \, \Delta^{-1} \delta^{ab} e_a{}^n B_{mn}  \, \bar \epsilon^A (C\Gamma_7)_{AB} \psi^B_b \nonumber \\[2pt]
&& \qquad +\tfrac{3i}{16}\, e^{-\frac34 \hat \phi} \, \Delta^{-1} e_a{}^n B_{mn}  \, \bar \epsilon^A (C\Gamma^a)_{AB} \lambda^B
+\tfrac{i}{8}\, e^{\frac12 \hat \phi} \, \Delta^{-1} e_m{}^a e_b{}^n A_n  \, \bar \epsilon^A (C\Gamma^b \Gamma_a)_{AB} \lambda^B \nonumber \\[2pt]
&& \qquad +\tfrac{i}{4\sqrt{2}} \, e^{\frac12 \hat \phi} \, \Delta^{-1} e_m{}^a e_c{}^n A_n   \, \bar \epsilon^A \big(C (\Gamma_a \delta^{cb} - \Gamma^c \delta_a^b + \delta_a^c \Gamma^b ) \Gamma_7 \big)_{AB} \psi_c^B \nonumber \\[2pt]
&& \qquad -\tfrac{i}{2\sqrt{2}} \, \Delta^{-1} e_a{}^n  e_b{}^p A_p B_{mn}    \, \bar \epsilon^A \big(C \Gamma^{(a} ( \delta^{b)c} +\tfrac12 \Gamma^{b)} \Gamma^c \big)_{AB} \psi_c^B \; . 
\end{eqnarray}
Adding these two expressions and simplifying further, we obtain
{\setlength\arraycolsep{0pt}
\begin{eqnarray} \label{eq:spin1/2combi}
&& 2 \big( V^{n8}{}_{AB} \,  \tilde{V}_{mn \, CD} +  V^{78}{}_{AB} \,  \tilde{V}_{m7 \, CD}  \big) \,  \bar{\epsilon}^{[A}  \chi^{BCD]}  =
\tfrac{3i}{2\sqrt{2}} \, e^{-\frac14 \hat \phi} \, \Delta^{-1} e_m{}^a \, \bar \epsilon^A \big( C( \delta_a^b-\Gamma_a\Gamma^b) \big)_{AB} \psi^B_b \nonumber \\[2pt]
&& \qquad -\tfrac{3i}{8} \, e^{-\frac14 \hat \phi} \, \Delta^{-1} e_m{}^a \, \bar \epsilon^A ( C \Gamma_a\Gamma_7)_{AB} \lambda^B 
+ \tfrac{3i}{2\sqrt{2}} \, e^{\frac12 \hat \phi} \, \Delta^{-1} A_m   \, \bar \epsilon^A (C \Gamma^a \Gamma_7 )_{AB} \psi_a^B \nonumber \\[2pt]
&& \qquad 
+\tfrac{3i}{4}\, e^{\frac12 \hat \phi} \, \Delta^{-1}  A_m  \, \bar \epsilon^A C_{AB} \lambda^B \; .
\end{eqnarray}
Note that all terms containing $B_{mn}$ cancel. With the help of (\ref{eq:spin1/2combi}), the 
$ \bar{\epsilon}^{[A} \gamma_{\mu \nu} \chi^{BCD]}$ terms of $\delta C_{\mu \nu \, m }{}^{8}$ in (\ref{susytensors4dSL6m}) can be seen to give rise to the spin-1/2 terms of $\delta A_{\mu \nu m} $ in (\ref{susyTwoForms}). Similarly, we can compute
\begin{equation}
2 \,V^{m8}{}_{AB} \,  \tilde{V}_{m7 \, CD} \, \bar{\epsilon}^{[A}  \chi^{BCD]}  =
-\tfrac{3i}{2\sqrt{2}} \, e^{\frac12 \hat \phi} \, \Delta^{-1}  \, \bar \epsilon^A ( C \Gamma^a\Gamma_7)_{AB} \psi^B_a - \tfrac{3i}{4} \, e^{\frac12 \hat \phi} \, \Delta^{-1}  \bar \epsilon^A C_{AB} \lambda^B \; ,
\end{equation}
which ensures that the spin 1/2 terms of $\delta B_{\mu \nu} $  in (\ref{susyTwoForms}) follow from $\delta C_{\mu \nu \, 7 }{}^{8}$ in (\ref{susytensors4dSL67}). 

Finally, the terms in vector times variation of vector in (\ref{susytensors4dSL6m}), (\ref{susytensors4dSL67}) can be easily seen to reproduce those in (\ref{susyTwoForms}) once the definitions (\ref{redefvectors})--(\ref{redefThreeForm}) are taken into account. 

\subsection{Supersymmetry transformations of the three-forms} \label{ap:susy3forms}

Similar manipulations allow us to compute 
\begin{eqnarray}
&& V^{I8}{}_{BD} \, \big( V^{J8}{}^{DC} \,  \tilde{V}_{IJ}{}_{AC} +  \tilde{V}_{IJ}{}^{DC}{} \,  V^{J8}{}_{AC} \big)  \nonumber \\
&& \qquad  = \quad V^{m8}{}_{BD} \, \big( V^{n8}{}^{DC} \,  \tilde{V}_{mn}{}_{AC}  + V^{78}{}^{DC} \,  \tilde{V}_{m7}{}_{AC} +  \tilde{V}_{mn}{}^{DC}{} \,  V^{n8}{}_{AC} +  \tilde{V}_{m7}{}^{DC}{} \,  V^{78}{}_{AC} \big) \nonumber \\
&& \qquad \quad \;  -V^{78}{}_{BD} \, \big( V^{m8}{}^{DC} \,  \tilde{V}_{m7}{}_{AC} +  \tilde{V}_{m7}{}^{DC}{} \,  V^{m8}{}_{AC} \big) \nonumber \\
&&  \qquad  = \quad -\tfrac{21 }{16}  \, e^{-\frac14 \hat \phi} \, \Delta^{-\frac32} \, C_{AB} \; , 
\end{eqnarray}
and 
\begin{eqnarray}
&& 2\, V^{I8 \, AE} \,   V^{J8}{}_{[EB|} \,  \tilde{V}_{IJ \, |CD]}  \,  \bar{\epsilon}_A \,  \chi^{BCD} \nonumber \\
&& \qquad  = \quad 2\, V^{m8 \, AE} \, \big(   V^{n8}{}_{[EB|} \,  \tilde{V}_{mn \, |CD]}   + V^{78}{}_{[EB|} \,  \tilde{V}_{m7 \, |CD]}   \big) \,  \bar{\epsilon}_A \,  \chi^{BCD}   \nonumber \\
&& \qquad \quad \;  -2\, V^{78 \, AE} \,   V^{n8}{}_{[EB|} \,  \tilde{V}_{n7 \, |CD]}  \,  \bar{\epsilon}_A \,  \chi^{BCD} \nonumber \\
&&  \qquad  = \quad  \tfrac{9i}{4\sqrt{2}} \, e^{-\frac14 \hat \phi} \, \Delta^{-\frac32}   \, \bar \epsilon_A (\Gamma^a)^A{}_B \,\psi_a^B  +  \tfrac{3 i}{8} \, e^{-\frac14 \hat \phi} \, \Delta^{-\frac32}   \,   \bar \epsilon_A (\Gamma_7)^A{}_B \,  \lambda^B \; .
\end{eqnarray}
These expressions respectively show that the $\bar{\epsilon}^{A} \gamma_{[\mu \nu} \psi_{\rho]}^B$ and $ \bar{\epsilon}_A \gamma_{\mu \nu \rho} \chi^{BCD} $ contributions of the supersymmetry variation of the SL(7)--singlet three-form (\ref{eq:susySL73form}) reproduces the spin 3/2 and 1/2 contributions of the variation (\ref{susyThreeForms}).


Finally, the terms in the last line of (\ref{susyThreeForms}) containing tensors times variations of tensors can be checked to follow from their counterparts in (\ref{eq:susySL73form}) when the redefinitions (\ref{redefvectors})--(\ref{redefThreeForm}) are used.

\section{Non-linear field redefinitions from group theory} \label{app:GroupThRedef}

We now discuss the group theory underlying the non-linear field redefinitions (\ref{redefvectors})--(\ref{redefThreeForm}) which cast the SL(6)--covariant type IIA fields (\ref{eq:SL6fieldcontent}) into the SL(7)--covariant fields (\ref{TensorsSL7}). In group-theoretic language, these redefinitions can be put down to the reallocation of Abelian charges associated to embeddings of SL(6) into $\textrm{SL}(8) \subset \textrm{E}_{7(7)}$ along two different branches. 

The SL(6) representations in (\ref{eq:SL6fieldcontent}) correspond to the IIA branch discussed in \cite{deWit:2003hq}. For the vectors, the relevant branching rules from SL(8) are
\begin{equation}
\begin{array}{ccccc}
\textrm{SL}(8) & \hspace{3mm} \supset  \hspace{3mm} & \textrm{SL}(6) \times \textrm{SL}(2) \times \mathbb{R}_{\textrm{A}} & \hspace{3mm} \supset \hspace{3mm} & \textrm{SL}(6) \times \mathbb{R}_{2} \times \mathbb{R}_{\textrm{A}} \\[2mm]
\textbf{28} & \rightarrow & \textbf{(1,1)}_{-6} + \textbf{(6,2)}_{-2} + \textbf{(15,1)}_{+2} & \rightarrow &  \textbf{1}_{(0,-6)} + \textbf{6}_{(\pm1,-2)} + \textbf{15}_{(0,+2)} \\[2mm]
\textbf{28}' & \rightarrow & \textbf{(1,1)}_{+6} + \textbf{(6$'$,2)}_{+2} + \textbf{(15$'$,1)}_{-2} & \rightarrow &  \textbf{1}_{(0,+6)} + \textbf{6}'_{(\pm1,+2)} + \textbf{15}'_{(0,-2)}
\end{array}
\end{equation}
where $\,\textbf{n}_{(q_{2},q_{A})}\,$ denotes an SL(6) representation $\,\textbf{n}\,$ with $\,\mathbb{R}_{2} \times \mathbb{R}_{\textrm{A}} \,$ charges $\,(q_{2},q_{A})\,$. The vectors in (\ref{eq:SL6fieldcontent}) can be assigned to representations of $\textrm{SL}(6) \times \mathbb{R}_{2} \times \mathbb{R}_{\textrm{A}} $ as
\begin{equation}
\label{app_Afields_1}
B_{\mu m} \equiv \textbf{6}_{(+1,-2)}
\hspace{3mm} , \hspace{3mm}
A_{\mu mn} \equiv \textbf{15}_{(0,+2)}
\hspace{5mm} \textrm{ and } \hspace{5mm} 
A_{\mu} \equiv \textbf{1}_{(0,+6)}
\hspace{3mm} , \hspace{3mm}
B_{\mu}{}^{m} \equiv \textbf{6}'_{(+1,+2)} \ .
\end{equation}
A similar analysis of the branching rules relevant for the scalars and tensor fields in (\ref{eq:SL6fieldcontent}) results in the additional identifications
\begin{equation}
\label{app_Afields_2}
\begin{array}{c}
A_{m} \equiv \textbf{6}_{(-1,+4)}
\hspace{3mm} , \hspace{3mm}
A_{mnp} \equiv \textbf{20}_{(-1,0)}
\hspace{3mm} , \hspace{3mm}
B_{mn} \equiv \textbf{15}_{(0,-4)} \ ,  \\[2mm]
A_{\mu\nu m} \equiv \textbf{6}_{(+1,+4)}
\hspace{3mm} ,  \hspace{3mm}
B_{\mu\nu} \equiv \textbf{1}_{(+2,0)} 
\hspace{3mm} ,  \hspace{3mm}
A_{\mu\nu\rho} \equiv \textbf{1}_{(+2,+6)} \, \ .
\end{array}
\end{equation}

A different branch, naturally related to the M-theory origin of type IIA supergravity, embeds SL(6) into SL(8) through SL(7). For the vectors, the relevant decompositions are
\begin{equation} \label{MtheoryBranch}
\begin{array}{ccccc}
\textrm{SL}(8) & \hspace{3mm} \supset  \hspace{3mm} & \textrm{SL}(7) \times \mathbb{R}_{\textrm{M}} & \hspace{3mm} \supset \hspace{3mm} & \textrm{SL}(6) \times \mathbb{R}_{7} \times \mathbb{R}_{\textrm{M}} \\[2mm]
\textbf{28} & \rightarrow & \textbf{7}_{-6} +  \textbf{21}_{+2} & \rightarrow &  \big( \,  \textbf{1}_{(-6,-6)} + \textbf{6}_{(+1,-6)} \, \big) \,\, + \,\,  \big( \, \textbf{6}_{(-5,+2)}  + \textbf{15}_{(+2,+2)} \, \big) \\[2mm]
\textbf{28}' & \rightarrow & \textbf{7}'_{+6} + \textbf{21}'_{-2} & \rightarrow &  \big( \, \textbf{1}_{(+6,+6)} + \textbf{6}'_{(-1,+6)} \, \big) \,\, + \,\,  \big( \, \textbf{6}'_{(+5,-2)}  + \textbf{15}'_{(-2,-2)} \, \big)
\end{array}
\end{equation}
with the same notation $\,\textbf{n}_{(q_{7},q_{M})}\,$ for the final states in the decomposition as before, altthough the subscripts now represent charges under $\mathbb{R}_{7} \times \mathbb{R}_{\textrm{M}}$. The identification between these representations and the vectors in (\ref{redefvectors}) proceeds as 
\begin{equation}
\label{app_Cfields_1}
\tilde{C}_{\mu m7} \equiv \textbf{6}_{(-5,+2)}
\hspace{2mm} , \hspace{2mm}
\tilde{C}_{\mu mn} \equiv \textbf{15}_{(+2,+2)}
\hspace{3mm} \textrm{ and } \hspace{3mm}
C_{\mu}{}^{78} \equiv \textbf{1}_{(+6,+6)}
\hspace{2mm} , \hspace{2mm}
C_{\mu}{}^{m8} \equiv \textbf{6}'_{(-1,+6)} \ .
\end{equation}
A similar analysis yields the identifications
\begin{equation}
\label{app_Cfields_2}
C_{\mu\nu m}{}^{8} \equiv \textbf{6}_{(+1,+8)}
\hspace{3mm} ,  \hspace{3mm}
C_{\mu\nu 7}{}^{8} \equiv \textbf{1}_{(-6,+8)}
\hspace{3mm} ,  \hspace{3mm}
C_{\mu\nu \rho}{}^{88} \equiv \textbf{1}_{(0,+14)} \ ,
\end{equation}
for the two-forms in (\ref{redefTwoForms}) and the three-form in (\ref{redefThreeForm}).

In agreement with the main text, the SL(6)--covariant fields (\ref{app_Afields_1}), (\ref{app_Afields_2}) can be non-linearly redefined into (\ref{app_Cfields_1}), (\ref{app_Cfields_2}). The latter are then naturally packed into SL(7)--covariant fields via (\ref{TensorsSL7}), namely, through the intermediate step in the embedding (\ref{MtheoryBranch}). This redefinition is controlled by the mapping between the $\,\mathbb{R}_{2} \times \mathbb{R}_{\textrm{A}}\,$ and $\,\mathbb{R}_{7} \times \mathbb{R}_{\textrm{M}}\,$ charges. This is given by
\begin{equation}
\label{Rot_charges}
\left( \begin{array}{c}
q_{7} \\ q_{\textrm{M}}
\end{array} \right)
=
\left( \begin{array}{rr}
-3 & \,\,\,\,1 \\
4 & 1
\end{array} \right)
\left( \begin{array}{c}
q_{2} \\ q_{\textrm{A}}
\end{array} \right) \ .
\end{equation}
The most general non-linear field redefinitions, compatible with the mapping (\ref{Rot_charges}), between the vectors in (\ref{app_Cfields_1}) and the field content in (\ref{app_Afields_1})--(\ref{app_Afields_2}) are thus
\begin{equation}
\label{app:vector_redef}
\begin{array}{c}
C_{\mu}{}^{m8} = B_{\mu}{}^{m} 
\hspace{5mm} , \hspace{5mm}
C_{\mu}{}^{78} = {A^{\textrm{\tiny{KK}}}}_{\mu} \ , \\[2mm]
\tilde{C}_{\mu mn} = {A^{\textrm{\tiny{KK}}}}_{\mu mn} \oplus {A^{\textrm{\tiny{KK}}}}_{\mu} \, B_{mn} \oplus  {B^{\textrm{\tiny{KK}}}}_{\mu [m} \, A_{n]}
\hspace{5mm} , \hspace{5mm}
\tilde{C}_{\mu m7} ={B^{\textrm{\tiny{KK}}}}_{\mu m} \ ,
\end{array}
\end{equation}
with the KK labels (which have been supressed in the main text) indicating the standard Kaluza-Klein redefinitions
\begin{equation}
{A^{\textrm{\tiny{KK}}}}_{\mu}  = A_{\mu} \oplus B_{\mu}{}^{p} \, A_{p}
\hspace{3mm} , \hspace{3mm}
{B^{\textrm{\tiny{KK}}}}_{\mu m}  = B_{\mu m} \oplus B_{\mu}{}^{p} \, B_{pm}
\hspace{3mm} , \hspace{3mm}
{A^{\textrm{\tiny{KK}}}}_{\mu mn}  = A_{\mu mn} \oplus B_{\mu}{}^{p} \, A_{pmn} \ .
\end{equation}
The field redefinitions in (\ref{app:vector_redef}) agree with those in (\ref{redefvectors}) obtained from a different approach based on rewriting the supersymmetry transformations in an $\mathcal{N}=8$, $D=4$ fashion. Notice, however, that the last term $\, {B^{\textrm{\tiny{KK}}}}_{\mu [m} \, A_{n]}\,$ in the vectors $\,\tilde{C}_{\mu mn}\,$ of (\ref{app:vector_redef}) is not present in (\ref{redefvectors}). The reason for this is our choice for a term $\hat{A}_{[M} \, \delta   \hat{B}_{NP]}$ rather than $\hat{B}_{[MN} \, \delta \hat{A}_{P]}$ in the supersymmetry transformation of $\hat{A}_{MNP}$ in (\ref{IIAsusyVars}). Both choices are related by a gauge transformation.

Finally, the non-linear redefinitions for the fields in (\ref{app_Cfields_2}) given in (\ref{redefTwoForms}), (\ref{redefThreeForm}) can be shown to agree with the present group theory analysis, too.

\section{Geometric structures on $S^6$} 
\label{ap:KVsSn}

This appendix collects some useful geometric facts about the six-sphere. We first review general formulae for the canonical embedding of $S^n$ in $\mathbb{R}^{n+1}$. We then discuss how this embedding determines, for $n=6$, the usual homogeneous nearly-K\"ahler structure on $S^6$.

\subsection{General features} \label{subset:GeneralSn}

Let $\mu^I$, $I=1, \ldots, n +1$, parametrise the unit radius $n$-sphere $S^n$ as the locus
\begin{eqnarray} \label{eqSn}
\delta_{IJ} \mu^I \mu^I = 1
\end{eqnarray}
 in $\mathbb{R}^{n+1}$. This equation can be explicitly solved by introducing $n$ angles $y^m$, $m=1,\ldots , n$, on $S^n$, so that the $\mathbb{R}^{n+1}$ coordinates, constrained as in  (\ref{eqSn}), become functions of them, $\mu^I = \mu^I (y^m)$. The line element of the round, SO$(n+1)$--invariant metric on an $n$-sphere of squared radius $g^{-2}$ ($g\neq 0$) is then
\begin{eqnarray} \label{RoundMet}
d\mathring{s}^2(S^n)  = g^{-2} \, \delta_{IJ} \,  d \mu^I \, d \mu^J =  g^{-2} \, \delta_{IJ} \  \partial_m \mu^I \  \partial_n \mu^J \ dy^m dy^n \equiv \mathring{g}_{mn}  \ dy^m dy^n \; .
\end{eqnarray}
This metric is Einstein, normalised so that the Ricci tensor equals $(n-1) g^2$ times the metric. 
The Killing vectors, with a lower world index, of the round metric (\ref{RoundMet}) and their covariant derivatives with respect to $y^m$ can be expressed in terms of $\mu^I(y^m)$ as 
\begin{eqnarray} \label{eq:relationsSn}
K_m{}^{IJ} = 2g^{-2} \, \mu^{[I} \partial_m \mu^{J]} \; , \qquad 
K_{mn}{}^{IJ} =  4g^{-2} \, \partial_{[m} \mu^I \partial_{n]} \mu^J \; , 
\end{eqnarray}
where we have used a convenient normalisation for $K_m{}^{IJ}$. The upper-index components of the Killing vectors are obtained by raising the index $m$ with  the inverse of (\ref{RoundMet}):
$
K^{m \, IJ} = \mathring{g}^{mn} K_m{}^{IJ}
$.
With the chosen normalisation, the SO$(n+1)$ commutation relations that these Killing vectors satisfy read
\begin{eqnarray}
[ K^{IJ}  , \, K_{KL} ] ^m= -4 \,  \delta^{[I}_{[K} K^{J]m}{}_{L]} \; .
\end{eqnarray}
This coincides with the normalisation for the SO(7) Lie algebra in (C.12) of \cite{Guarino:2015qaa}.

Other useful relations include
\begin{equation} \label{eq:propsS6}
g^{-2} \,  \mathring{g}^{mn} \partial_m \mu^I \partial_n \mu^J = \delta^{IJ} - \mu^I \mu^J \; , \;
K^m_{IJ} \,  \partial_m \mu^K = 2 \mu_{[I} \delta^K_{J]} \; , \;
K^m_{IJ} K_{mn}^{KL} = 8 g^{-2}  \mu_{[I} \delta^{[K}_{J]} \partial_n \mu^{L]} .
\end{equation}
Indices $I,J$ on $\mu^I$ and their derived quantities can be raised and lowered with $\delta_{IJ}$.

\subsection{Homogeneous nearly-K\"ahler structure} \label{subset:NKS6}

Particularising now to $n=6$, we will determine how the canonical, homogeneous nearly-K\"ahler structure on $S^6$ is inherited from $\mathbb{R}^7$ when the latter is endowed with its usual G$_2$--holonomy structure. More generally, recall that a nearly-K\"ahler six-dimensional manifold $M_6$ is an SU(3)-structure manifold, thus equipped with a real two-form ${\cal J}$ and a complex decomposable three-form $\Upomega$, which are $(1,1)$ and $(3,0)$ with respect to the natural complex structure defined by $\Upomega$. These forms are subject to the  following algebraic,
\begin{eqnarray} \label{SU3str}
\Upomega \wedge \bar \Upomega = -\tfrac{4i}{3} {\cal J} \wedge {\cal J} \wedge {\cal J} \neq 0 \; , \quad {\cal J} \wedge \Upomega = 0  \ ,
\end{eqnarray}
and differential relations
\begin{eqnarray} \label{SU3strDif}
d {\cal J} = 3 \,  \textrm{Re} \,  \Upomega   \; , \quad d \, \textrm{Im} \, \Upomega =  -2 \,  {\cal J} \wedge {\cal J} \ . 
\end{eqnarray}
The metric $ds^2(M_6)$  that this SU(3)--structure specifies is Einstein, normalised so that the Ricci tensor is $5$ times the metric, and its cone $C(M_6) = \mathbb{R}^+ \times M_6$ has G$_2$-holonomy. The closure of the  associative three-form $\psi$ and co-associative four-form $\tilde{\psi}$,
\begin{eqnarray} \label{AssociativeNK6}
\psi = r^2 dr\wedge {\cal J} + r^3 \, \textrm{Re} \, \Upomega \; , \qquad 
\tilde{\psi} = * \psi = \tfrac12 \, r^4 {\cal J} \wedge {\cal J} -r^3 dr \wedge \textrm{Im} \, \Upomega \; , 
\end{eqnarray}
on  $C(M_6)$ leads to the intrinsic-torsion constraints (\ref{SU3strDif}) on $M_6$. The Hodge dual in (\ref{AssociativeNK6}) is taken with respect to the canonical metric, $ds^2 =  dr^2 + r^2 ds^2(M_6)$, on  $C(M_6)$.

Now, when $M_6$ is taken to be $S^6 = $G$_2/$SU(3), the nearly-K\"ahler SU(3)-structure is invariant under the transitive action of G$_2$, and the structure can be specified in terms of the embedding (\ref{eqSn}) of $S^6$ into its cone $C(S^6) = \mathbb{R}^7$. In order to see this, we first relate the radial coordinate $r$ and the constrained coordinates $\mu^I$, $I=1, \ldots, 7$, to unconstrained coordinates $x^I = r \mu^I$ on $\mathbb{R}^7$, and then compare to (\ref{AssociativeNK6}) the generic expressions
 $\psi = \frac{1}{3!} \psi_{IJK} dx^I \wedge dx^J \wedge dx^K$
and
 $\tilde \psi = \frac{1}{4!} \tilde{\psi}_{IJKL} dx^I \wedge dx^J \wedge dx^K \wedge dx^L$. The nearly-K\"ahler structure $({\cal J}, \Upomega)$ on $S^6$ thus becomes related to the G$_2$--holonomy forms $\psi, \tilde{\psi}$ on $\mathbb{R}^7$ via 
\begin{eqnarray} \label{JOmegaintermsofmu}
{\cal J} = \tfrac12 \, \psi_{IJK} \,  \mu^I d\mu^J \wedge d\mu^K \; , \quad 
  \Upomega  = \tfrac16 \left( \psi_{JKL} -i \, \tilde{\psi}_{IJKL} \, \mu^I \right)  d \mu^J \wedge d\mu^K \wedge d\mu^L \; . 
\end{eqnarray}
These forms can be double-checked to satisfy the differential conditions (\ref{SU3strDif}). The metric $ds^2(S^6)$ that this nearly-K\"ahler structure specifies coincides with the round metric (\ref{RoundMet}) (with $g=1$). Its isometry is thus enhanced from G$_2$ to SO(7).

\bibliography{mIIAonS6refs}
\end{document}